\documentclass[trackchanges]{aastex701}

\usepackage{amsmath}
\usepackage{graphicx}
\usepackage{subcaption}
\usepackage{natbib}

\begin{document}

\title{A broadband search for coherent radio emission in cataclysmic variables}

\author[orcid=0000-0002-3968-2403]{Margaret E. Ridder}
\affiliation{U.S. Naval Research Laboratory, 4555 Overlook Ave SW, Washington, DC 20375}
\affiliation{University of Alberta Physics Dept., CCIS 4-183, Edmonton, AB, T6G 2E1, Canada}
\email[show]{margaret.e.ridder.ctr@us.navy.mil} 

\author[orcid=0000-0002-8456-1424]{Paul E. Barrett}
\affiliation{The George Washington University, Corcoran Hall Room 404C, 725 21st St NW, Washington, DC 20052}
\email{pebarrett@gwu.edu}

\author[orcid=0000-0003-3944-6109]{Craig O. Heinke}
\affiliation{University of Alberta Physics Dept., CCIS 4-183, Edmonton, AB, T6G 2E1, Canada}
\email{heinke@ualberta.ca}

\author[orcid=0000-0001-6682-916X]{Gregory R. Sivakoff}
\affiliation{University of Alberta Physics Dept., CCIS 4-183, Edmonton, AB, T6G 2E1, Canada}
\email{sivakoff@ualberta.ca}

\begin{abstract}
    Radio observations of cataclysmic variables have revealed a variety of behavior. From some systems, we see bright unpolarized radio flares occurring during dwarf nova outbursts, consistent with synchrotron emission from jets. In others, we see highly polarized emission, restricted in frequency, superimposed on a flat-spectrum continuum, suggesting a coherent emission process. Here, we present spectro-temporal analysis of 2--4 GHz and 8--12 GHz VLA observations of 6 cataclysmic variables. Our results show both broad- and narrow-band, highly polarized, variable radio emission. We suggest that this emission is consistent with electron-cyclotron maser emission or plasma radiation. This could be from an isolated emission region in the case of the narrow-band emission, or a region with varying magnetic field strength or density in the case of the broad-band emission. In one target, V2400 Oph, we see largely unpolarized emission changing on minute timescales, that may coincide with interactions between the white dwarf's magnetosphere and diamagnetic blobs.
\end{abstract}


\section{Introduction}

The nature of radio emission from cataclysmic variables (CVs) is not yet clear, with evidence for multiple mechanisms, including synchrotron emission from jets \citep{Coppejans20} and coherent plasma emission processes \citep{chanmugam1982,Barrett2017,barrett2020,ridder2025}. Wideband studies of their radio spectra, polarization, and variability are capable of constraining their radio emission mechanisms. Polarization measurements of a few dozen magnetic CVs suggest a high fractional circular polarization \citep{Barrett2017,barrett2020}, but often with large uncertainties. Currently published data on radio polarization of magnetic CVs have generally not tested whether the polarized flux is produced in short-duration flares, in frequency-limited spectral windows, or both. This gap in the literature makes it difficult to ascertain the exact radio emission mechanism responsible in magnetic CVs, also known as polars and intermediate polars (IPs). 

However, a few magnetic CVs have been shown to exhibit high variability and large fractional polarization. For example, historic observations of AM Her have seen it as bright as $\sim$10 mJy with $\sim$100\% circular polarization \citep{chanmugam1982}. Multiple mechanisms have been proposed for the flaring and quiescent emission, including gyrosynchrotron emission \citep{chanmugam1982} and electron-cyclotron maser emission \citep[ECME; ][]{Dulk83}. Later observations by \citet{barrett2020} uncovered high degrees of circular polarization in a number of CVs such as MR Ser and ST LMi, although many CVs in their sample did not have sufficient S/N to clearly determine their polarization fraction. 

Currently, there are only two confident detections of IPs at radio frequencies in the literature, V1323 Her \citep{Barrett2017} and V2400 Oph \citep{ridder2023}, but there are over two dozen known detections \citep{barrettprivcomm}. V2400 Oph is of particular interest due to its unusual accretion geometry. In normal IPs, matter is accreted from the donor star into an accretion disk around the WD. From there, ionized matter in the evacuated inner disk follows the WD's magnetic field lines onto the magnetic poles. However, V2400 Oph is thought to be a ``diskless'' IP that undergoes ``diamagnetic blob accretion,'' where blobs of material orbit around the WD and then interact with the WD's magnetic field, either threading onto magnetic field lines to allow accretion, or instead being accelerated and pushed away by the interaction, likely back into the Roche lobe of the donor \citep{langford_2022_blob}. Optical photometry revealed multiple features in V2400 Oph's light curve that could be associated with blobs interacting with the WD or with themselves. For example, low-frequency optical QPOs were attributed to high-energy blobs that migrated outwards, creating a beat frequency between the blob's orbit and the WD's spin. Higher-frequency ``flickering'' might be caused by blobs interacting viscously \citep{langford_2022_blob}. Jet production appears to require disk accretion down to small radii \citep{LyndenBell96}, so it seems unlikely that such an accretion mechanism would permit a jet.

Thermal free-free emission cannot explain the observed radio emission for many CVs \citep[e.g.][]{chanmugam1982}. Gyrosynchrotron emission may be common among CVs, but at a luminosity too dim to detect at the distances of most CVs with typical current radio observations. Taking QS Vir as an example, its gyrosynchrotron flux density \citep[$\sim$500 $\mu$Jy; ][]{ridder2025} if it were placed at 500 pc would be 5 $\mu$Jy, which would need several hours on source to detect. The emission detectable at larger distances should be coherent, given that it is intense and, in the cases where polarimetry is available, highly polarized. For example, V2400 Oph is 15 times more radio luminous than the brightest isolated M dwarf stars observed by \citet{yiu2024}. Uncovering what powers V2400 Oph's radio emission is crucial to understanding the generation of radio emission in magnetic CVs, as well as our understanding of coherent plasma emission processes in flare stars by examining a CV's fast-rotating donor.

In this paper, we present the results from spectral, polarization, and variability analyses on a VLA sample of mostly magnetic CVs in X and S bands, with prior radio observations reported by \citet{barrett2020}: the polars MR Ser, ST LMi, UZ For, and EF Eri, the IP V2400 Oph, and the old nova V603 Aql \citep[also studied by][]{coppejans2015}.

The frequency of emission in ECME probes the magnetic field strength of the emission region and constrains its location within binaries. The \textit{S}- and \textit{X}-band observations of these CVs probe regions where the magnetic field would be several kG (assuming the emission is at the fundamental frequency), which may include the lower corona of the donor star \citep{dulk1985}. An interesting prospect is the possibility of a flux tube connecting the donor and WD in a similar manner to the Jupiter-Io system. A gradually varying magnetic field strength would result multiple fundamental frequencies of emission in the case of ECME and could result in a broad-spectrum, highly polarized signal \citep[see e.g. ][]{treumann2006}.

\section{Observations and analysis}

Table \ref{tab: targets} gives positions and distances of the 6 CVs studied in this work using the Jansky Very Large Array (VLA). They were each observed twice in \textit{X}-band (8--12 GHz) in 1-hour blocks (PI Barrett; ObsIDs 17B-144, 19A-166). All but one source, V603 Aql, were also observed in \textit{S}-band (2--4 GHz; PI Ridder, ObsID 25A-404).

\begin{table*} 
    \setlength{\tabcolsep}{1.5pt}
    \centering
    \caption{The list of targets. Coordinates and distances are from \textit{Gaia} EDR3 \citep{bailer-jones_2021}.}
    \label{tab: targets}    
    \begin{tabular}{c c c c c}
    \hline \hline
         Name & RA & Dec & Distance & Classification \\
          & (deg) & (deg) & (pc) & \\
        \hline
         EF Eri  & 48.555884 & $-$22.595499 & $160.0\pm4.0$ & Polar \\  
         UZ For  & 53.869471 & $-$25.739386 & $238.0\pm3.0$ & Polar \\  
         ST LMi & 166.415697 & 25.107827 & $114.2\pm0.9$ & Polar \\ 
         MR Ser & 238.196448 & 18.941621 & $131.2^{+0.5}_{-0.7}$ & Polar \\
         V2400 Oph & 258.151755 & $-$24.245759 & $700.0^{+10.0}_{-11.0}$ & IP \\
         V603 Aql  & 282.227705 & 0.584085 & $315.0^{+3.0}_{-4.0}$ & Old nova \\ 
         \hline
    \end{tabular}
\end{table*}

\begin{table*} 
    \setlength{\tabcolsep}{1.5pt}
    \centering
    \caption{The list of primary, secondary, and (for S) leakage calibrators, and VLA configurations, used for each source.}
    \label{tab: calibrators x}    
    \begin{tabular}{c | c c c | c c c c}
    \hline \hline
         Name & Primary  & Secondary  &  VLA Config & Primary  & Secondary  & Leakage  & VLA Config \\
         \hline
         & \multicolumn{3}{c}{X (8-12 GHz)} & \multicolumn{4}{|c}{S (2-4 GHz)} \\
        \hline
         EF Eri  & 3C138 & J0329$-$2357 & B & 3C138 & J0240-2309 & J0319+4130 & D \\  
         UZ For  & 3C138 & J0329$-$2357 & B & 3C138 & J0416-1851 & J0319+4130 & D \\  
         ST LMi & 3C286 & J1125+2610 & B & 3C286 & J1125+2610 & J1407+2827 & D \\ 
         MR Ser & 3C286 & J1520+2016 & B & 3C286 & J1553+1256 & J1407+2827 & D \\
         V2400 Oph & 3C286 & J1744$-$3116 & B & 3C286 & J1751-2524 & J1407+2827 & C \\
         V603 Aql  & 3C286 & J1856+0610 & B & \nodata & \nodata & \nodata & \nodata \\ 
         \hline
    \end{tabular}
\end{table*}


Parallel-hand calibration, as well as automatic and manual flagging, was conducted with the Common Astronomy Software Applications \citep[CASA;][]{casa} pipeline version 6.5.4. We did not perform self-calibration on these observations. A summary of the primary (flux/bandpass/polarization angle) calibrators and secondary (complex gain) calibrators is given in Table \ref{tab: calibrators x}, including leakage calibrators for the \textit{S}-band observations. Leakage calibration was not conducted as part of the \textit{X}-band observations. While this raises the noise for Stokes V X-band measurements, the leakage moduli for the VLA have been shown to be around 5\% \citep{vla_memo201}. We assume this value as a lower limit to our 1$\sigma$ uncertainty for X-band Stokes V in Table \ref{tab: x-band flux}. 

2--4 GHz and 8--12 GHz images in Stokes I, Q, U, and V were produced by WSClean \citep{wsclean}. In addition to imaging the full band, we also produced images with the bandwidth of a VLA spectral window (128 MHz). Each target scan from the observations was split into roughly 10 time bins to produce light curves and dynamic SEDs (shown in the next section), to search for variable emission features and potential flares. Often the spectral windows nearest to 2 GHz are contaminated with radio frequency interference (RFI). For data that is not flagged, this will manifest as large uncertainties (e.g. the SEDs in Fig. \ref{fig: stlmi 25a} and Fig. \ref{fig: mrser 25a}).

Assuming each target was a point source, each was fit with a 2D Gaussian using their \textit{Gaia} EDR3 positions \citep{bailer-jones_2021}. The peak intensity normalization and source position were left as free parameters in the full-band Stokes I images. This position was fixed for full band Stokes V images and all other images produced for the SEDs and light curves, while leaving the peak intensity normalization free.

\section{Results}\label{res p3}

The Stokes I and Stokes V fluxes as well as circular polarization fractions in \textit{X}- and \textit{S}-band are shown in Tables \ref{tab: x-band flux} and \ref{tab: s-band flux}, respectively. MR Ser, ST LMi, and V2400 Oph were detected at 2--4 GHz in Stokes I and (except for one observation of V2400 Oph) in Stokes V. All sources in our sample were detected at 8--12 GHz in Stokes I, but only EF Eri, MR Ser, and V603 Aql were detected in Stokes V at $\geq3\sigma$. No sources were detected in Stokes Q and U. The full-band SEDs, light curves, and dynamic SEDs are presented in Figs.\ \ref{fig: eferi 17b} to \ref{fig: v603 17b}. Alternative, 2-dimensional versions of the  1-dimensional dynamic SEDs therein are available in an online repository\footnote{\url{https://doi.org/10.5281/zenodo.18460656}}. Uncertainties presented in this section are the RMS of a background region, defined as an annulus with 35--100$\times$ the area of a circle with the same radius as the beam major axis (depending on how crowded the field). The inner radius of this annulus was equal to the beam major axis.

\begin{table*} 
    \setlength{\tabcolsep}{1.5pt}
    \centering
    \caption{The \textit{X}-band (8--12 GHz) flux in Stokes I and Stokes V for each target in our sample. Stokes V fluxes were produced by a forced fit at the Stokes I position. We impose a lower limit on the polarization fraction uncertainty of 5\% due to the absence of a leakage calibrator. Sources where the uncertainty was raised to this level are marked with an asterisk. The polarization handedness is denoted either left-circularly polarized (LCP) for negative values of Stokes V or right-circularly polarized (RCP) for positive values, following the IAU convention.}
    \label{tab: x-band flux}    
    \begin{tabular}{c c c c c}
    \hline \hline
         Name & Date & Stokes I flux & Stokes V flux & Pol. fraction \\
          & & ($\mu$Jy) & ($\mu$Jy) & (\%)\\
        \hline 
         EF Eri & 2017 September 28 & $157\pm6$ & $-124\pm6$ & $79\pm5$ LCP \\ 
         EF Eri & 2017 October 2 & $67\pm5$ & $-48\pm5$ & $71\pm9$ LCP \\  
         UZ For & 2017 October 17 & $30\pm5$ & $<\left|15\right|$ & $<\left|50\right|$ \\ 
         UZ For & 2017 October 19 & $35\pm5$ & $<\left|15\right|$ & $<\left|43\right|$ \\ 
         ST LMi & 2017 November 1 & $80\pm5$ & $<\left|15\right|$ & $<\left|19\right|$ \\ 
         ST LMi & 2017 November 5 & $40\pm5$ & $<\left|15\right|$ & $\mathbf<\left|38\right|$ \\ 
         MR Ser & 2017 October 11 & $42\pm6$ & $<\left|18\right|$ & $<\left|43\right|$ \\ 
         MR Ser & 2017 November 25 & $308\pm8$ & $256\pm8$ & $83\pm5$* RCP\\
         V2400 Oph & 2019 April 25 & $266\pm8$ & $<\left|21\right|$ & $<\left|8\right|$ \\
         V2400 Oph & 2019 May 11 & $320\pm8$ & $<\left|9\right|$ & $<\left|5\right|$* \\
         V603 Aql & 2017 December 18 & $90\pm10$ & $30\pm10$ & $33\pm12$ RCP \\ 
         V603 Aql & 2018 January 25 & $71\pm7$ & $<\left|18\right|$ & $<\left|25\right|$ \\ 
         \hline
    \end{tabular}
\end{table*}

\begin{table*} 
    \setlength{\tabcolsep}{1.5pt}
    \centering
    \caption{The \textit{X}-band (8--12 GHz) 3$\times$RMS upper limits on linear polarization.}
    \label{tab: x-band lin pol}    
    \begin{tabular}{c c c c}
    \hline \hline
         Name & Date & Lin. pol. flux & Pol. fraction \\
          & & ($\mu$Jy) & (\%)\\
        \hline 
         EF Eri & 2017 September 28 & $<$23 & $<$15 \\ 
         EF Eri & 2017 October 2 & $<$22 & $<$33 \\  
         UZ For & 2017 October 17 & $<$24 & $<$80 \\ 
         UZ For & 2017 October 19 & $<$22 & $<$63 \\ 
         ST LMi & 2017 November 1 & $<$21 & $<$26 \\ 
         ST LMi & 2017 November 5 & $<$21 & $<$53 \\ 
         MR Ser & 2017 October 11 & $<$25 & $<$60 \\ 
         MR Ser & 2017 November 25 & $<$26 & $<$8 \\
         V2400 Oph & 2019 April 25 & $<$34 & $<$13 \\ 
         V2400 Oph & 2019 May 11 & $<$37 & $<$12 \\
         V603 Aql & 2017 December 18 & $<$39 & $<$43 \\ 
         V603 Aql & 2018 January 25 & $<$27 & $<$38 \\ 
         \hline
    \end{tabular}
\end{table*}

\begin{table*} 
    \centering
    \caption{The \textit{S}-band (2--4 GHz) fluxes in Stokes I and Stokes V for each target in our sample. Upper limits on the Stokes I and V fluxes for EF Eri, UZ For, and V2400 Oph are equal to $3\times$RMS. The absolute value of the polarization fraction is quoted for these targets because there is no directionality in the error estimates.}
    \label{tab: s-band flux}    
    \begin{tabular}{c c c c c}
    \hline \hline
         Name & Date & Stokes I flux & Stokes V flux & Pol. fraction \\
          & & ($\mu$Jy) & ($\mu$Jy) & (\%)\\
        \hline
         EF Eri & 2025 March 19 & $<$100 & $<\left|26\right|$ & $<\left|26\right|$ \\ 
         UZ For & 2025 March 22 & $<$39 & $<\left|21\right|$ & $<\left|54\right|$ \\ 
         ST LMi & 2025 March 23 & $160\pm25$ & $-87\pm6$ & $54\pm9$ LCP \\  
         MR Ser & 2025 April 5 & $87\pm16$ & $59\pm5$ & $67\pm14$ RCP \\ 
         V2400 Oph & 2025 August 9 & $519\pm15$ & $<\left|18\right|$ & $<\left|3\right|$ \\ 
         V2400 Oph & 2025 August 15 & $637\pm15$ & $247\pm7$ & $39\pm1$ RCP \\
         \hline
    \end{tabular}
\end{table*}

\begin{table*} 
    \centering
    \caption{The \textit{S}-band (2--4 GHz) 3$\times$RMS upper limits on linear polarization.}
    \label{tab: s-band lin pol}    
    \begin{tabular}{c c c c}
    \hline \hline
         Name & Date & Lin. pol. flux & Pol. fraction \\
          & & ($\mu$Jy) & (\%)\\
        \hline
         EF Eri & 2025 March 19 & $<$61 & $<$61 \\ 
         UZ For & 2025 March 22 & $<$30 & $<$77 \\ 
         ST LMi & 2025 March 23 & $<$30 & $<$19 \\  
         MR Ser & 2025 April 5 & $<$31 & $<36$ \\ 
         V2400 Oph & 2025 August 9 & $<$28 & $<$5 \\ 
         V2400 Oph & 2025 August 15 & $<$28 & $<$4 \\
         \hline
    \end{tabular}
\end{table*}

\subsection{EF Eri}


\begin{figure}[ht!]
    \centering

    \begin{subfigure}[b]{0.32\textwidth}
        \centering
        \includegraphics[width=\textwidth]{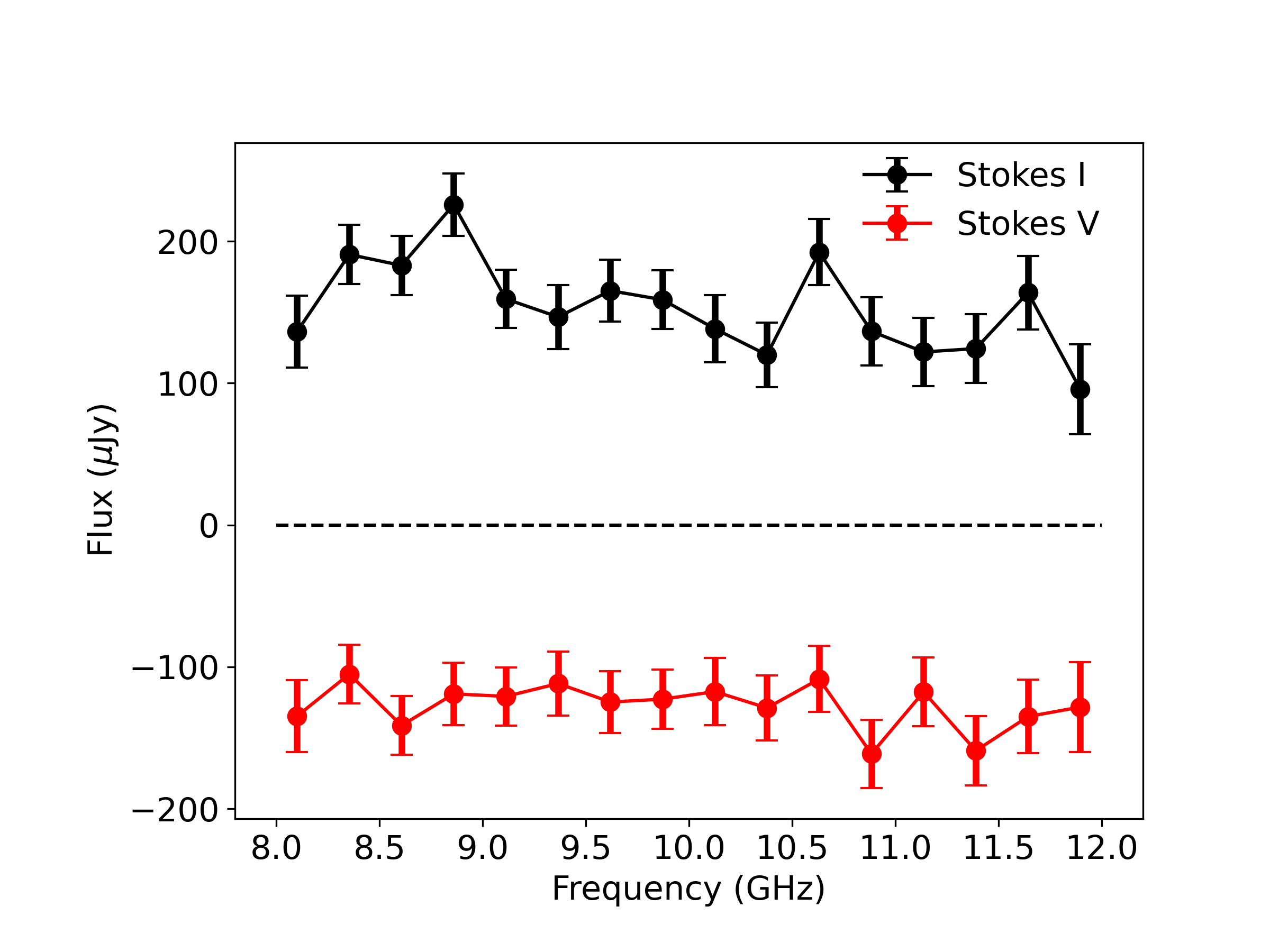}
        \caption{The 8--12 GHz SED of EF Eri on September 28.}
        \label{fig:1a}
    \end{subfigure}
    \hspace{0.2em}
    \begin{subfigure}[b]{0.32\textwidth}
        \centering
        \includegraphics[width=\textwidth]{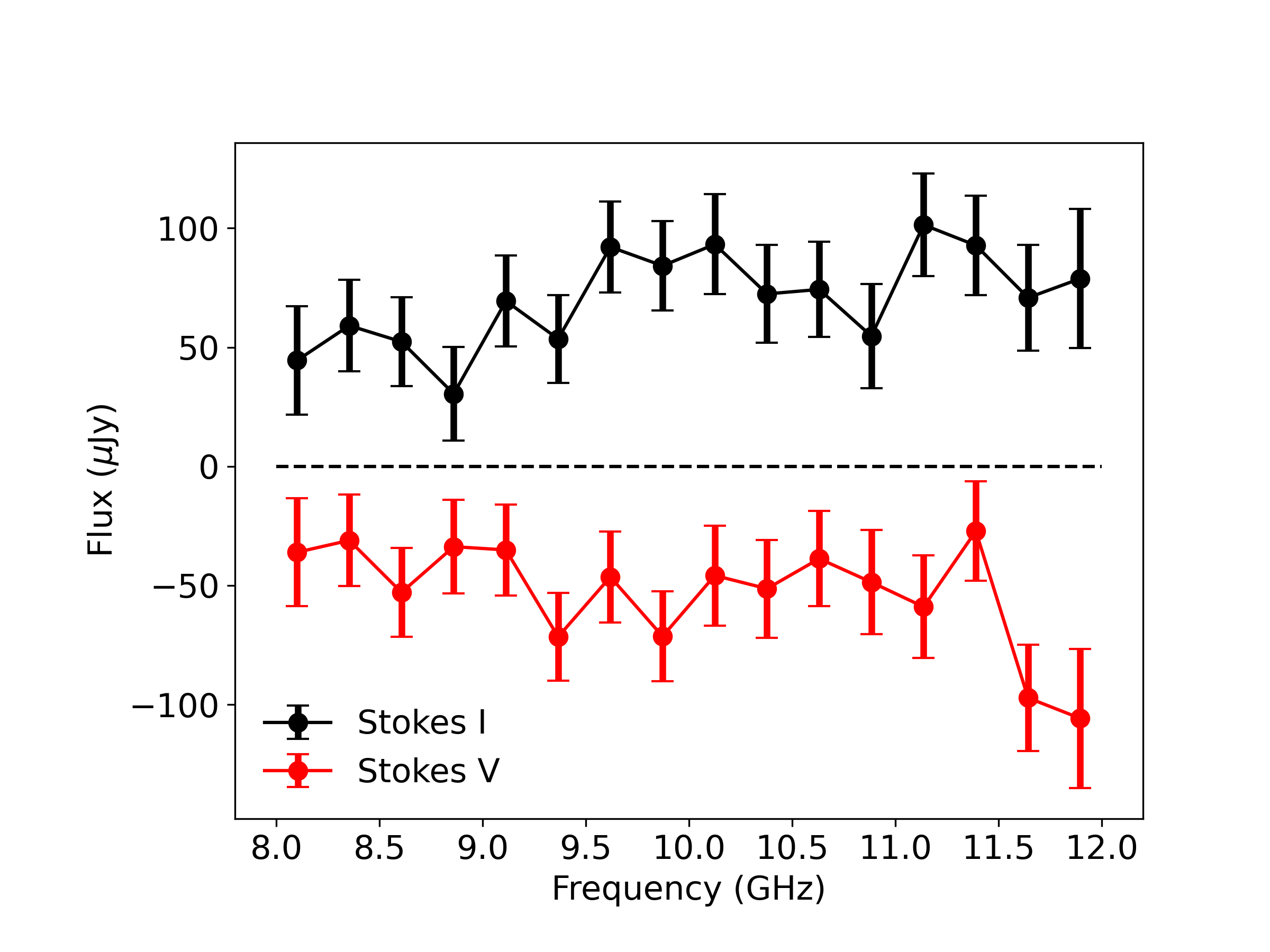}
        \caption{The 8--12 GHz SED of EF Eri on October 2.}
        \label{fig:1b}
    \end{subfigure}

    \begin{subfigure}[b]{0.32\textwidth}
        \centering
        \includegraphics[width=\textwidth]{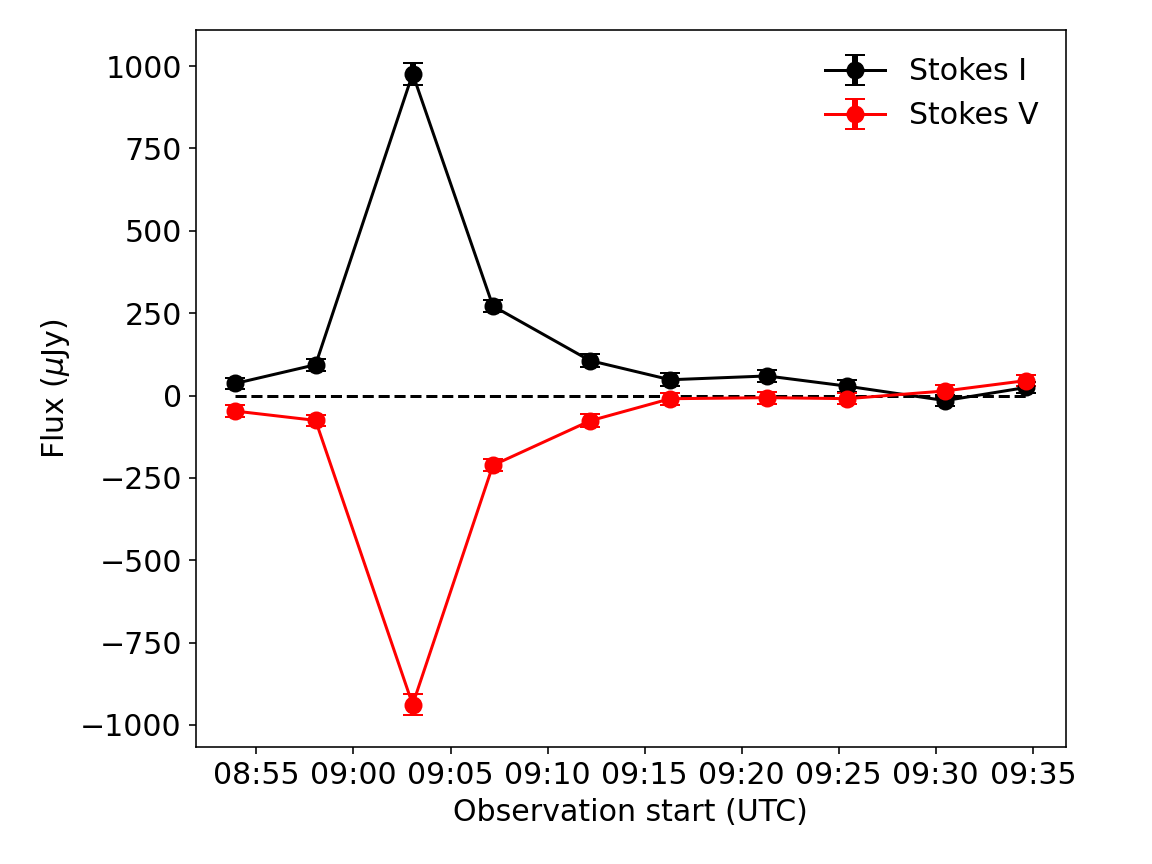}
        \caption{The 8--12 GHz light curve of EF Eri on September 28.}
        \label{fig:2a}
    \end{subfigure}
    \hspace{0.2em}
    \begin{subfigure}[b]{0.32\textwidth}
        \centering
        \includegraphics[width=\textwidth]{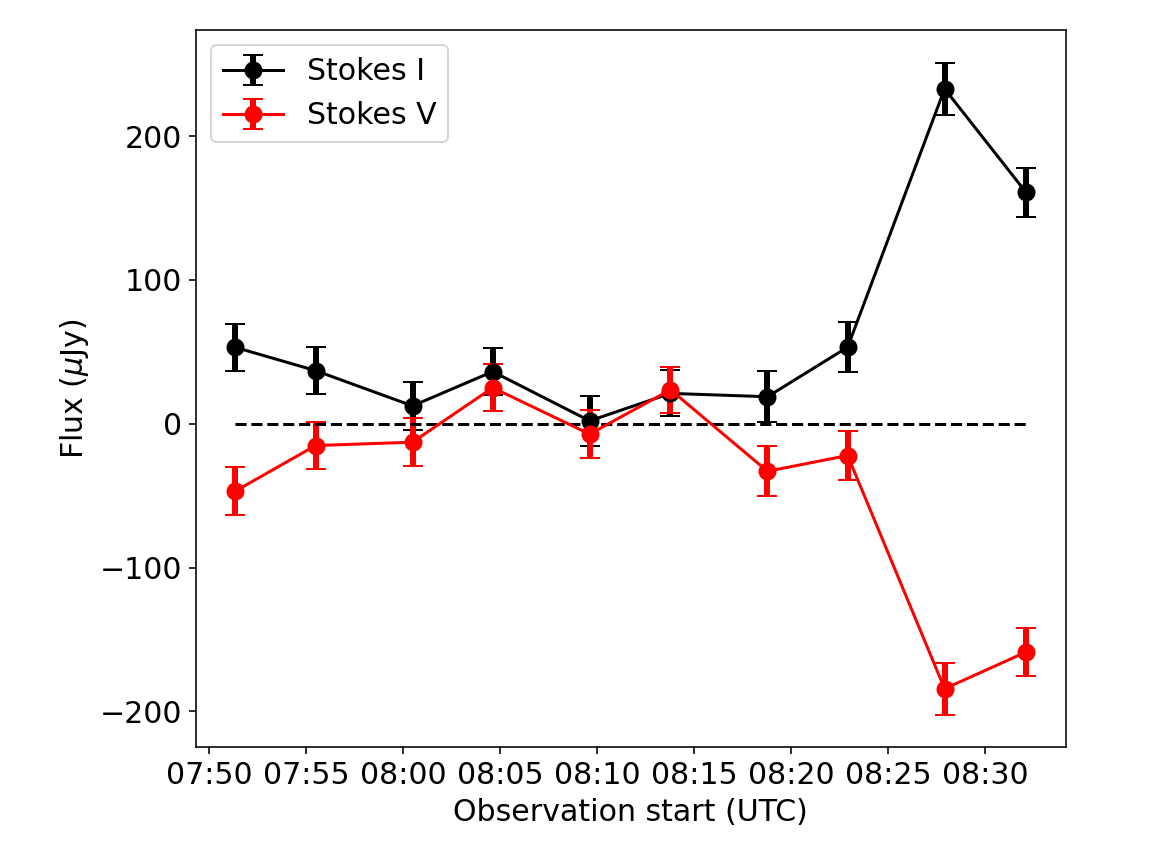}
        \caption{The 8--12 GHz light curve of EF Eri on October 2.}
        \label{fig:2b}
    \end{subfigure}

    \begin{subfigure}[b]{0.32\textwidth}
        \centering
        \includegraphics[width=\textwidth]{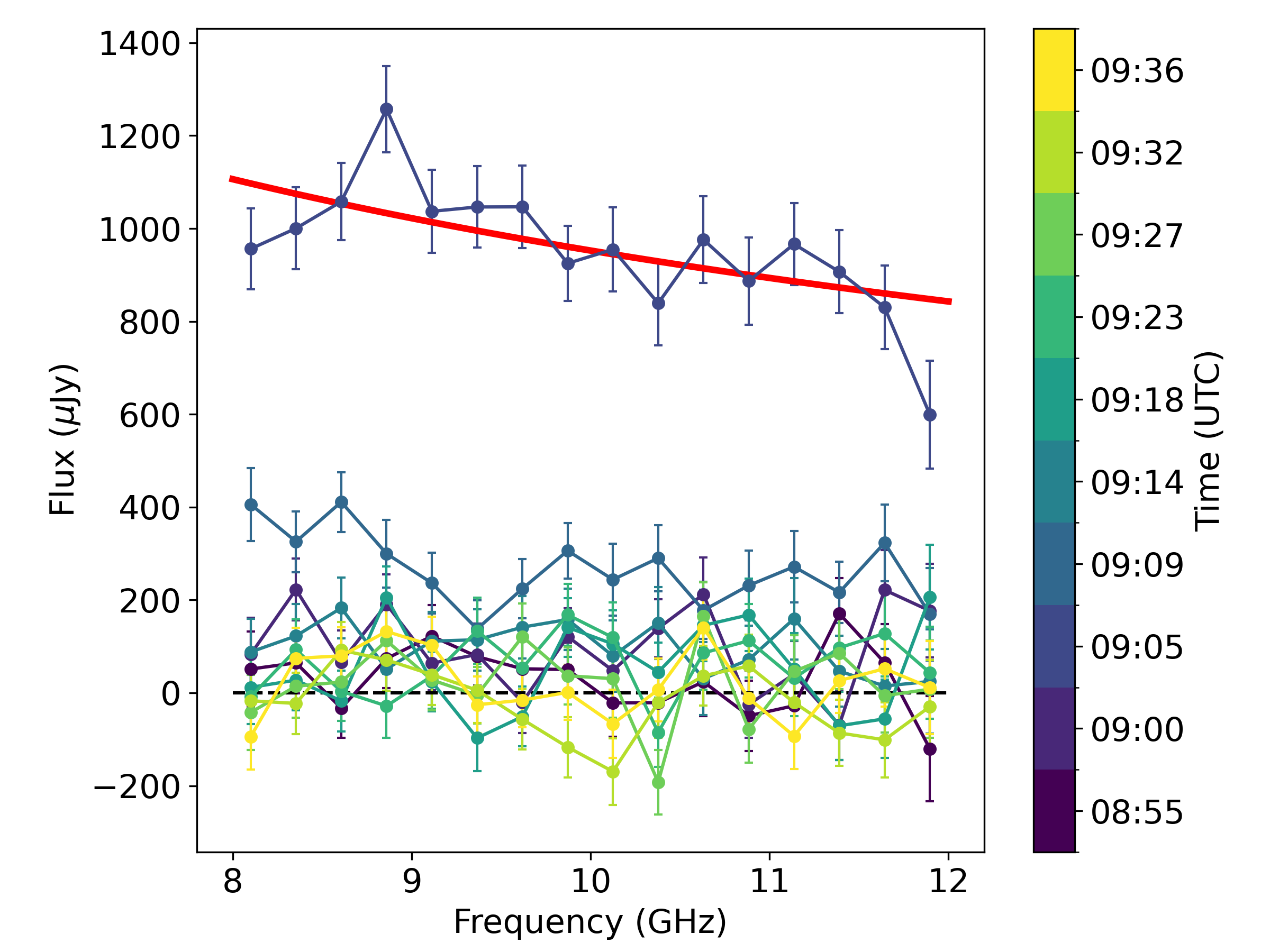}
        \caption{The 8--12 GHz dynamic SED of EF Eri in Stokes I on September 28.}
        \label{fig:3a}
    \end{subfigure}
    \hspace{0.2em}
    \begin{subfigure}[b]{0.32\textwidth}
        \centering
        \includegraphics[width=\textwidth]{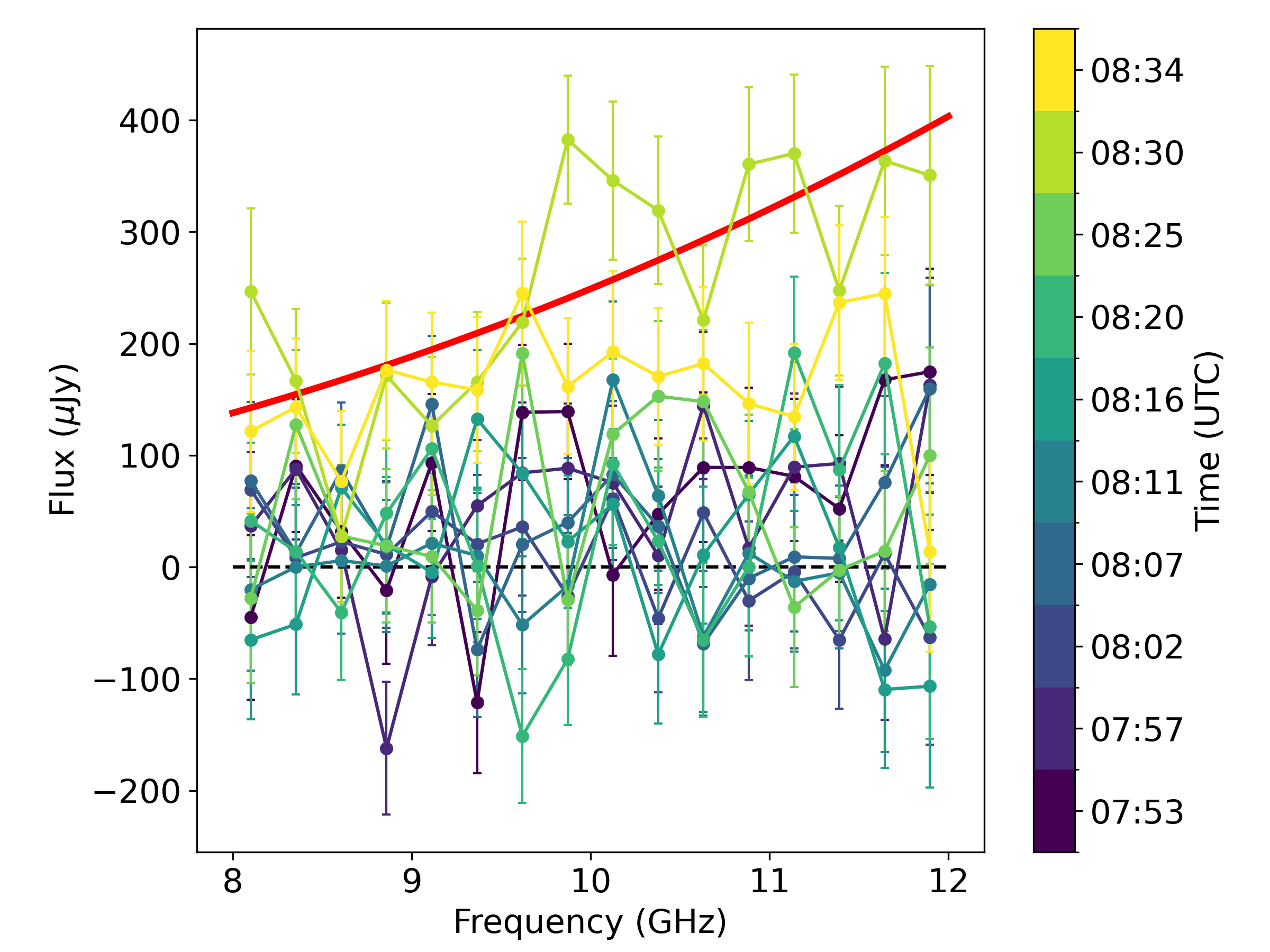}
        \caption{The 8--12 GHz dynamic SED of EF Eri in Stokes I on October 2.}
        \label{fig:3b}
    \end{subfigure}

    \begin{subfigure}[b]{0.32\textwidth}
        \centering
        \includegraphics[width=\textwidth]{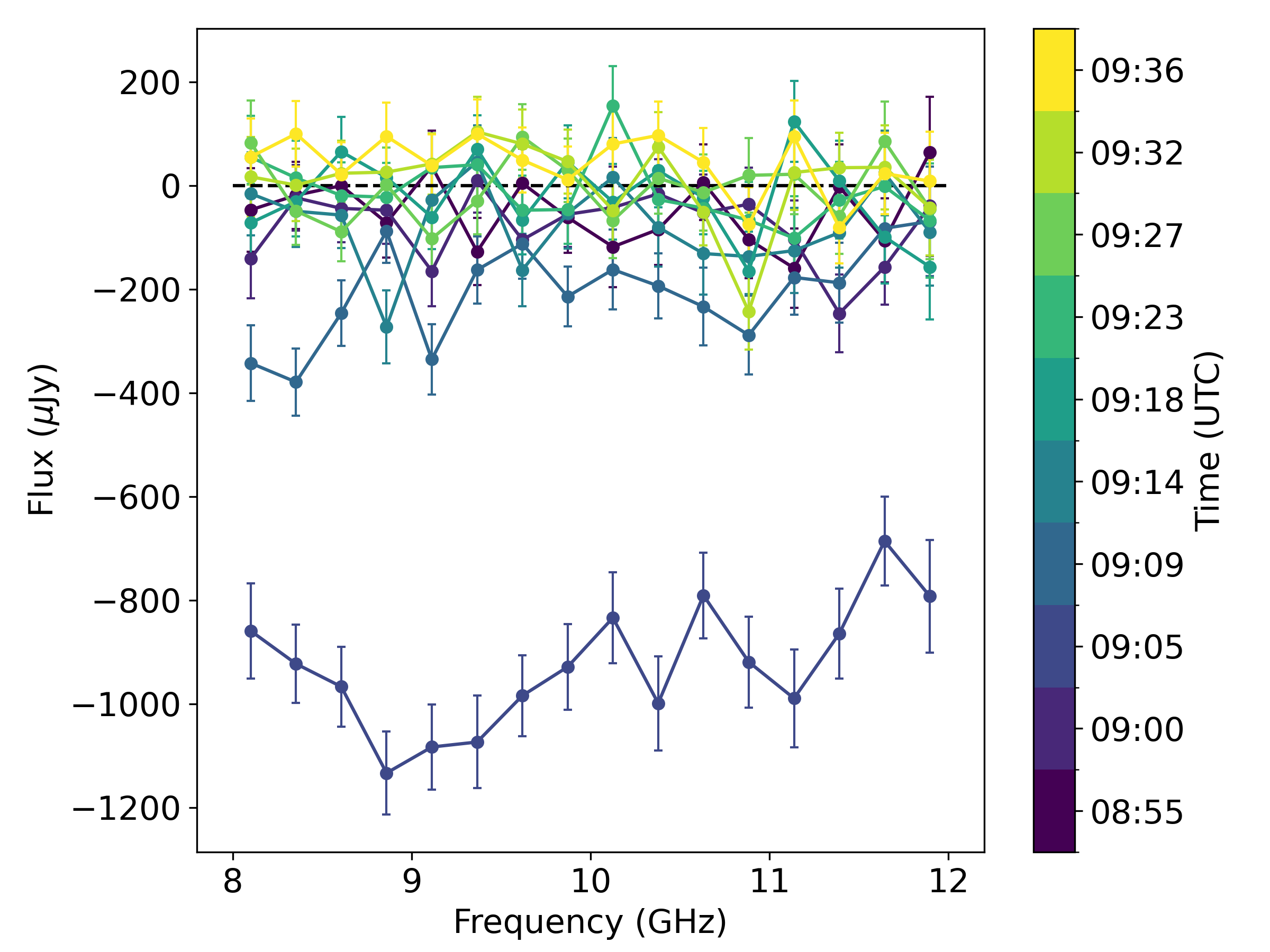}
        \caption{The 8--12 GHz dynamic SED of EF Eri in Stokes V on September 28.}
        \label{fig:4a}
    \end{subfigure}
    \hspace{0.2em}
    \begin{subfigure}[b]{0.32\textwidth}
        \centering
        \includegraphics[width=\textwidth]{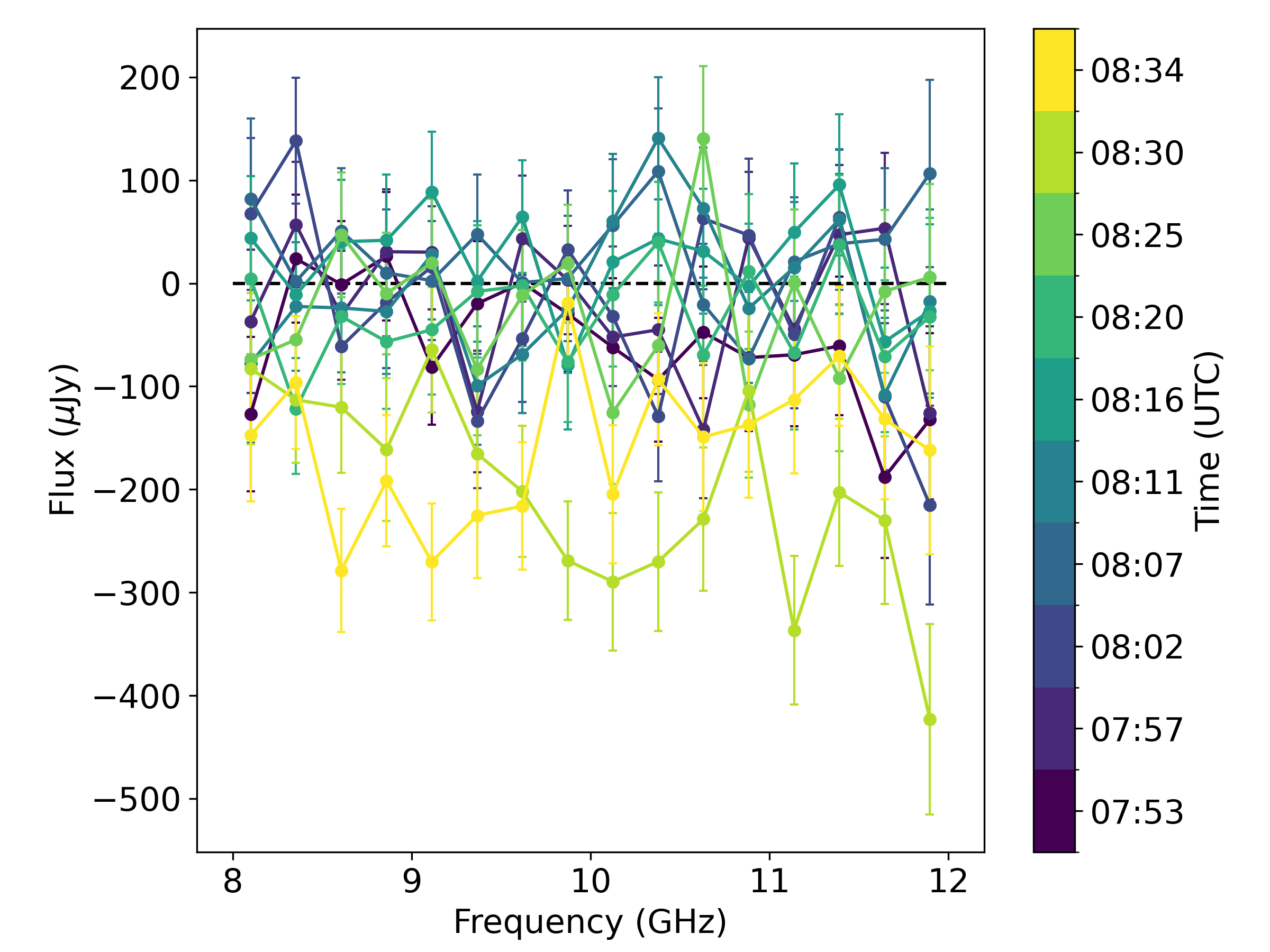}
        \caption{The 8--12 GHz dynamic SED of EF Eri in Stokes V on October 2.}
        \label{fig:4b}
    \end{subfigure}

    \caption{The 8--12 GHz data for EF Eri on 2017 September 28 (\textit{left column}) and 2017 October 2 (\textit{right column}). The top panels show the full-band SEDs, the second panels show the full-band light curves, the third panels show the Stokes I dynamic SEDs, and the bottom panels show the Stokes V dynamic SEDs. One flare was detected in each 8--12 GHz observation of EF Eri. Both produced broadband LCP emission. The SEDs at the peak of each flare were fit in Stokes I to calculate the spectral index, shown in red on panels \textit{(e)} and \textit{(f)}. These are $\alpha=-0.7\pm0.2$ for September 28 and $\alpha=2.7\pm0.8$ for October 2.}
    \label{fig: eferi 17b}
\end{figure}

EF Eri was detected in both epochs in Stokes I and Stokes V. Two minute-scale flares (see Fig.\ \ref{fig: eferi 17b}) were detected, one in each epoch, each corresponding to an increase in circular polarization. At the peak of the flares, the polarization fraction was $96\pm5$\% LCP on 2017 September 28 and $79\pm5$\% LCP on 2017 October 2. In the corresponding dynamic SEDs, a broadband spectral feature appears at the peak of both flares (shown in red in Fig. \ref{fig: eferi 17b}), which has a slightly negative fit to the spectral index in the first observation ($\alpha=-0.7\pm0.2$), and a somewhat positive index in the second($\alpha=2.7\pm0.8$).

EF Eri was not detected in the 2--4 GHz observations on 2025 March 19. The $3\times$RMS upper limit on this observation, 96 $\mu$Jy, is consistent with the quiescent emission (that which is outside the flare) observed at 8--12 GHz.

\subsection{UZ For}


\begin{figure}[ht!]
    \centering

    \begin{subfigure}[b]{0.32\textwidth}
        \centering
        \includegraphics[width=\textwidth]{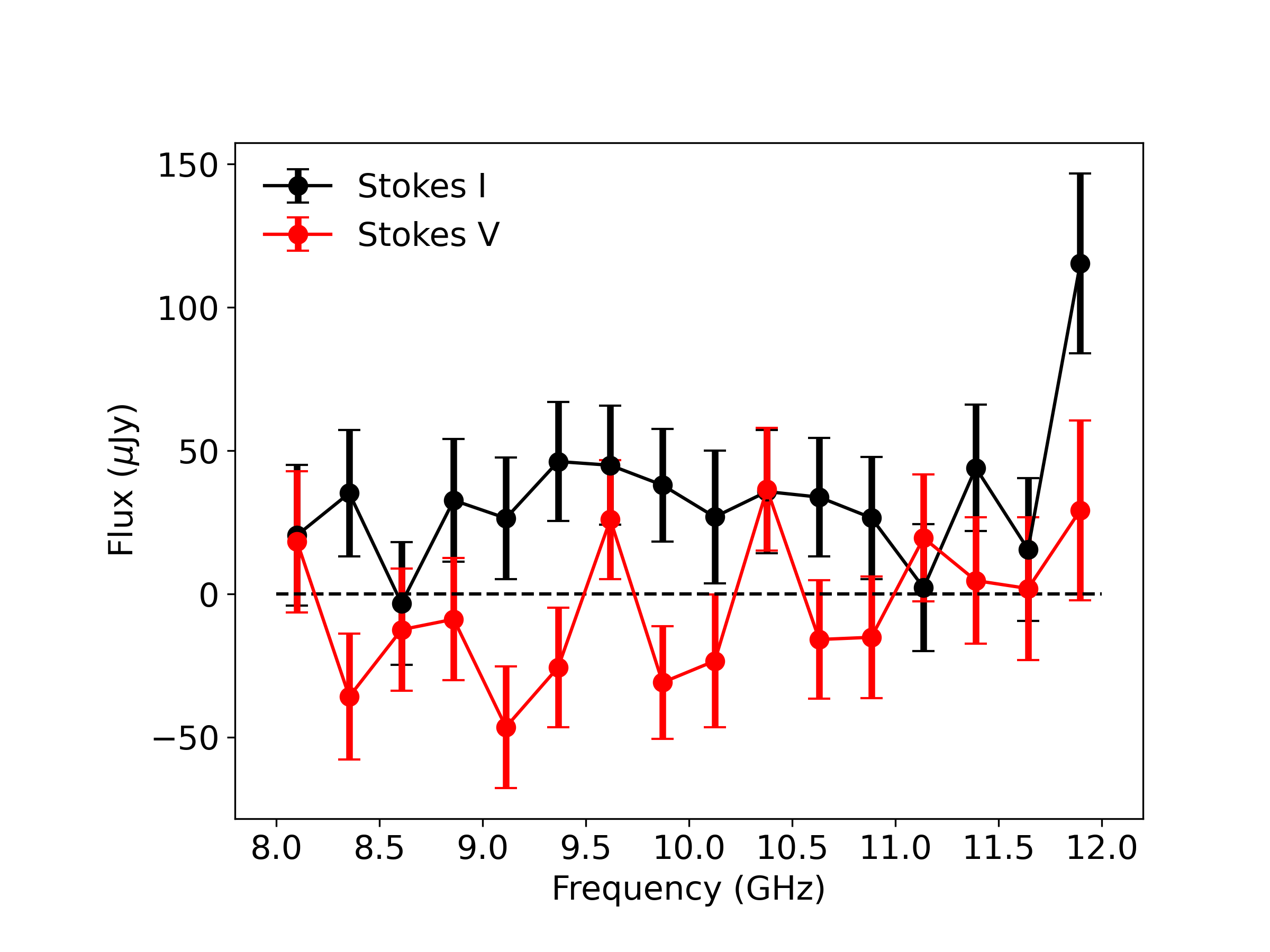}
        \caption{The 8--12 GHz SED of UZ For on October 17.}
        \label{fig:1a}
    \end{subfigure}
    \hspace{0.2em}
    \begin{subfigure}[b]{0.32\textwidth}
        \centering
        \includegraphics[width=\textwidth]{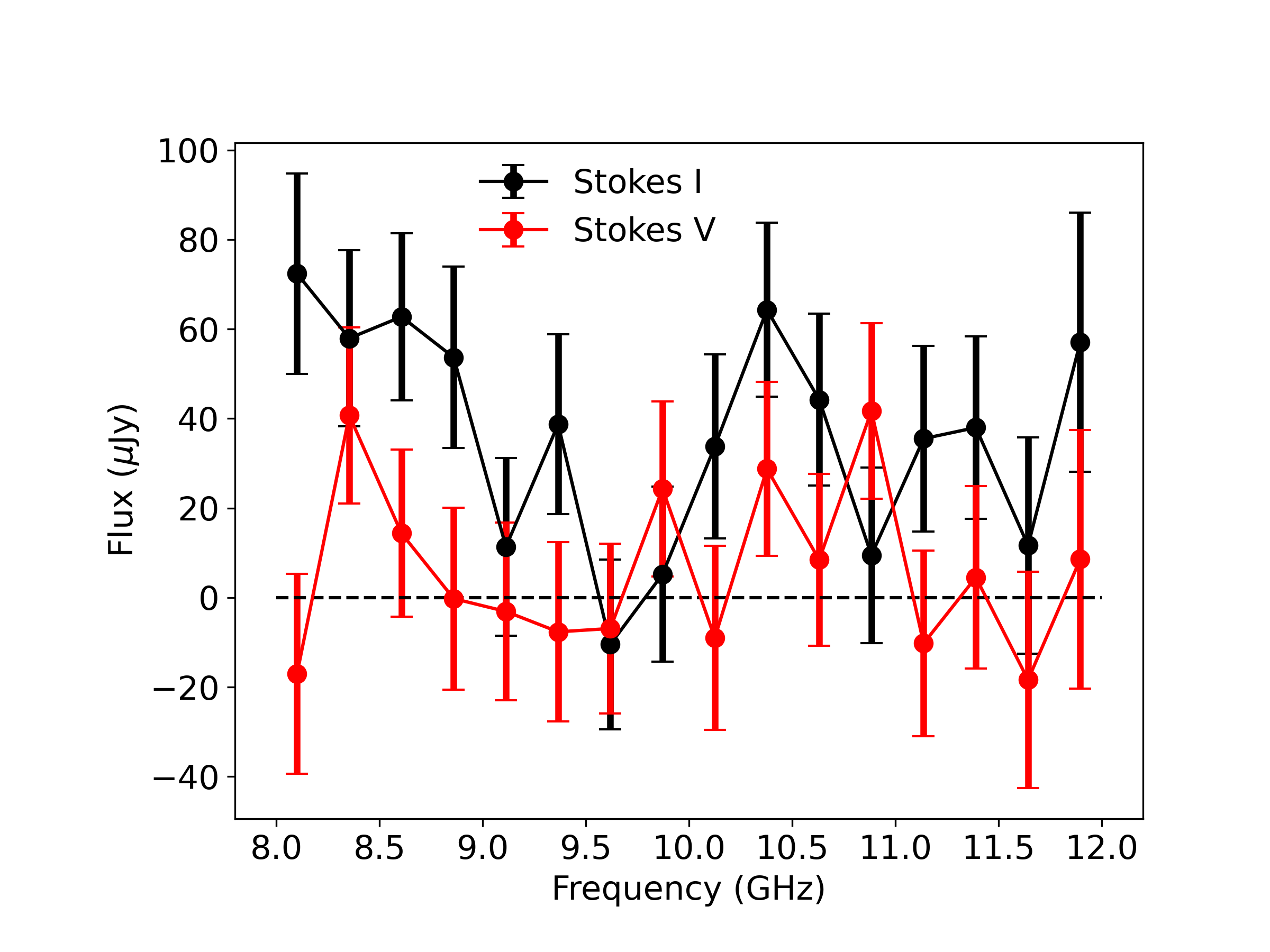}
        \caption{The 8--12 GHz SED of UZ For on October 19.}
        \label{fig:1b}
    \end{subfigure}

    \begin{subfigure}[b]{0.32\textwidth}
        \centering
        \includegraphics[width=\textwidth]{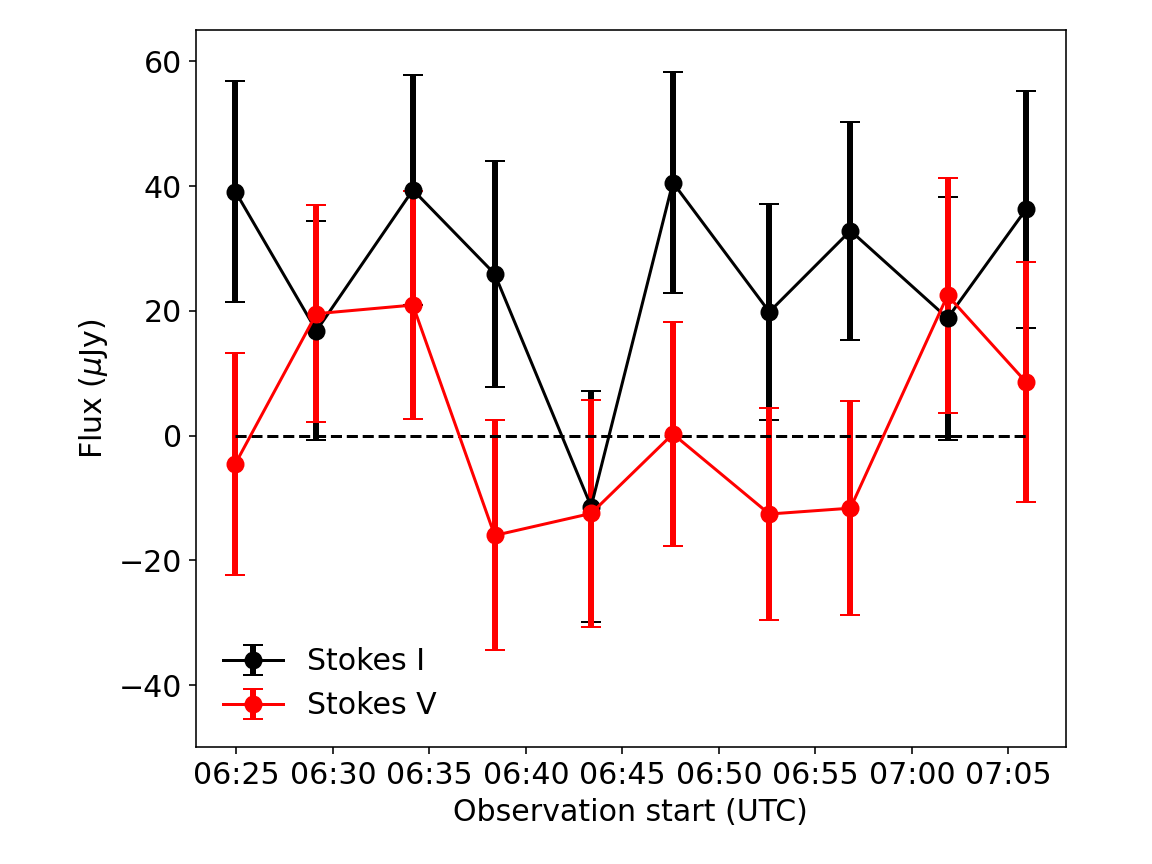}
        \caption{The 8--12 GHz light curve of UZ For on October 17.}
        \label{fig:2a}
    \end{subfigure}
    \hspace{0.2em}
    \begin{subfigure}[b]{0.32\textwidth}
        \centering
        \includegraphics[width=\textwidth]{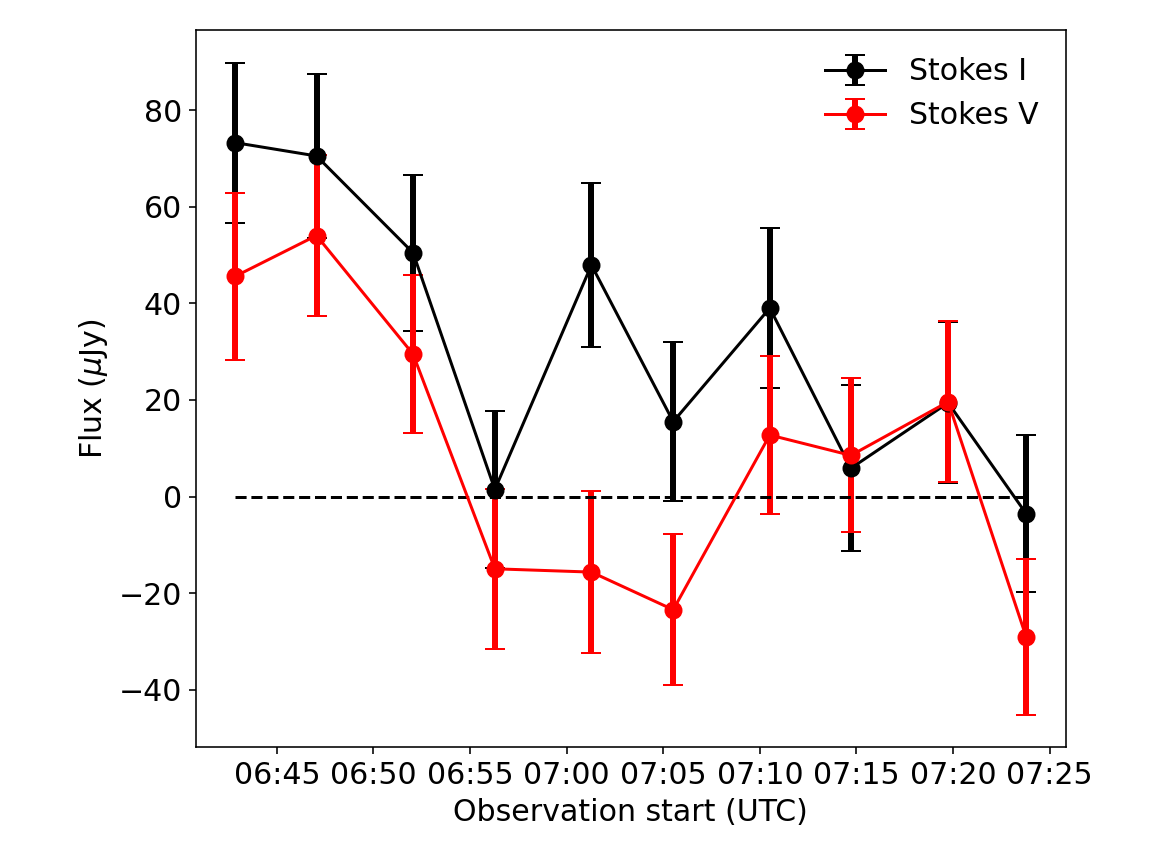}
        \caption{The 8--12 GHz light curve of UZ For on October 19.}
        \label{fig:2b}
    \end{subfigure}

    \caption{The 8--12 GHz data of UZ For on 2017 October 17 (\textit{left column}) and 2017 October 19 (\textit{right column}). The top panel shows the full-band SEDs, while the bottom panel shows the full-band light curves. There are neither clear flares nor circular polarization detected.}
    \label{fig: uzfor 17b}
\end{figure}

While UZ For, the only eclipsing system in our sample, was detected in both epochs in Stokes I, it was undetected in Stokes V, and there are no obvious emission features in either SED (Fig.\ \ref{fig: uzfor 17b}). The light curve for 2017 October 19, suggests some amount of variation that corresponds with increasing circular polarization, but the measurement is marginal. Averaging over the first three and the following three data points, UZ For decreases in flux by about a factor of 2. This fractional change is similar to what was observed in QS Vir during previous observations \citep{ridder2025}. The S/N is unfortunately too low for further analysis, including the examination of its dynamic SEDs.

UZ For was not detected in the 2--4 GHz observations on 2025 March 22.

\subsection{ST LMi}


\begin{figure}[ht!]
    \centering

    \begin{subfigure}[b]{0.32\textwidth}
        \centering
        \includegraphics[width=\textwidth]{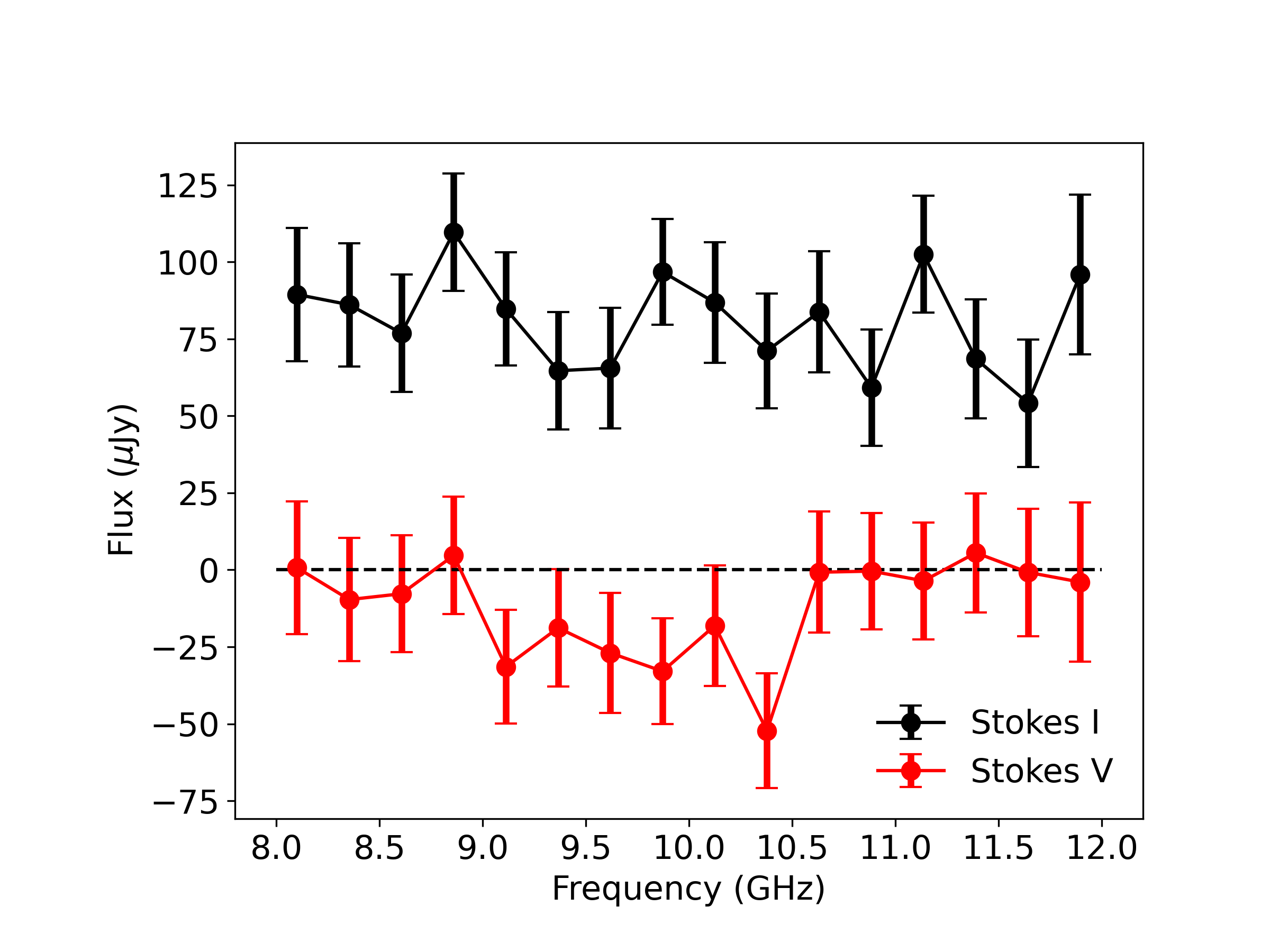}
        \caption{The 8--12 GHz SED of ST LMi on November 1.}
        \label{fig:1a}
    \end{subfigure}
    \hspace{0.2em}
    \begin{subfigure}[b]{0.32\textwidth}
        \centering
        \includegraphics[width=\textwidth]{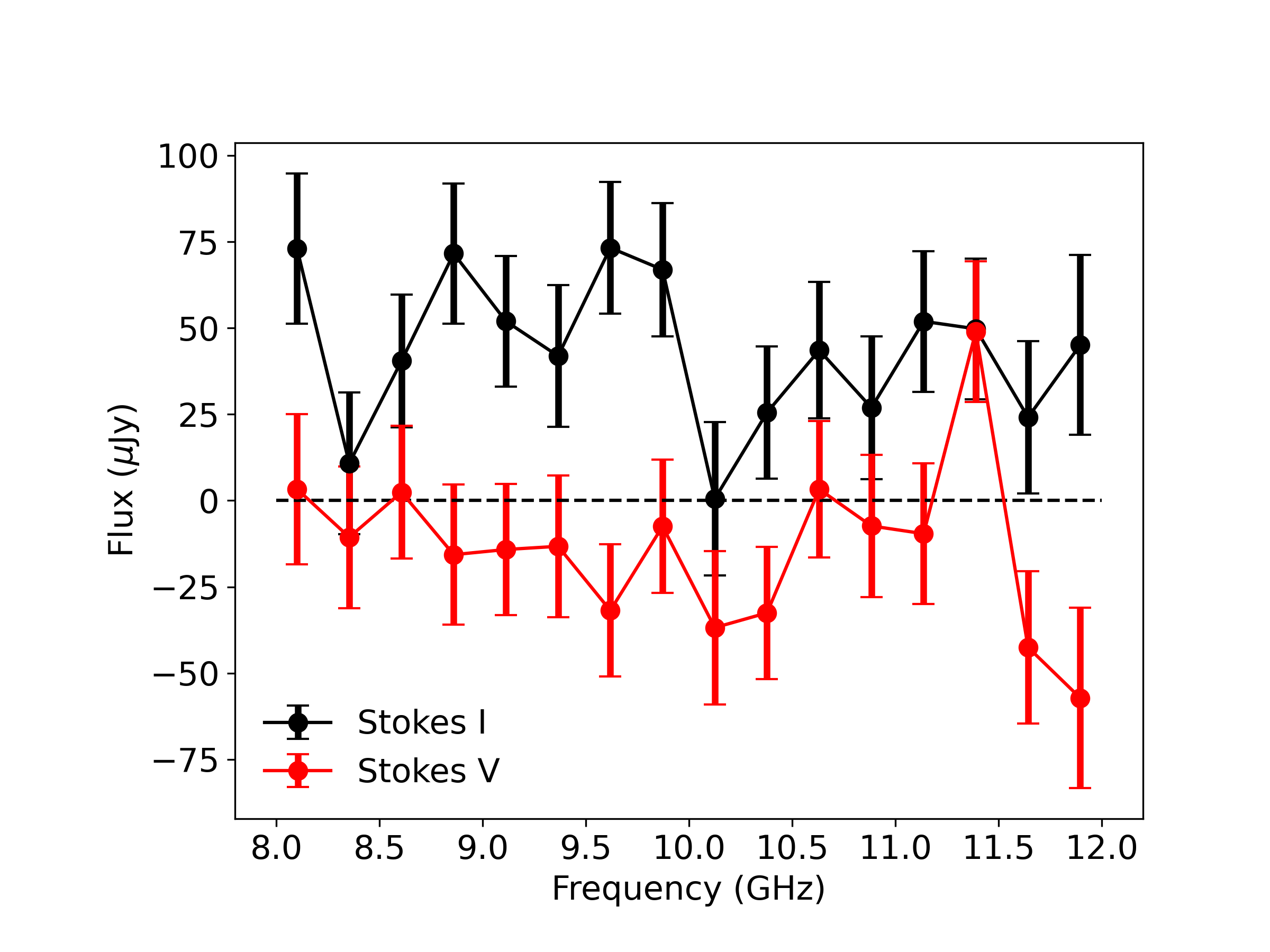}
        \caption{The 8--12 GHz SED of ST LMi on November 5.}
        \label{fig:1b}
    \end{subfigure}

    \begin{subfigure}[b]{0.32\textwidth}
        \centering
        \includegraphics[width=\textwidth]{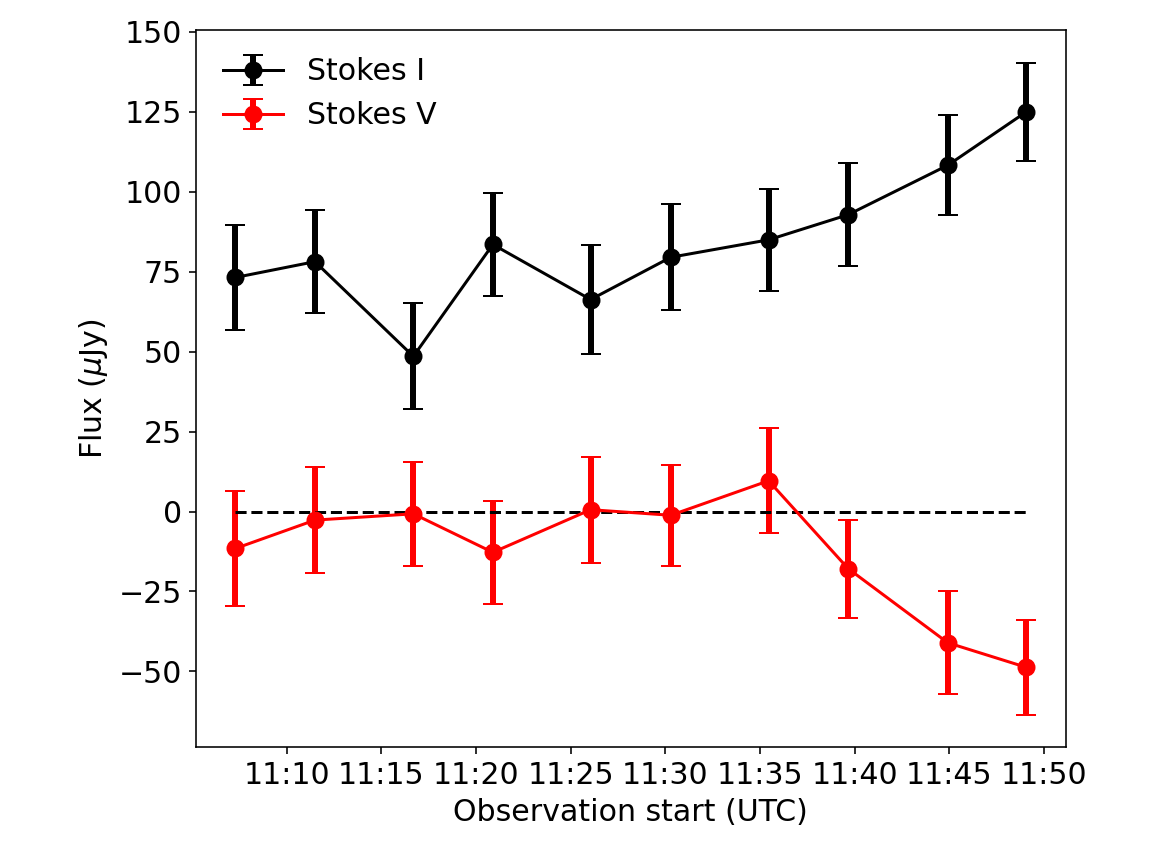}
        \caption{The 8--12 GHz light curve of ST LMi on November 1.}
        \label{fig:2a}
    \end{subfigure}
    \hspace{0.2em}
    \begin{subfigure}[b]{0.32\textwidth}
        \centering
        \includegraphics[width=\textwidth]{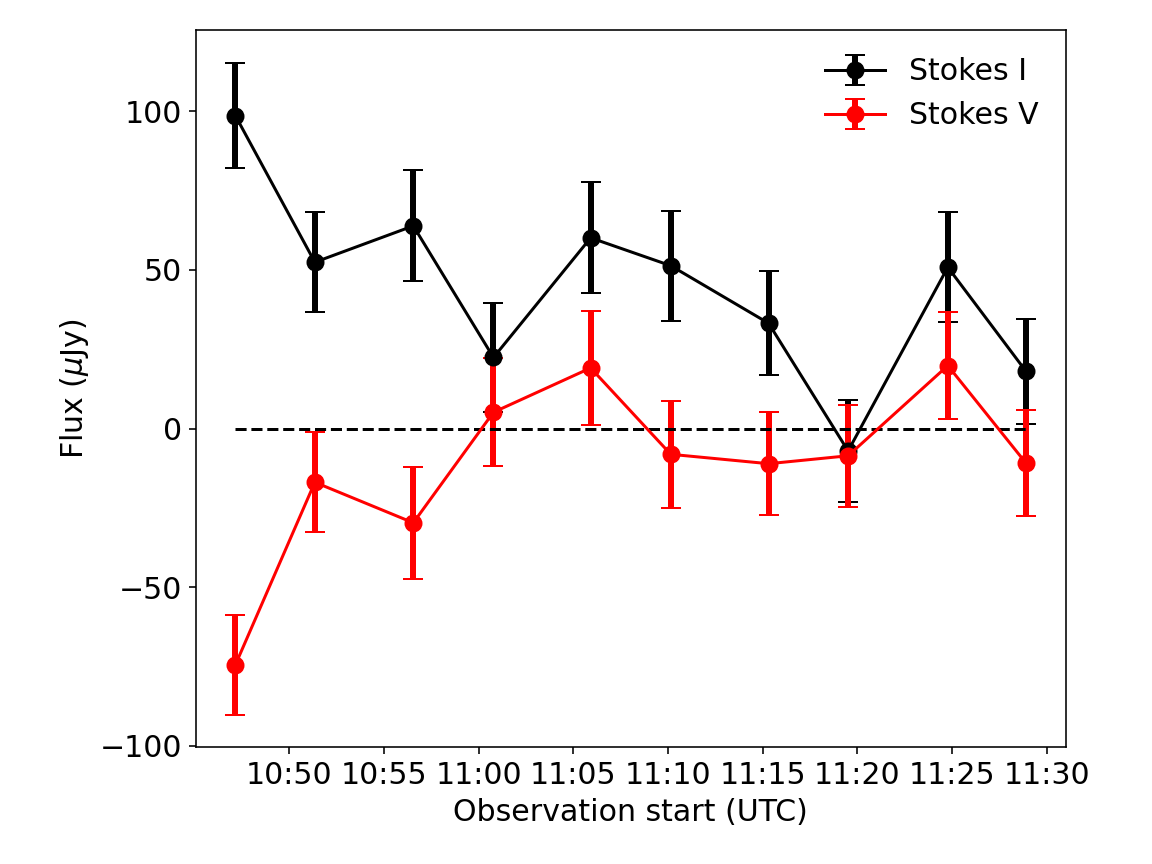}
        \caption{The 8--12 GHz light curve of ST LMi on November 5.}
        \label{fig:2b}
    \end{subfigure}

    \caption{The 8--12 GHz data for ST LMi on 2017 November 1 (\textit{left column}) and 2017 November 5 (\textit{right column}). The top panels show the full-band SEDs, while the bottom panels show the full-band light curves. Detections of circular polarization are marginal, but there is an indication of increasing polarization with flux at the end of the first observation and the beginning of the first.}
    \label{fig: stlmi 17b}
\end{figure}


\begin{figure}[ht!]
    \centering

    \begin{subfigure}[b]{0.32\textwidth}
        \centering
        \includegraphics[width=\textwidth]{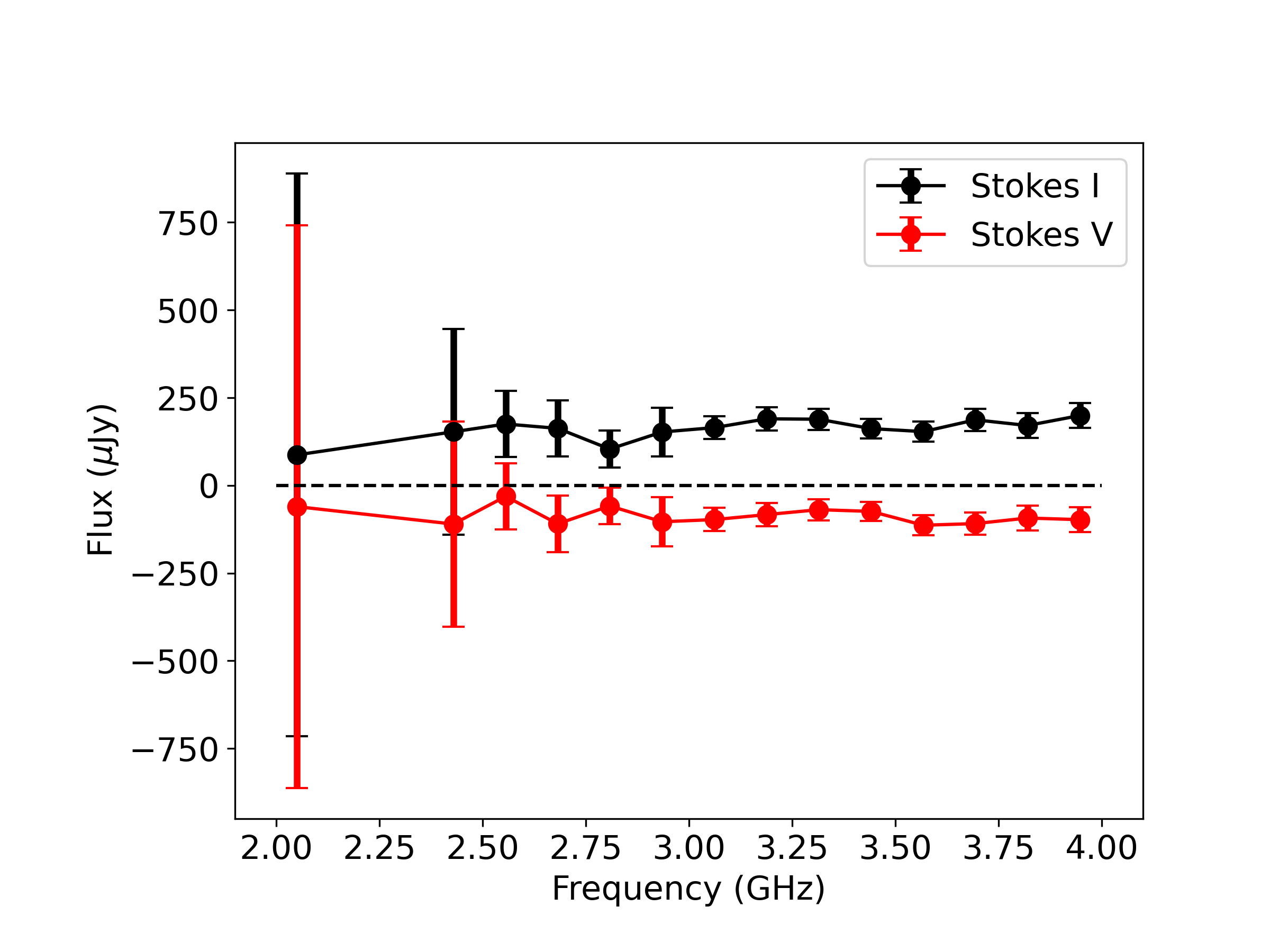}
        \caption{The 2--4 GHz SED of ST LMi on March 23.}
        \label{fig:1a}
    \end{subfigure}

    \begin{subfigure}[b]{0.32\textwidth}
        \centering
        \includegraphics[width=\textwidth]{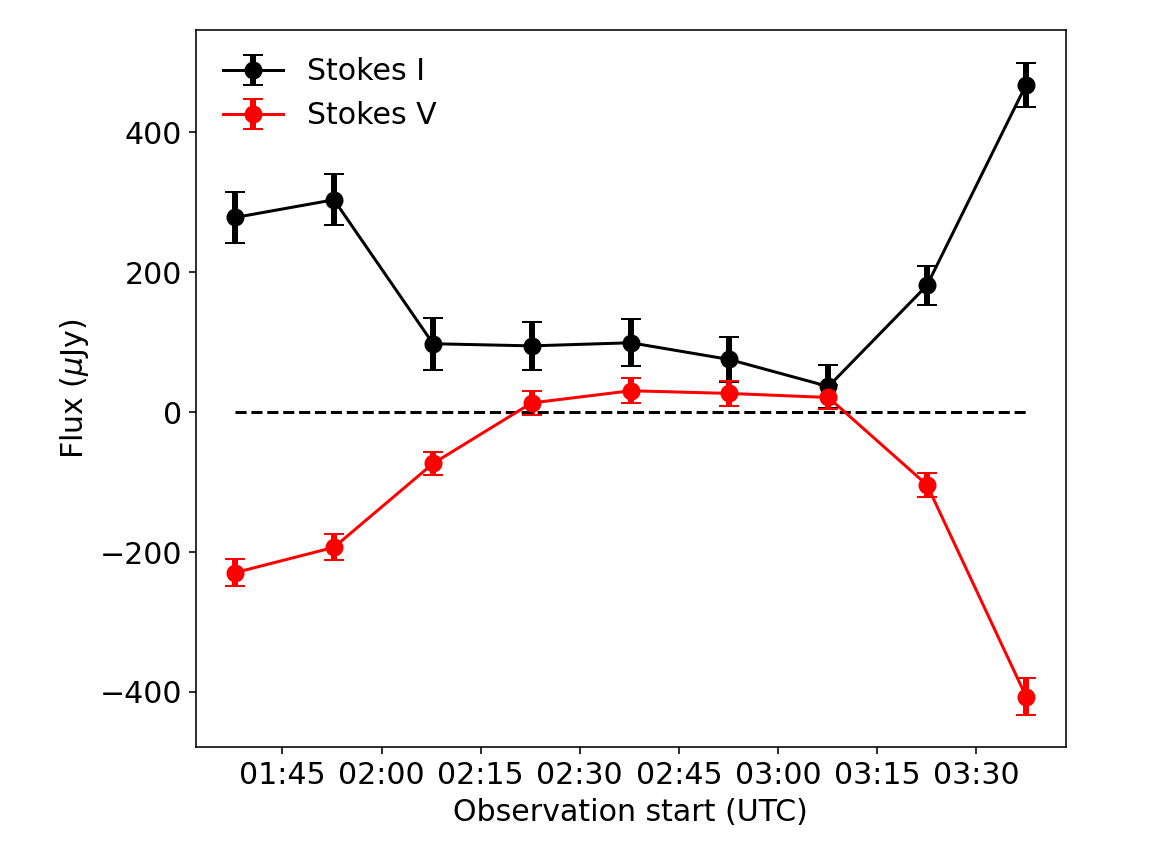}
        \caption{The 2--4 GHz light curve of ST LMi on March 23.}
        \label{fig:2a}
    \end{subfigure}

    \begin{subfigure}[b]{0.32\textwidth}
        \centering
        \includegraphics[width=\textwidth]{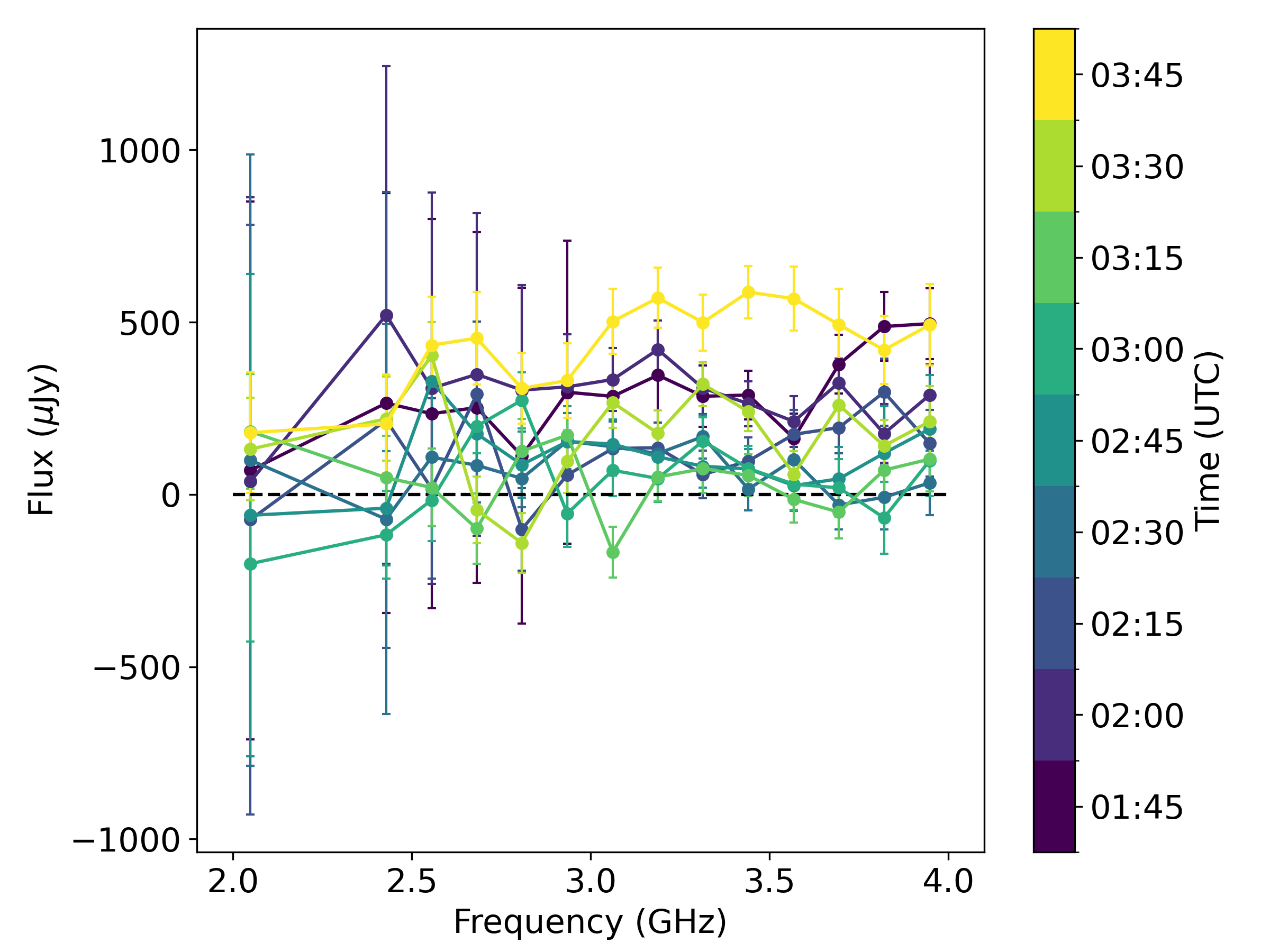}
        \caption{The 2--4 GHz dynamic SED of ST LMi in Stokes I on March 23.}
        \label{fig:3a}
    \end{subfigure}

    \begin{subfigure}[b]{0.32\textwidth}
        \centering
        \includegraphics[width=\textwidth]{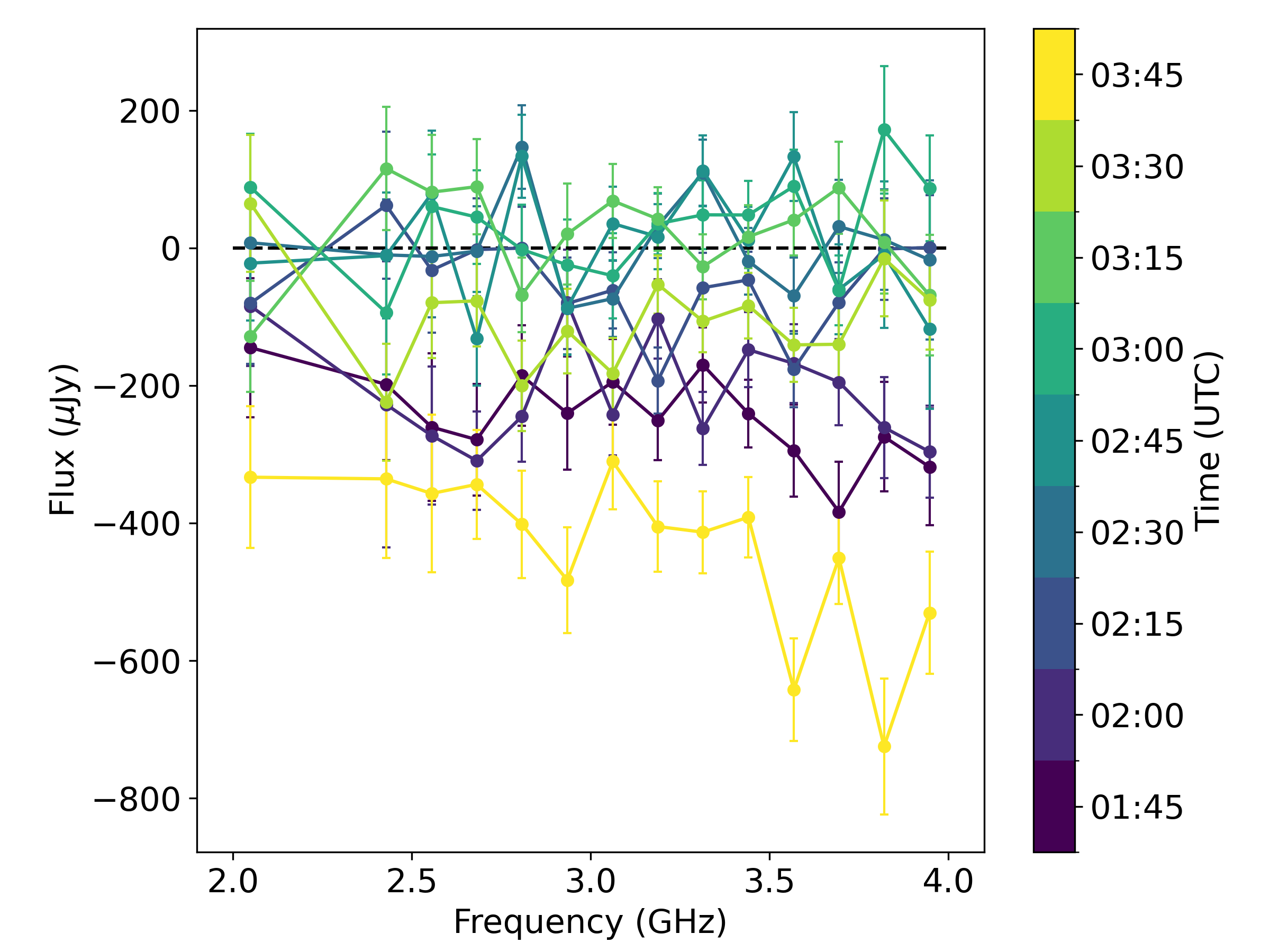}
        \caption{The 2--4 GHz dynamic SED of ST LMi in Stokes V on March 23.}
        \label{fig:4a}
    \end{subfigure}

    \caption{The 2--4 GHz data for ST LMi on 2025 March 23. The top panel shows the full-band SED, the second panel shows the full-band light curve, the third panel shows the Stokes I dynamic SED, and the bottom panel shows the Stokes V dynamic SED. There are two flares detected at the beginning and end of the observation that correspond with clear detections of LCP emission.}
    \label{fig: stlmi 25a}
\end{figure}

Most of the data on ST LMi lack clear signals of variation or circular polarization, but near the end of the first observation (Fig.\ \ref{fig: stlmi 17b}), there is an increase suggestive of a flare, with a significant increase in circular polarization. Over the last 15 minutes of the observation, the fractional polarization is $38\pm12$\% LCP. Due to the relatively low S/N ratio, we do not present 8--12 GHz dynamic SEDs for ST LMi.

At 2--4 GHz, ST LMi appears to be have a flat spectrum with a large fractional polarization (Fig.\ \ref{fig: stlmi 25a}). The origin of the polarized flux seems to be two bright flares at early and late times during the observation. The only two clear detections in Stokes I with acceptable uncertainties are in the first and final time bins, which suggest a broadband signal. These features are much clearer in the Stokes V dynamic SED, which shows that these flares produced large fractional LCP.

\subsection{MR Ser}


\begin{figure}[ht!]
    \centering

    \begin{subfigure}[b]{0.32\textwidth}
        \centering
        \includegraphics[width=\textwidth]{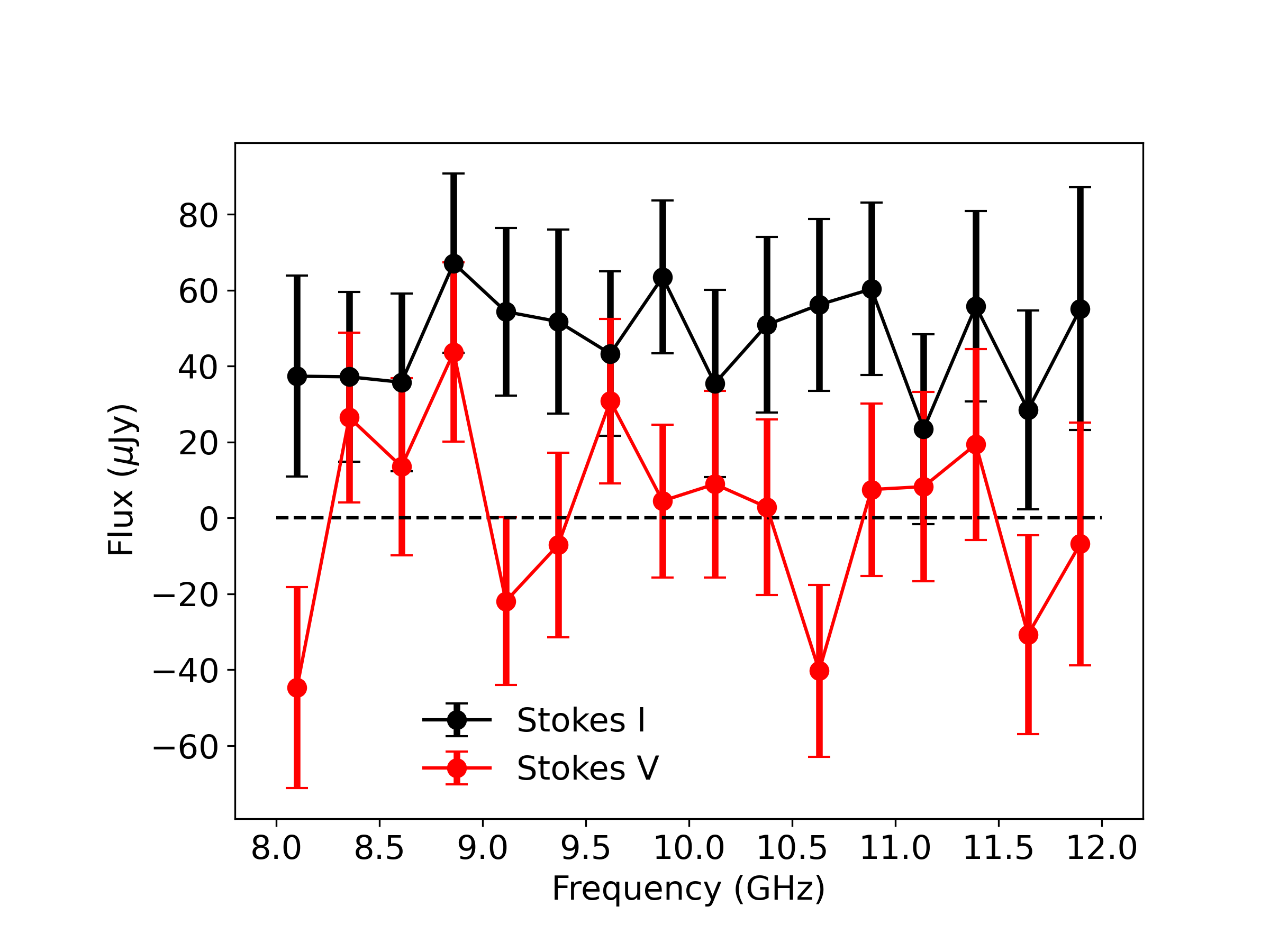}
        \caption{The 8--12 GHz SED of MR Ser on October 11.}
        \label{fig:1a}
    \end{subfigure}
    \hspace{0.2em}
    \begin{subfigure}[b]{0.32\textwidth}
        \centering
        \includegraphics[width=\textwidth]{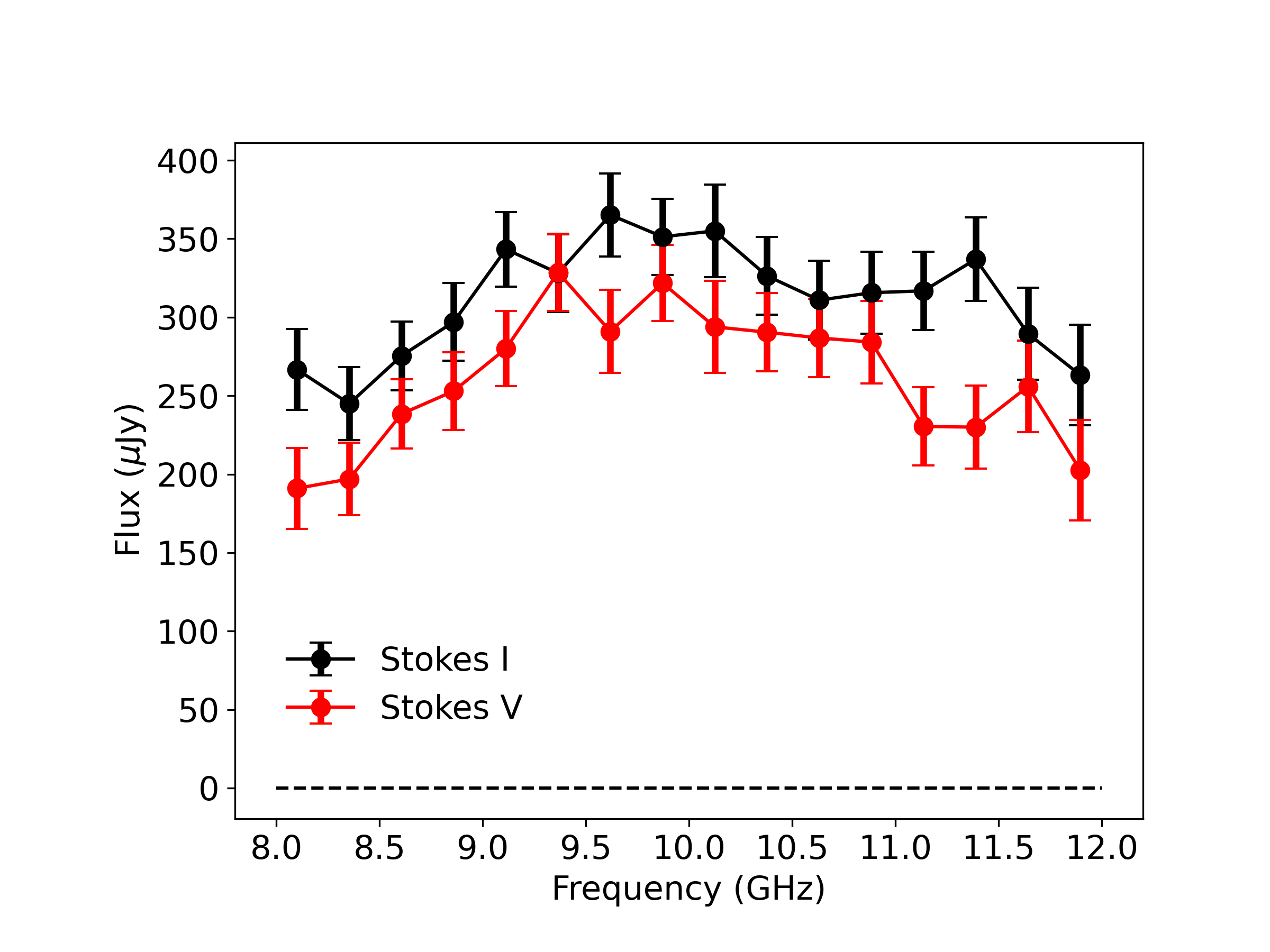}
        \caption{The 8--12 GHz SED of MR Ser on November 25.}
        \label{fig:1b}
    \end{subfigure}

    \begin{subfigure}[b]{0.32\textwidth}
        \centering
        \includegraphics[width=\textwidth]{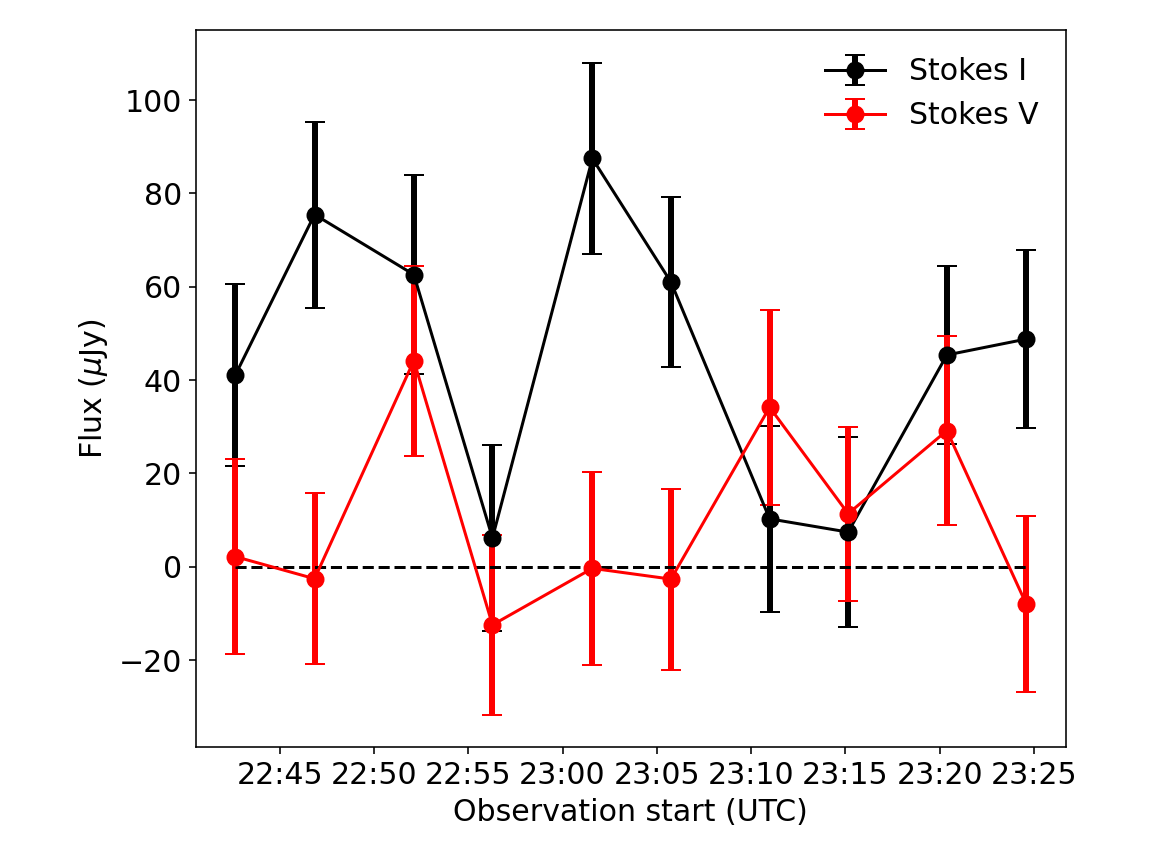}
        \caption{The 8--12 GHz light curve of MR Ser on October 11.}
        \label{fig:2a}
    \end{subfigure}
    \hspace{0.2em}
    \begin{subfigure}[b]{0.32\textwidth}
        \centering
        \includegraphics[width=\textwidth]{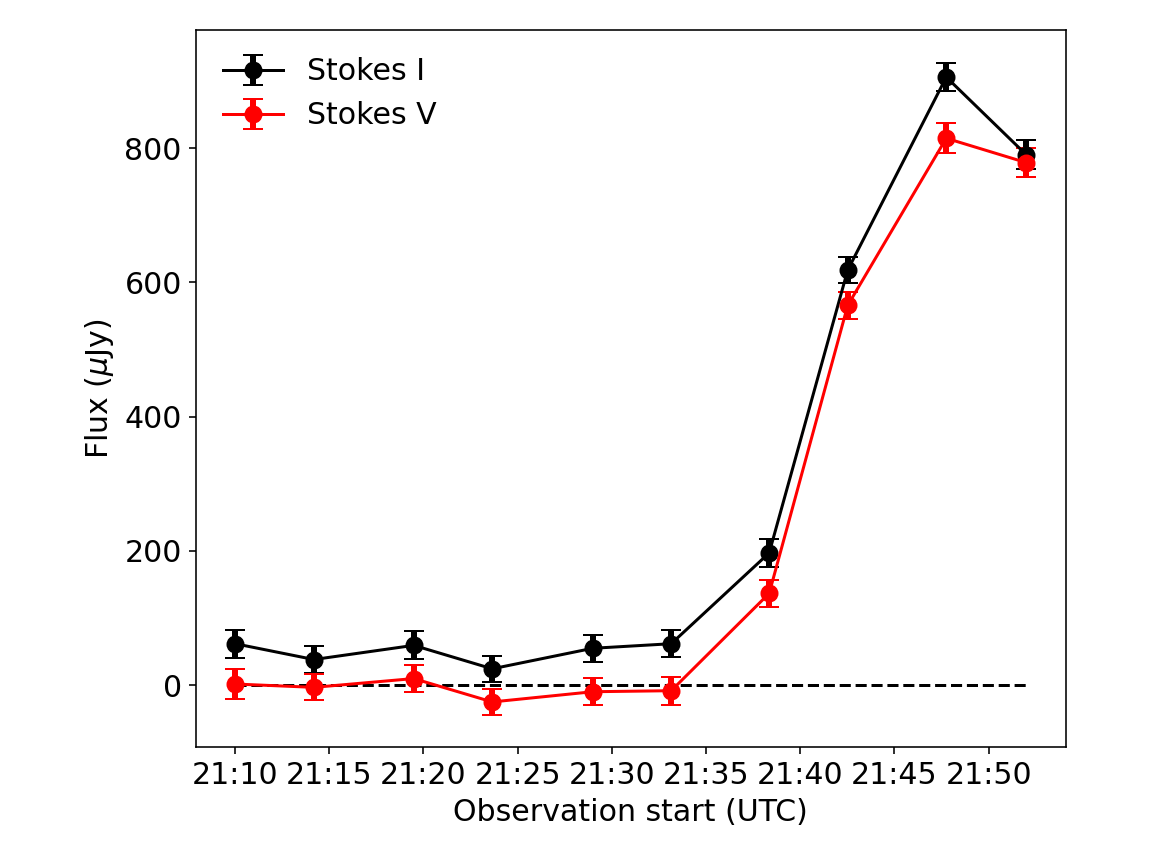}
        \caption{The 8--12 GHz light curve of MR Ser on November 25.}
        \label{fig:2b}
    \end{subfigure}

    \begin{subfigure}[b]{0.32\textwidth}
        \centering
        \includegraphics[width=\textwidth]{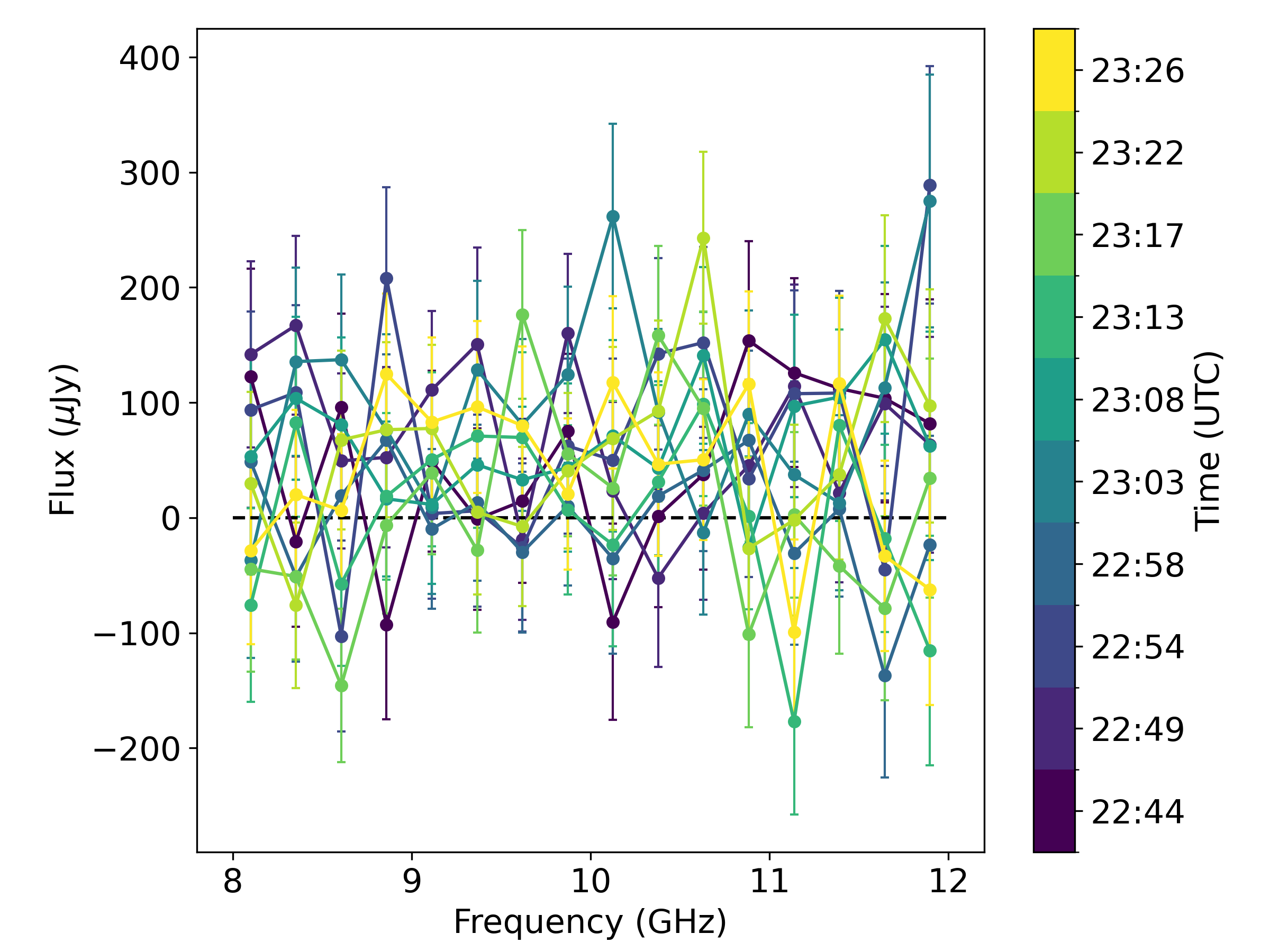}
        \caption{The 8--12 GHz dynamic SED of MR Ser in Stokes I on October 11.}
        \label{fig:3a}
    \end{subfigure}
    \hspace{0.2em}
    \begin{subfigure}[b]{0.32\textwidth}
        \centering
        \includegraphics[width=\textwidth]{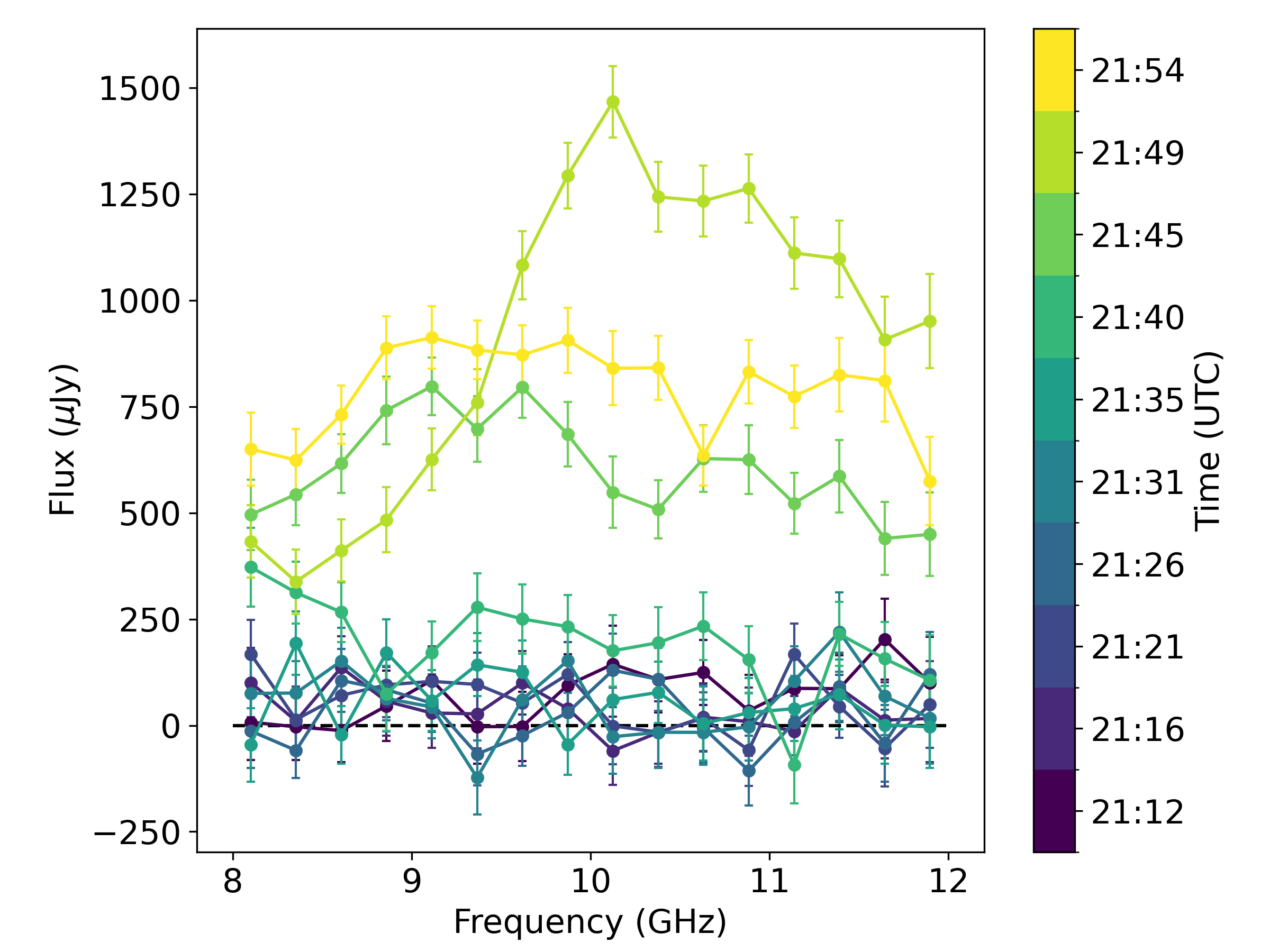}
        \caption{The 8--12 GHz dynamic SED of MR Ser in Stokes I on November 25.}
        \label{fig:3b}
    \end{subfigure}

    \begin{subfigure}[b]{0.32\textwidth}
        \centering
        \includegraphics[width=\textwidth]{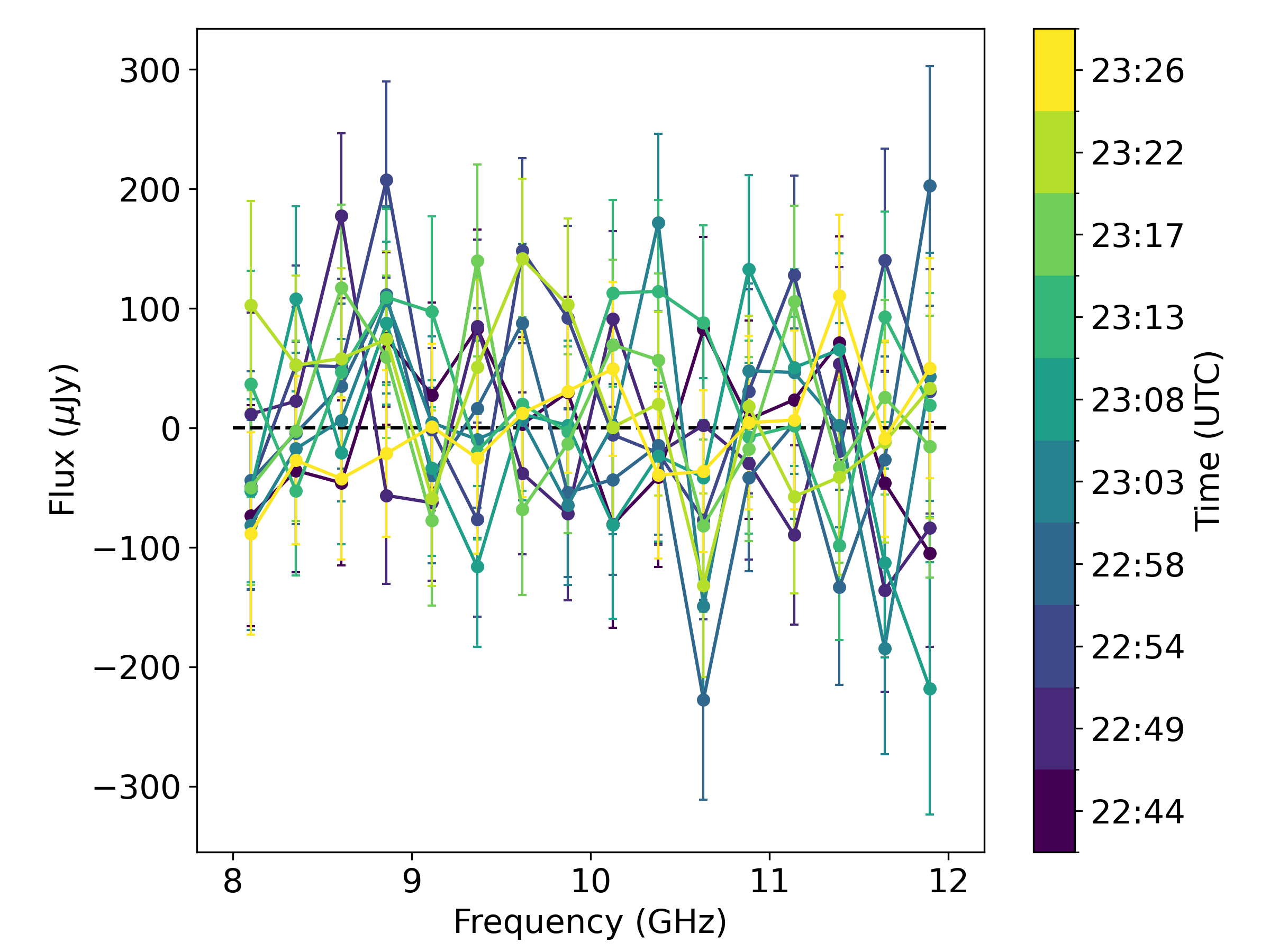}
        \caption{The 8--12 GHz dynamic SED of MR Ser in Stokes V on October 11.}
        \label{fig:4a}
    \end{subfigure}
    \hspace{0.2em}
    \begin{subfigure}[b]{0.32\textwidth}
        \centering
        \includegraphics[width=\textwidth]{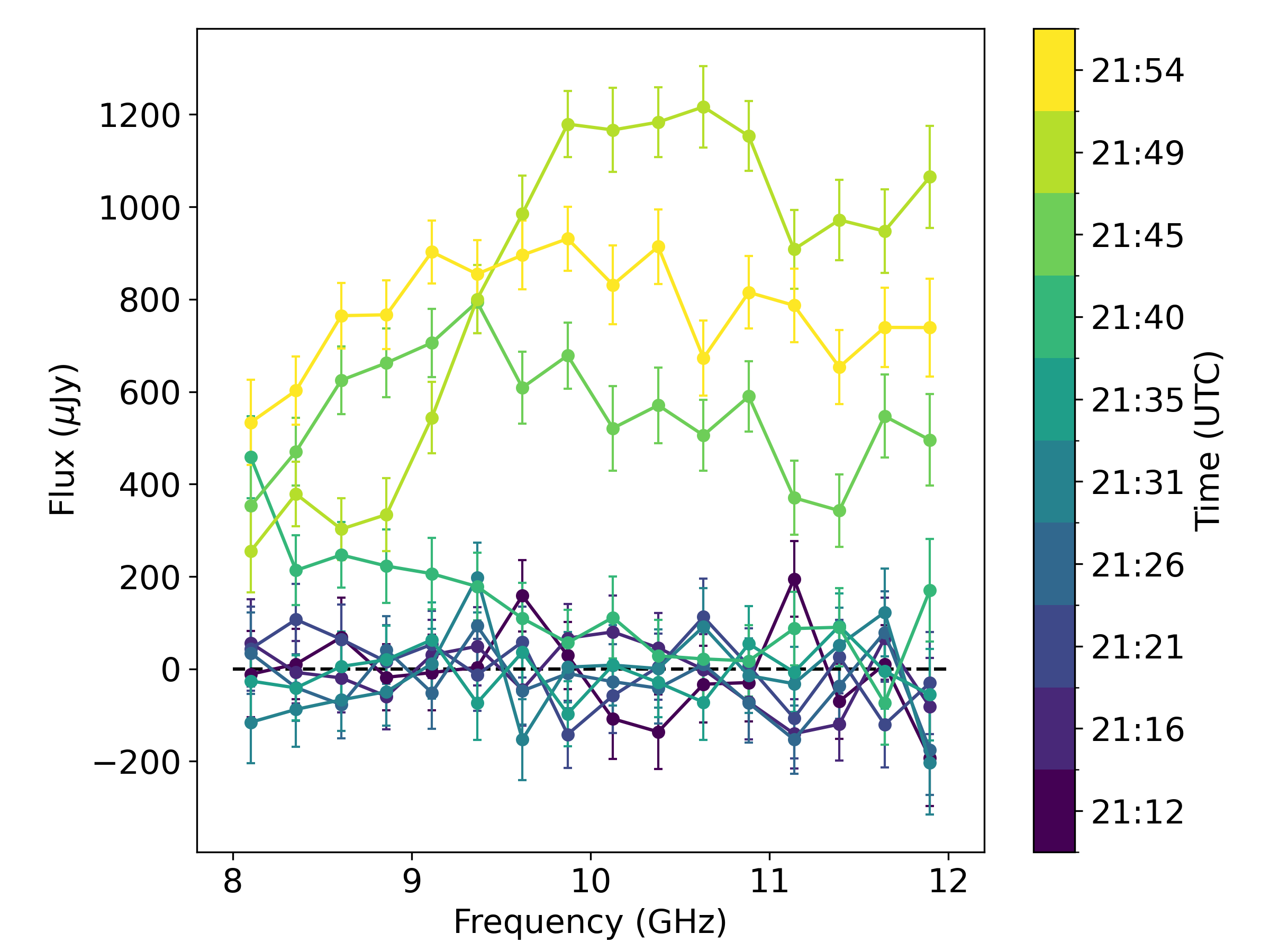}
        \caption{The 8--12 GHz dynamic SED of MR Ser in Stokes V on November 25.}
        \label{fig:4b}
    \end{subfigure}

    \caption{The 8--12 GHz data for MR Ser on 2017 October 11 (\textit{left column}) and 2017 November 25 (\textit{right column}). The top panels show the full-band SEDs, the second panels show the full-band light curves, the third panels show the Stokes I dynamic SEDs, and the bottom panels show the Stokes V dynamic SEDs. One flare was detection in the second observation, which was highly circularly polarized.}
    \label{fig: mrser 17b}
\end{figure}


\begin{figure}[ht!]
    \centering

    \begin{subfigure}[b]{0.32\textwidth}
        \centering
        \includegraphics[width=\textwidth]{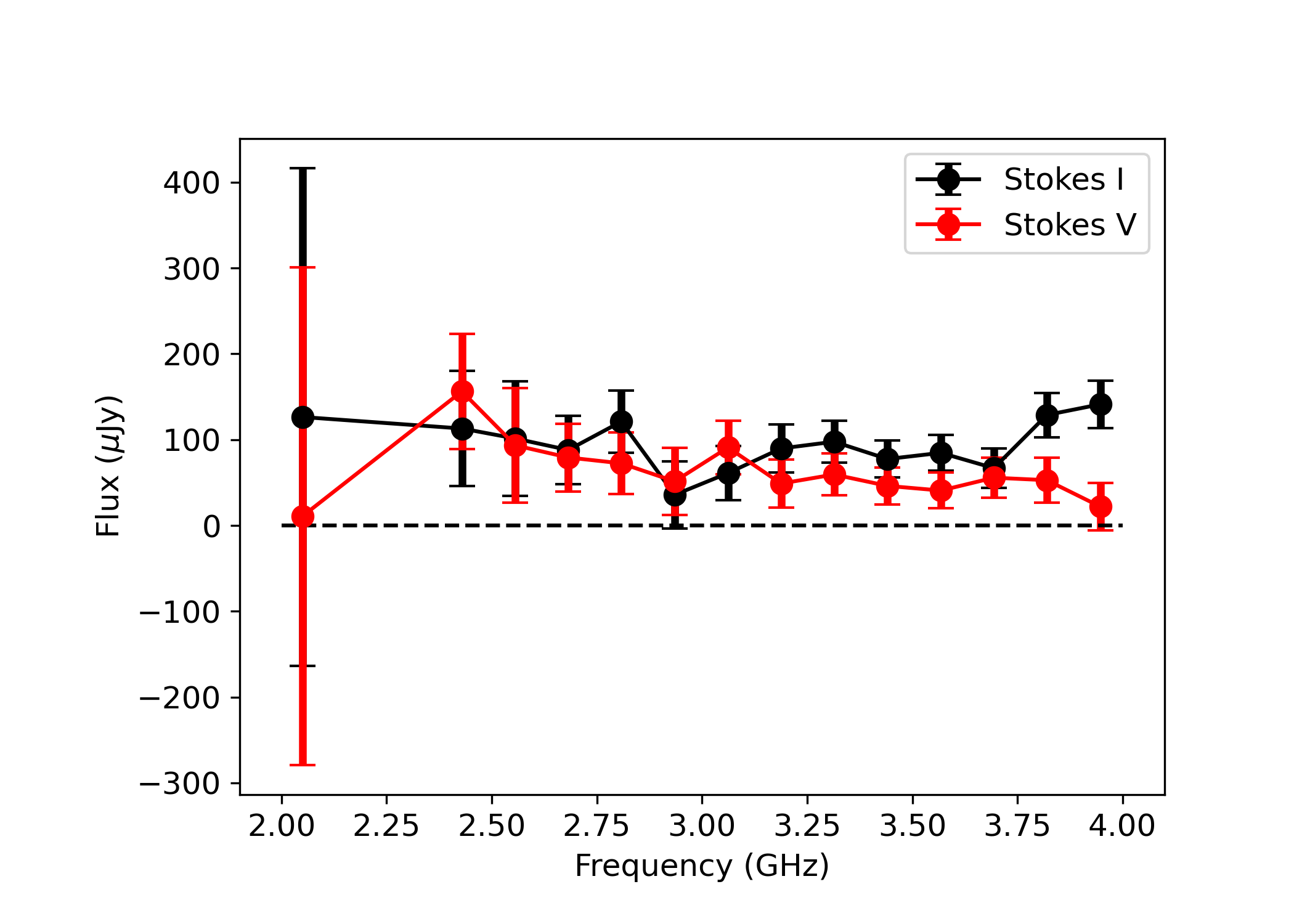}
        \caption{The 2--4 GHz SED of MR Ser on April 5.}
        \label{fig:1a}
    \end{subfigure}

    \begin{subfigure}[b]{0.32\textwidth}
        \centering
        \includegraphics[width=\textwidth]{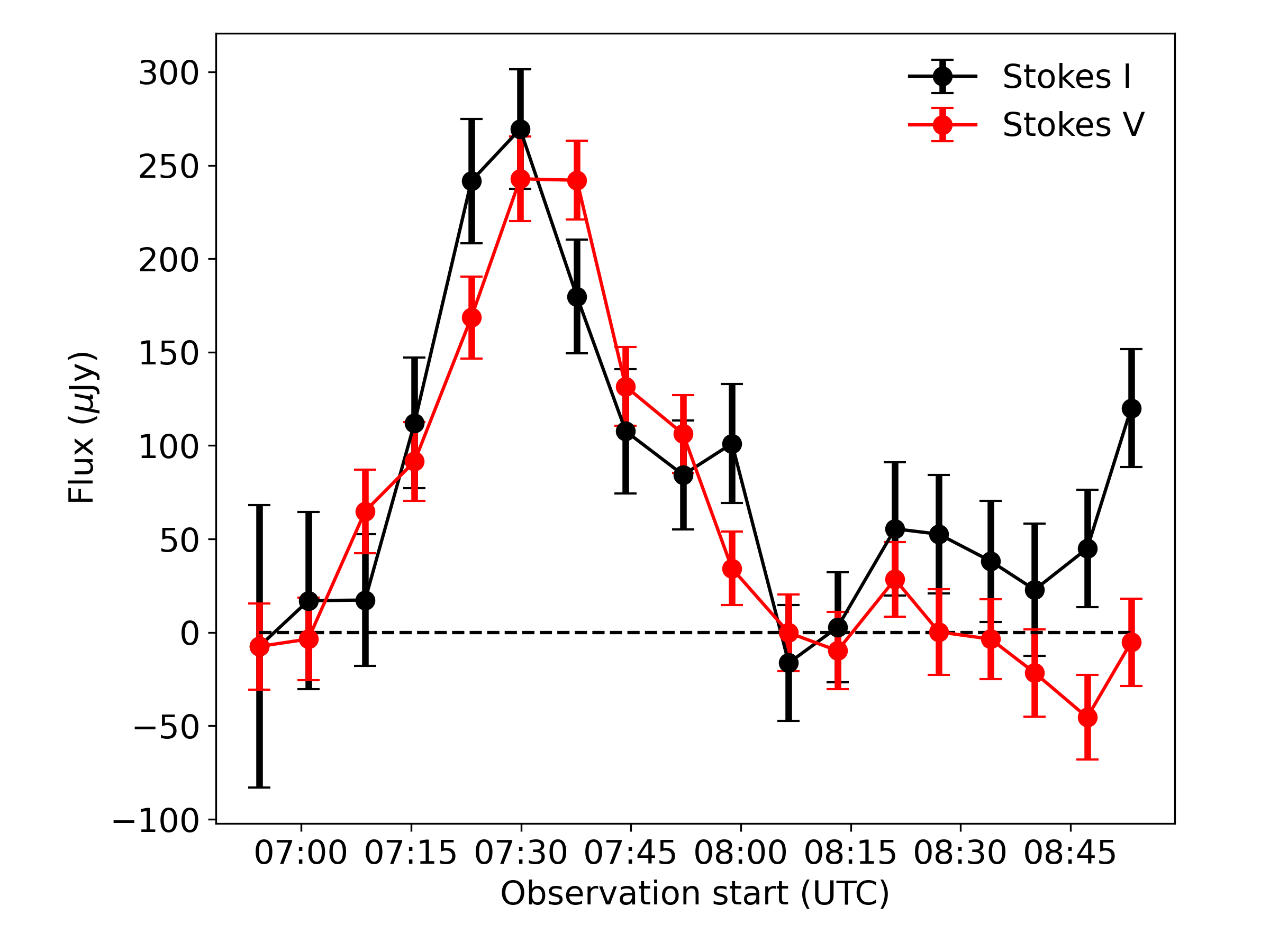}
        \caption{The 2--4 GHz light curve of MR Ser on April 5.}
        \label{fig:2a}
    \end{subfigure}

    \begin{subfigure}[b]{0.32\textwidth}
        \centering
        \includegraphics[width=\textwidth]{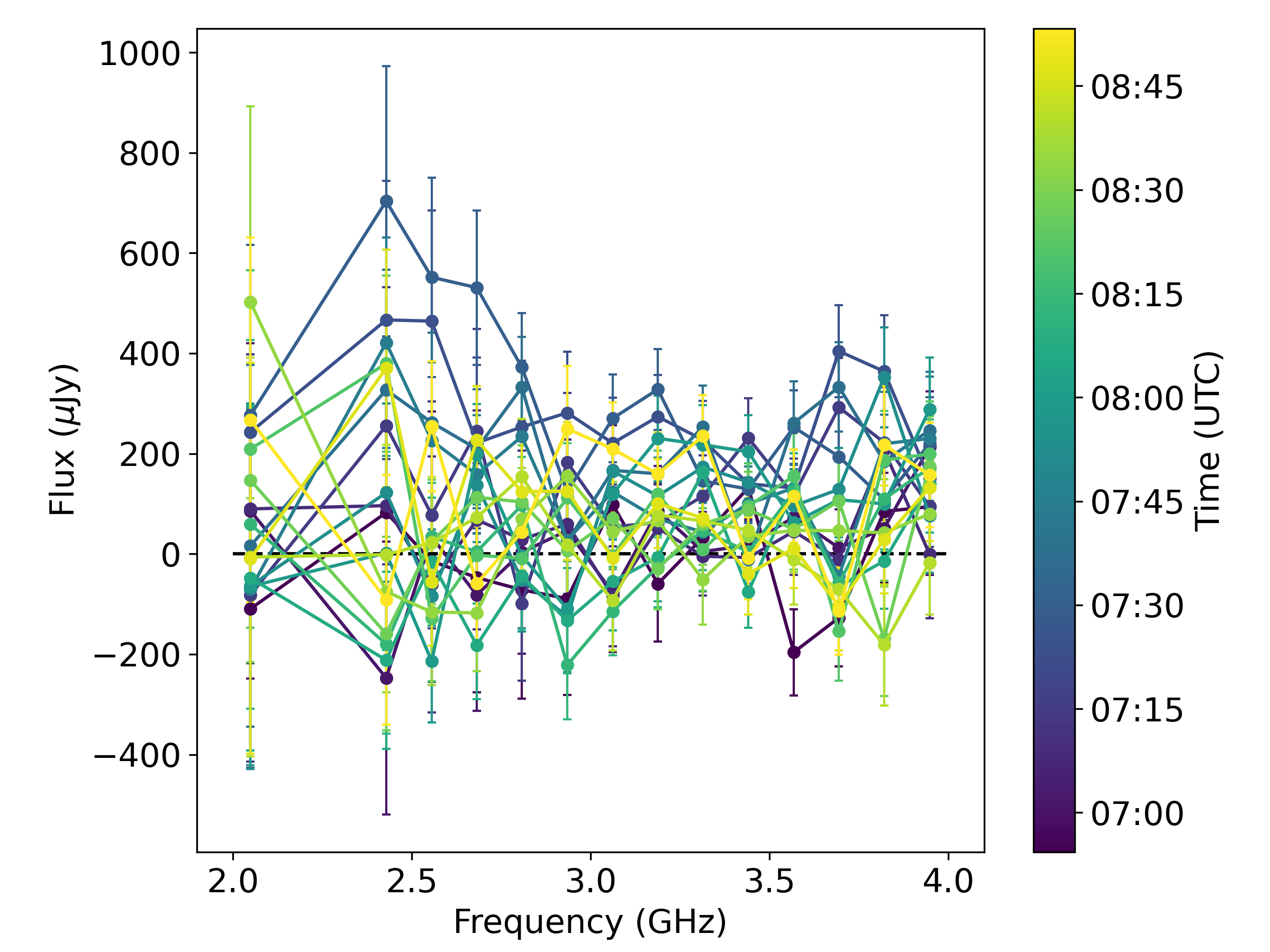}
        \caption{The 2--4 GHz dynamic SED of MR Ser in Stokes I on April 5.}
        \label{fig:3a}
    \end{subfigure}

    \begin{subfigure}[b]{0.32\textwidth}
        \centering
        \includegraphics[width=\textwidth]{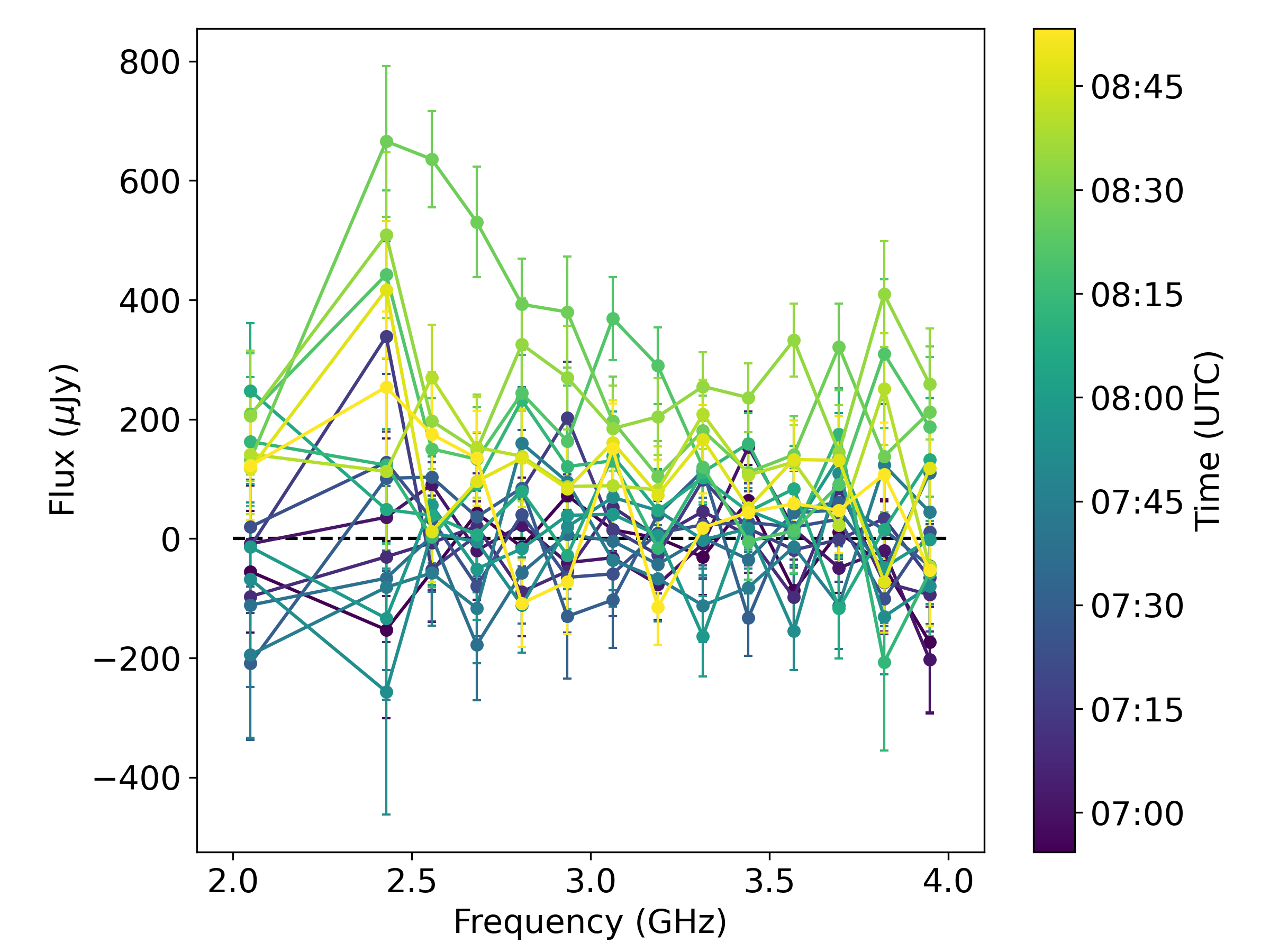}
        \caption{The 2--4 GHz dynamic SED of MR Ser in Stokes V on April 5.}
        \label{fig:4a}
    \end{subfigure}

    \caption{The 2--4 GHz data for MR Ser on 2025 April 5. The top panel shows the full-band SED, the second panel shows the full-band light curve, the third panel shows the Stokes I dynamic SED, and the bottom panel shows the Stokes V dynamic SED. One flare was detected in this observation. It is difficult to see any clear signal in the Stokes I data, but the shape is somewhat clearer in Stokes V.}
    \label{fig: mrser 25a}
\end{figure}

In the first observation of MR Ser on 2017 October 11 (Fig.\ \ref{fig: mrser 17b}), there was no clear Stokes V detection, no variation, and no emission features in the SED. However, on 2017 November 25, there is a strong flare (factor of $\sim$18) that peaks at 21:47 UTC. We only observe the rise of the flare, lasting $\sim$10 minutes. This flare corresponds with a broadband emission feature that then becomes more localized near 10--10.5 GHz before it begins to decay. At the peak, it is $90\pm5$\% RCP.

The behavior at 2--4 GHz is somewhat different by comparison. Stokes V is indeed detected (Fig. \ref{fig: mrser 25a}) and there is a flare at 17:30 UTC on 2025 April 5, but the spectral shape of the flare is difficult to determine in the Stokes I dynamic SED. This may be due to the crowded nature of the field raising the background noise in each time bin. If we assume that Stokes V is representative of the shape of the spectrum, the circularly polarized flare from MR Ser in \textit{S}-band is narrow-band, centered on 2.4 GHz with a FWHM of $\sim$0.5 GHz.

\subsection{V2400 Oph}


\begin{figure}[ht!]
    \centering

    \begin{subfigure}[b]{0.32\textwidth}
        \centering
        \includegraphics[width=\textwidth]{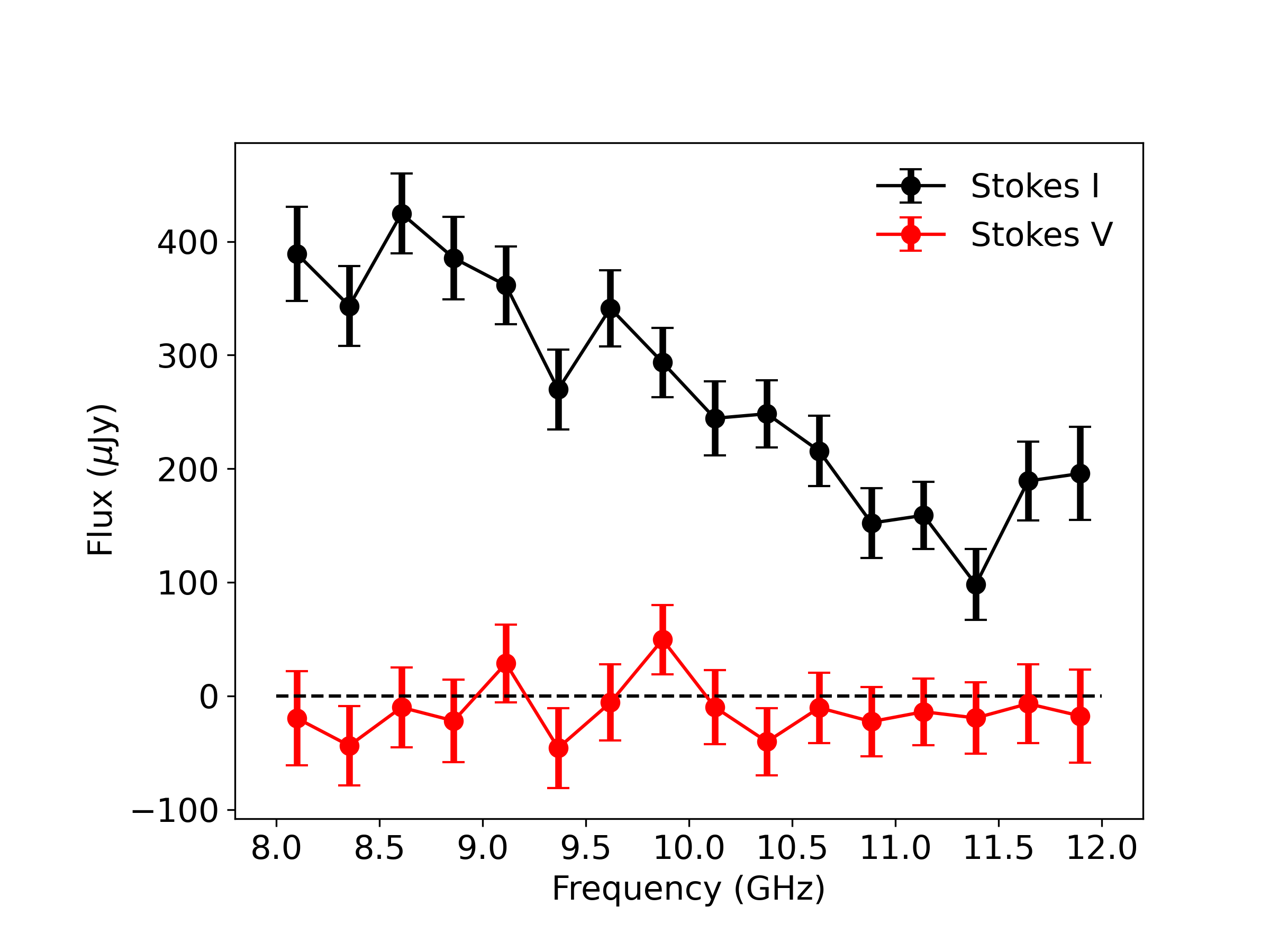}
        \caption{The 8--12 GHz SED of V2400 Oph on April 25.}
        \label{fig:1a}
    \end{subfigure}
    \hspace{0.2em}
    \begin{subfigure}[b]{0.32\textwidth}
        \centering
        \includegraphics[width=\textwidth]{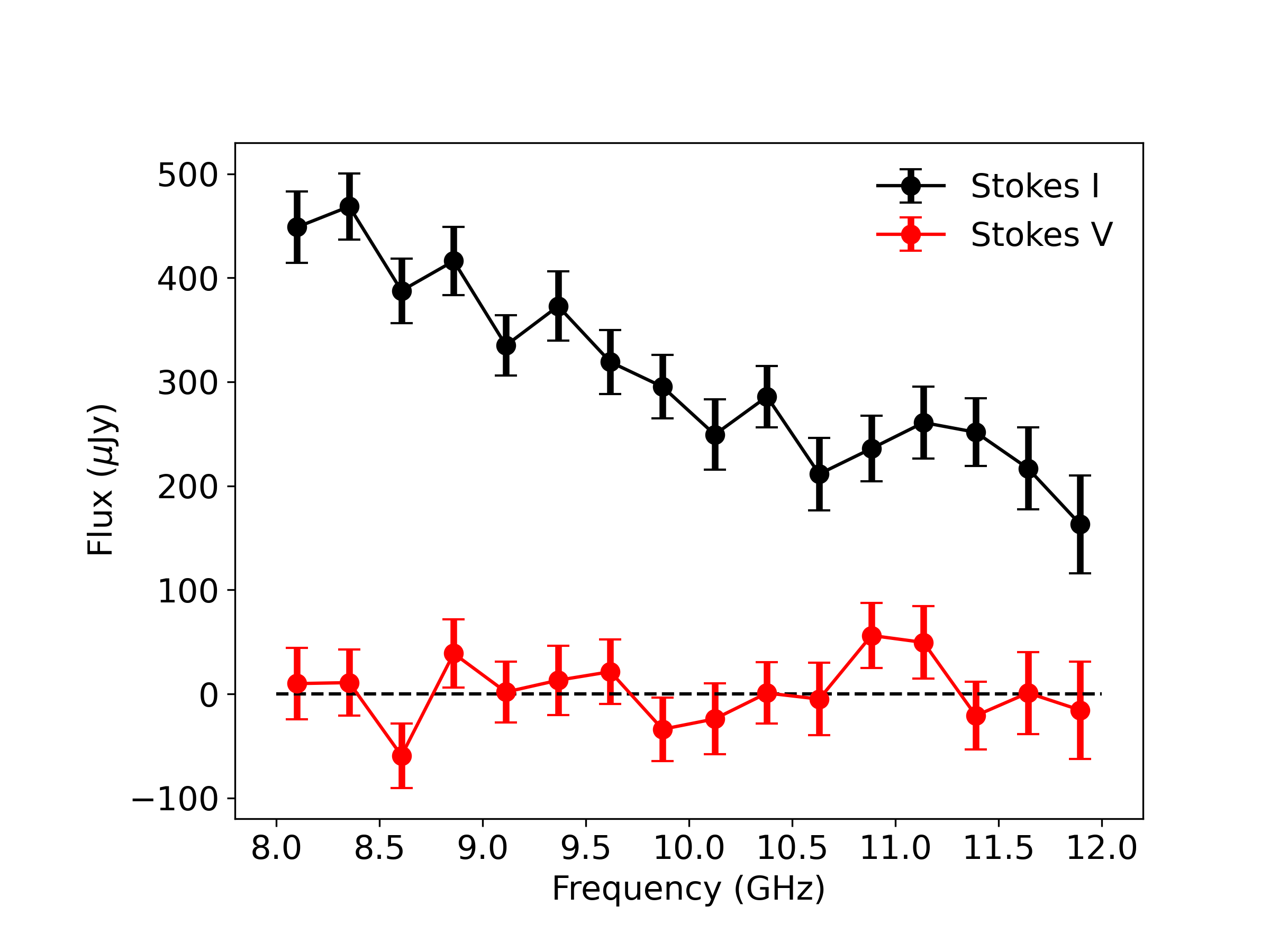}
        \caption{The 8--12 GHz SED of V2400 Oph on May 11.}
        \label{fig:1b}
    \end{subfigure}

    \vspace{0.5em}

    \begin{subfigure}[b]{0.32\textwidth}
        \centering
        \includegraphics[width=\textwidth]{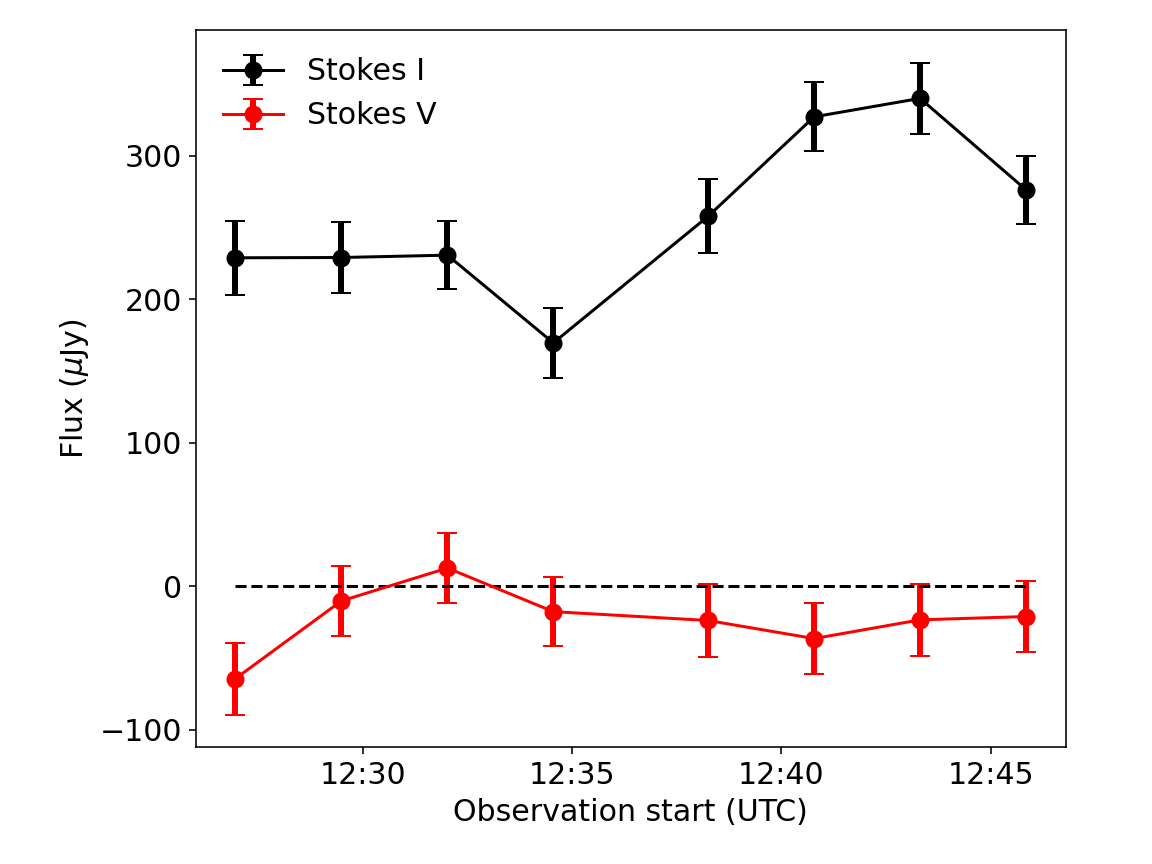}
        \caption{The 8--12 GHz light curve of V2400 Oph on April 25.}
        \label{fig:2a}
    \end{subfigure}
    \hspace{0.2em}
    \begin{subfigure}[b]{0.32\textwidth}
        \centering
        \includegraphics[width=\textwidth]{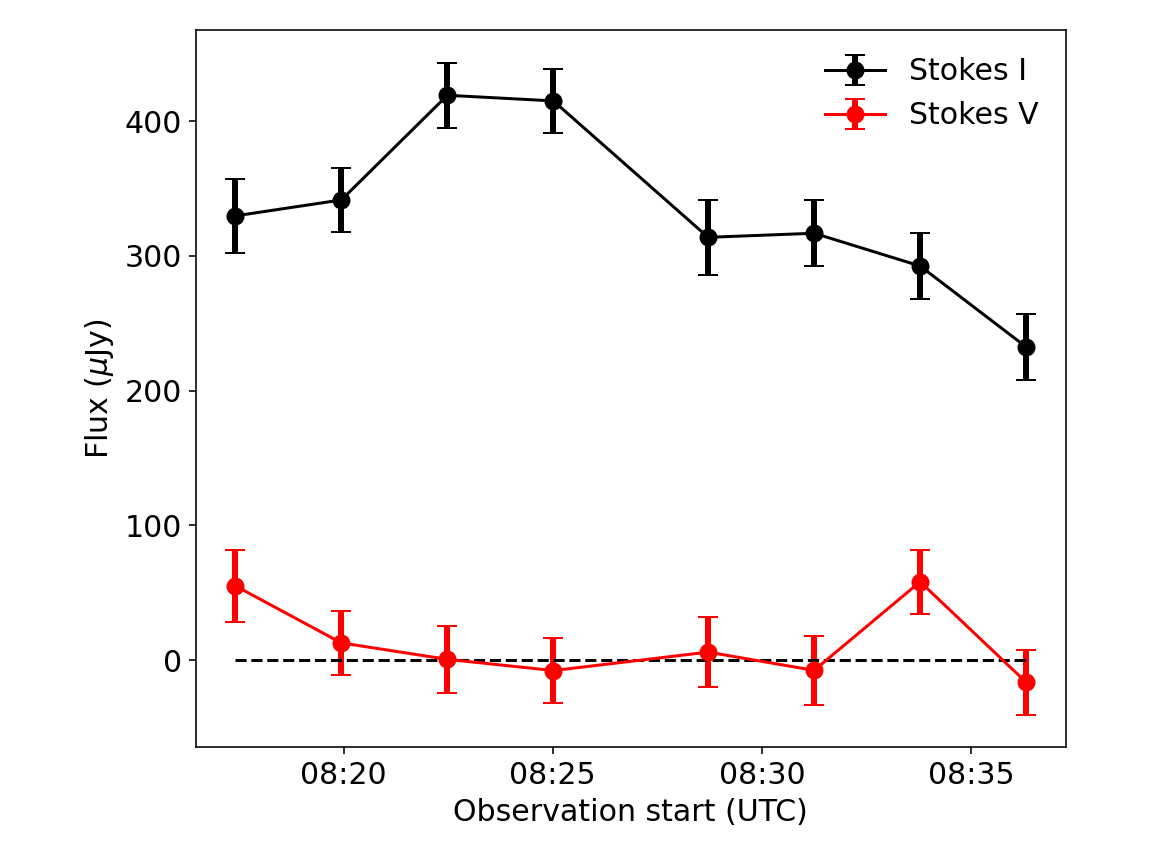}
        \caption{The 8--12 GHz light curve of V2400 Oph on May 11.}
        \label{fig:2b}
    \end{subfigure}

    \vspace{0.5em}

    \begin{subfigure}[b]{0.32\textwidth}
        \centering
        \includegraphics[width=\textwidth]{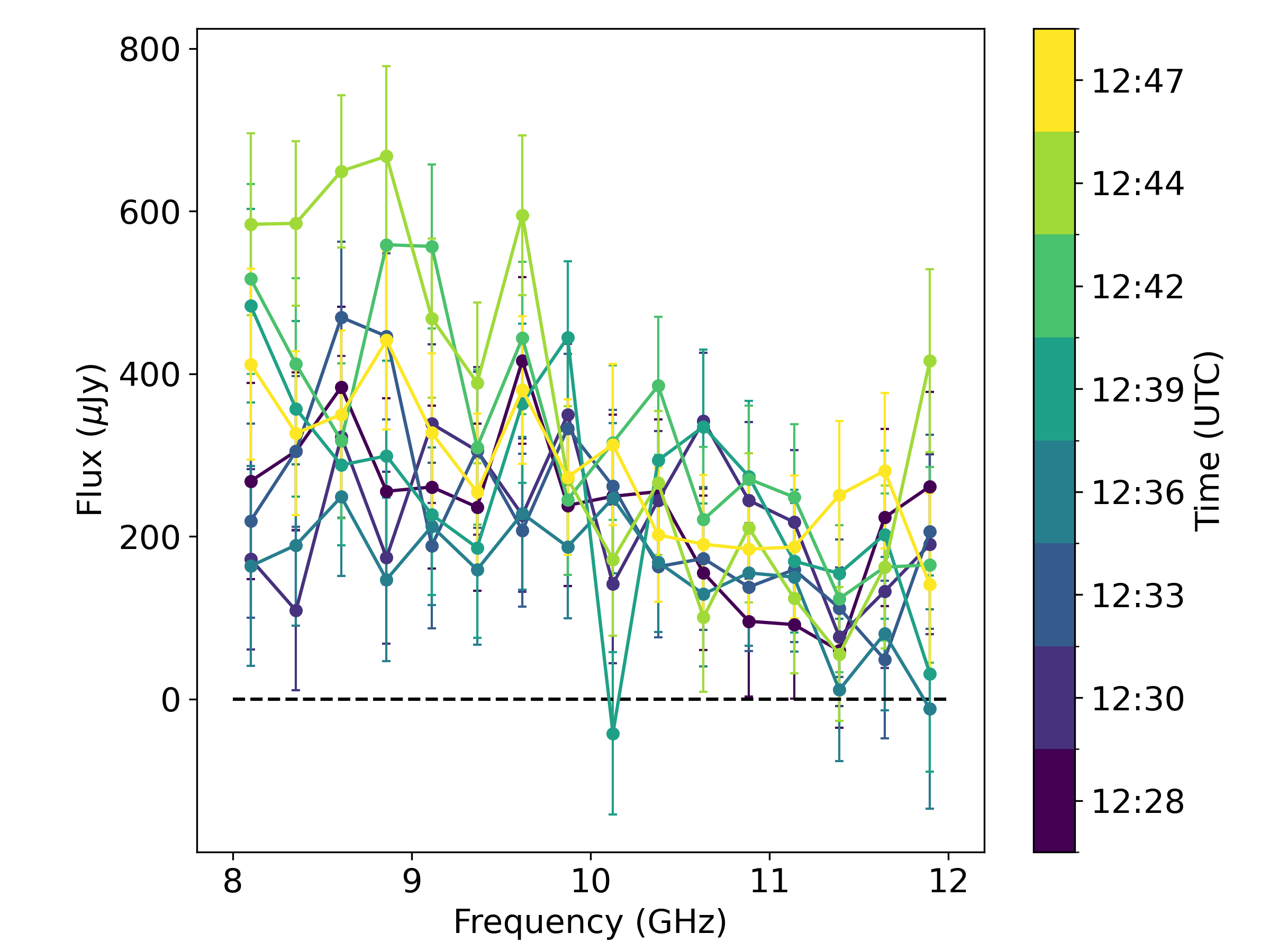}
        \caption{The 8--12 GHz dynamic SED of V2400 Oph in Stokes I on April 25.}
        \label{fig:3a}
    \end{subfigure}
    \hspace{0.2em}
    \begin{subfigure}[b]{0.32\textwidth}
        \centering
        \includegraphics[width=\textwidth]{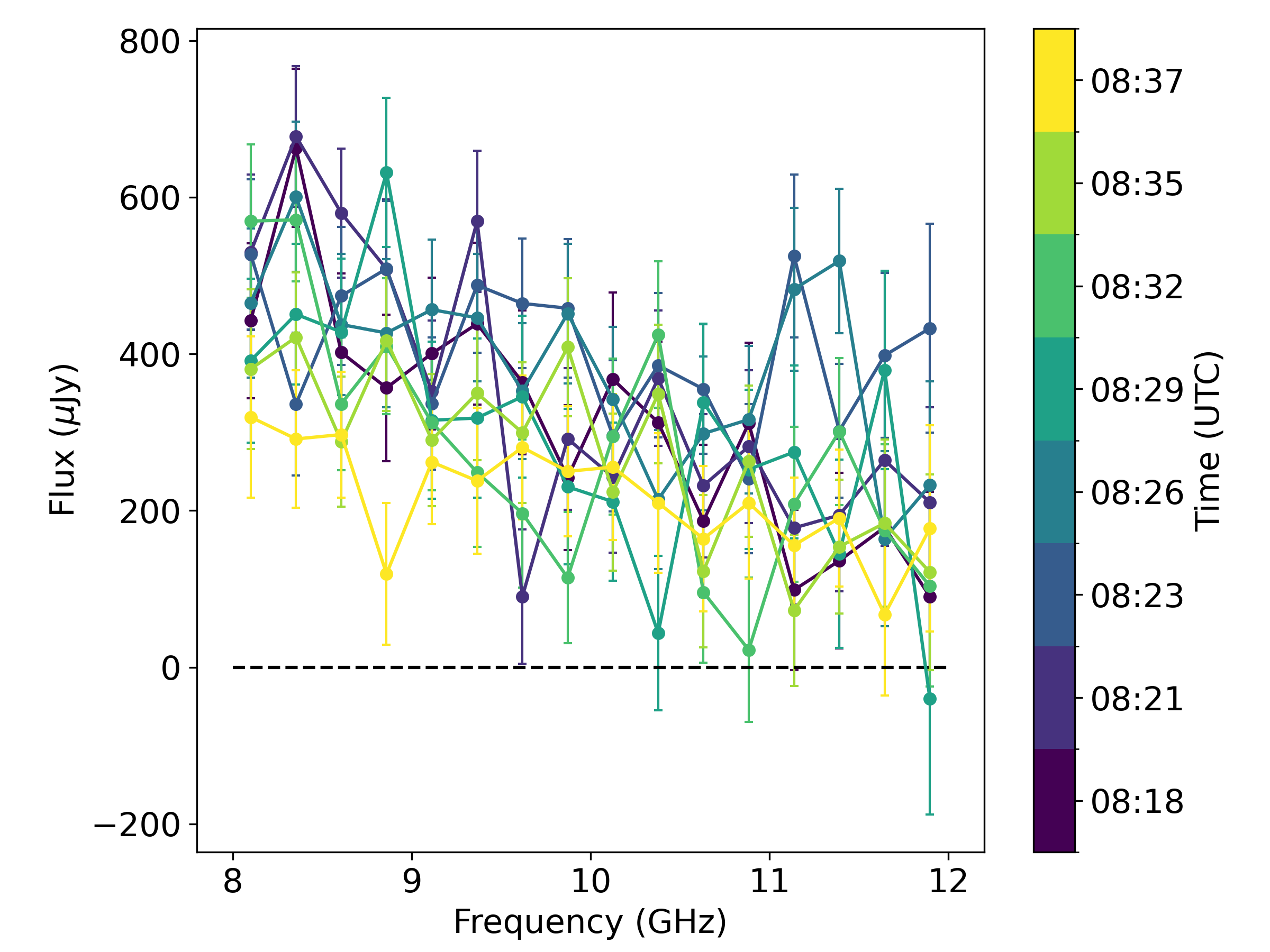}
        \caption{The 8--12 GHz dynamic SED of V2400 Oph in Stokes I on May 11.}
        \label{fig:3b}
    \end{subfigure}

    \caption{The 8--12 GHz data for V2400 Oph on 2019 April 25 (\textit{left column}) and 2019 May 11 (\textit{right column}). The top panels show the full-band SEDs, the middle panels show the full-band light curves, and the bottom panels show the dynamic SEDs in Stokes I. There is some moderate variation in each observation, but no obvious flares were detected.}
    \label{fig: v2400 19a}
\end{figure}

\begin{figure}[ht!]
     \centering
     \begin{subfigure}{0.47\textwidth}
         \centering
         \includegraphics[width=\textwidth]{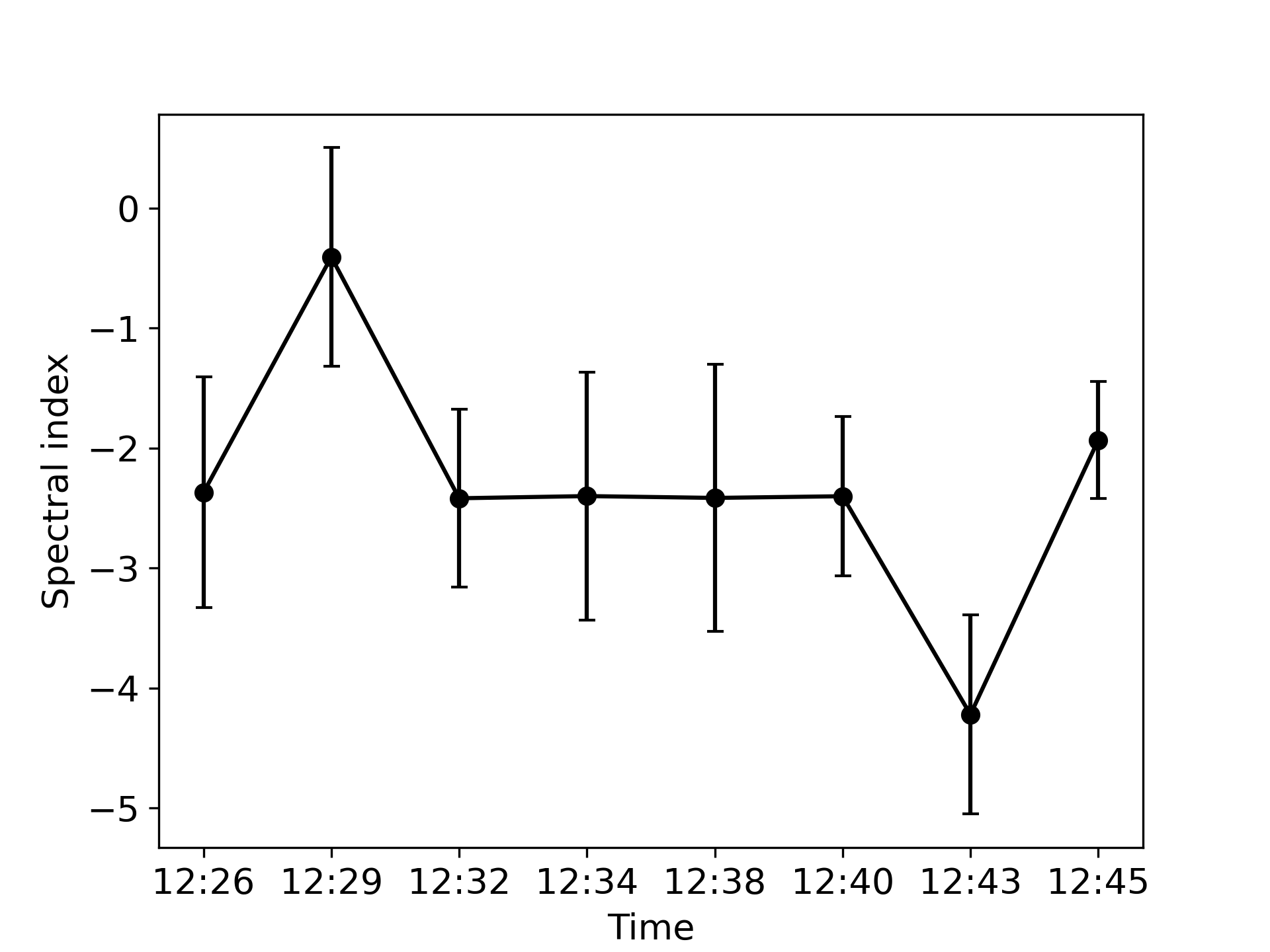}
         \caption{2019 April 25}
         \label{fig: v2400 oph spectral index 19-1}
     \end{subfigure}
     \begin{subfigure}{0.47\textwidth}
         \centering
         \includegraphics[width=\textwidth]{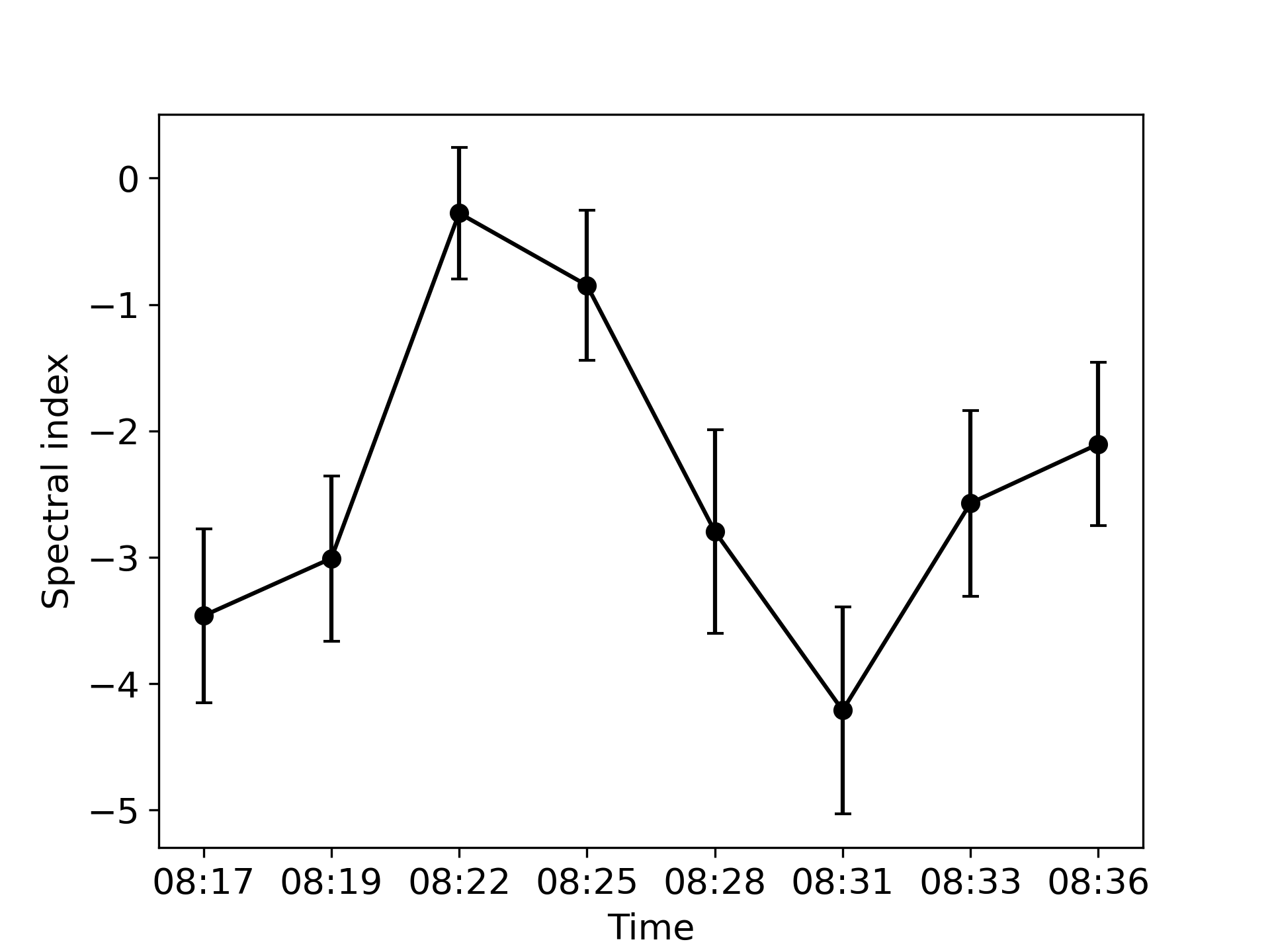}
         \caption{2019 May 11}
         \label{fig: v2400 oph spectral index 19-2}
     \end{subfigure}
        \caption{The spectral index of V2400 Oph on 2019 April 25 and 2019 May 11, produced by fits of a power law to the X-band dynamic SEDs in Fig. \ref{fig: v2400 19a}. In the first epoch \textit{(left)}, the spectral index is mostly stable, but in the second epoch \textit{(right)}, there is a clear evolution from flat to steep over $\sim$6 minutes.}
        \label{fig: v2400 oph spectral index 19}
\end{figure}


\begin{figure}[ht!]
    \centering

    \begin{subfigure}[b]{0.32\textwidth}
        \centering
        \includegraphics[width=\textwidth]{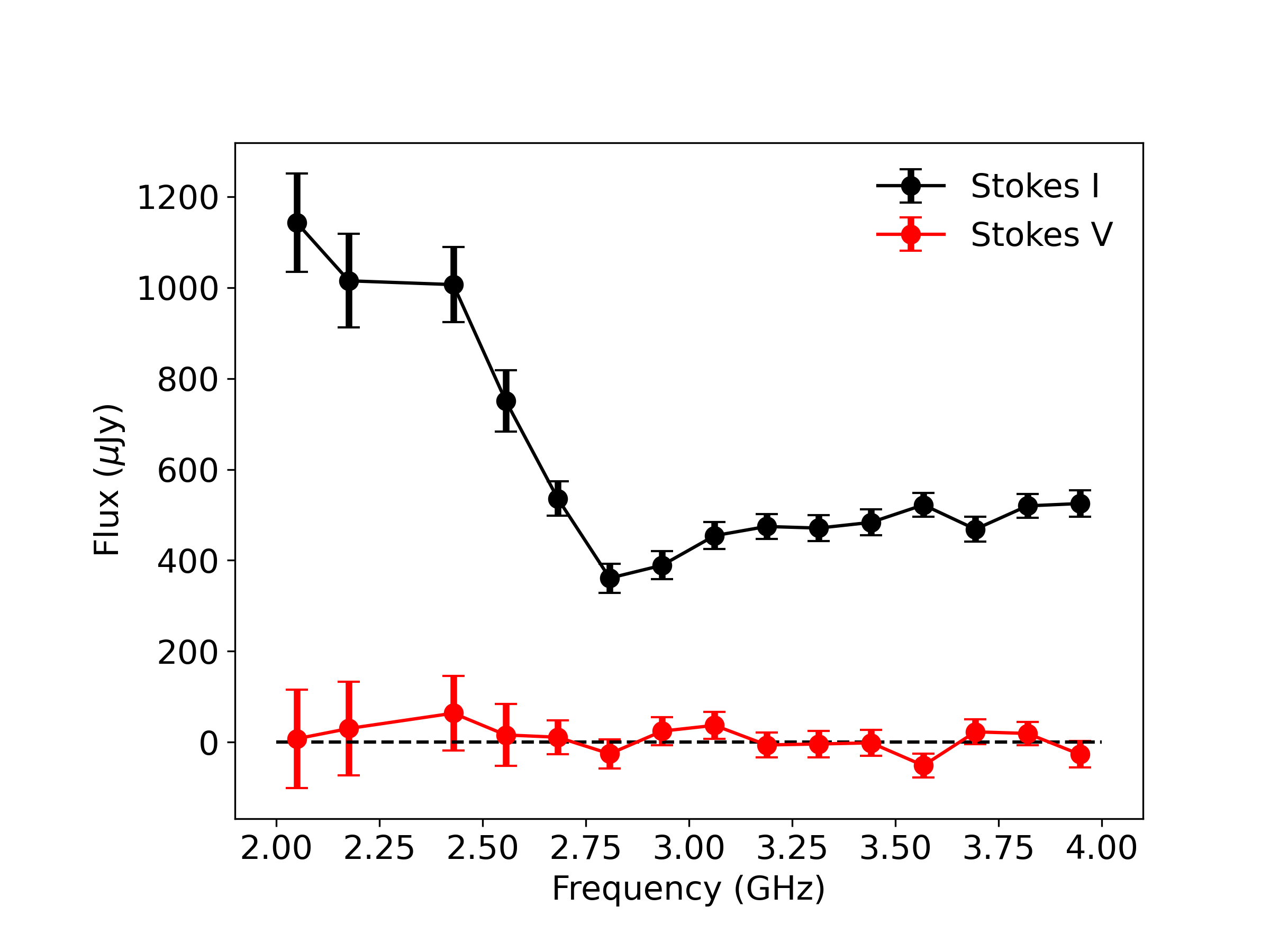}
        \caption{The 2--4 GHz SED of V2400 Oph on August 9.}
        \label{fig:1a}
    \end{subfigure}
    \hspace{0.2em}
    \begin{subfigure}[b]{0.32\textwidth}
        \centering
        \includegraphics[width=\textwidth]{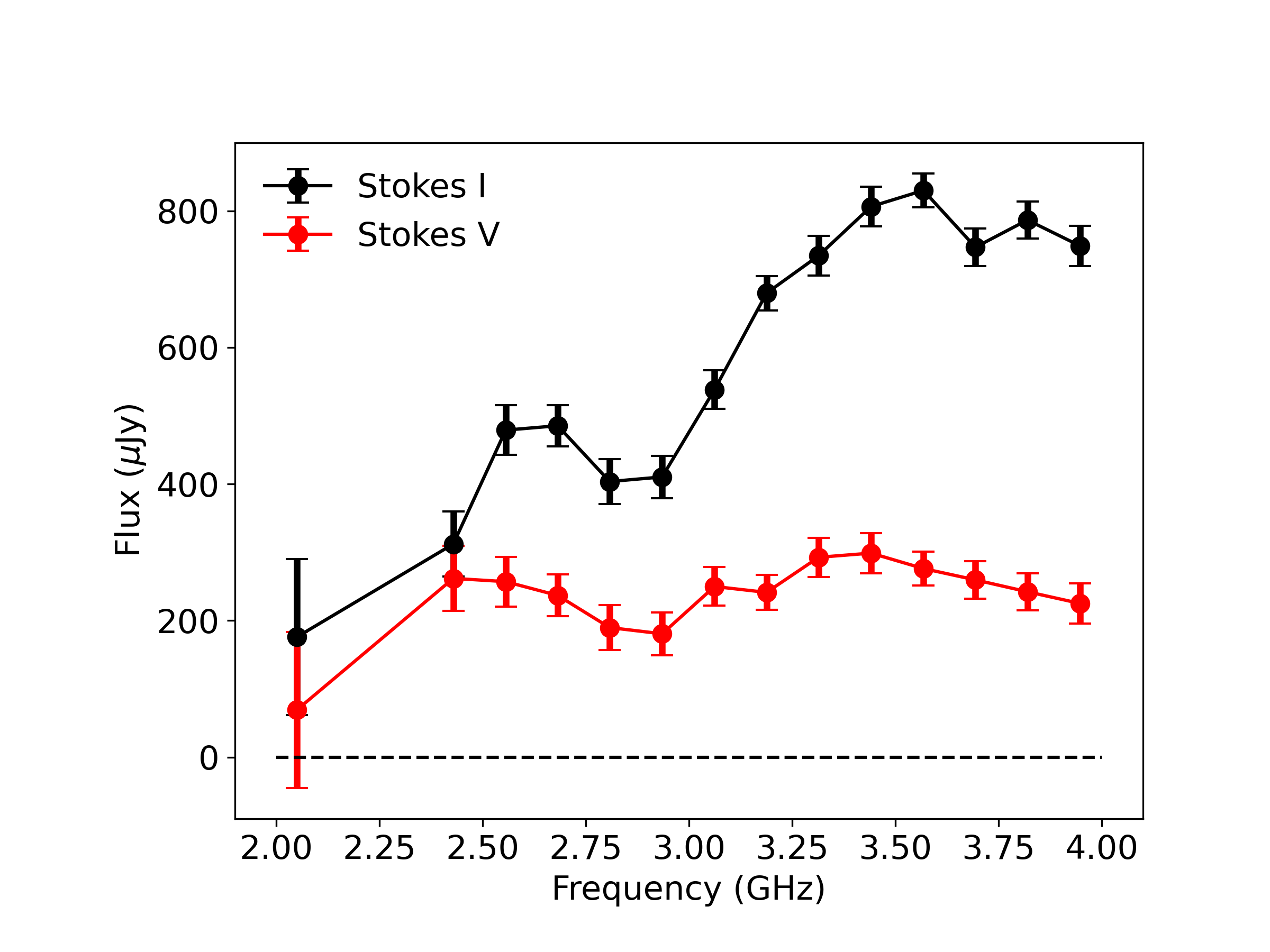}
        \caption{The 2--4 GHz SED of V2400 Oph on August 15.}
        \label{fig:1b}
    \end{subfigure}

    \vspace{0.5em}

    \begin{subfigure}[b]{0.32\textwidth}
        \centering
        \includegraphics[width=\textwidth]{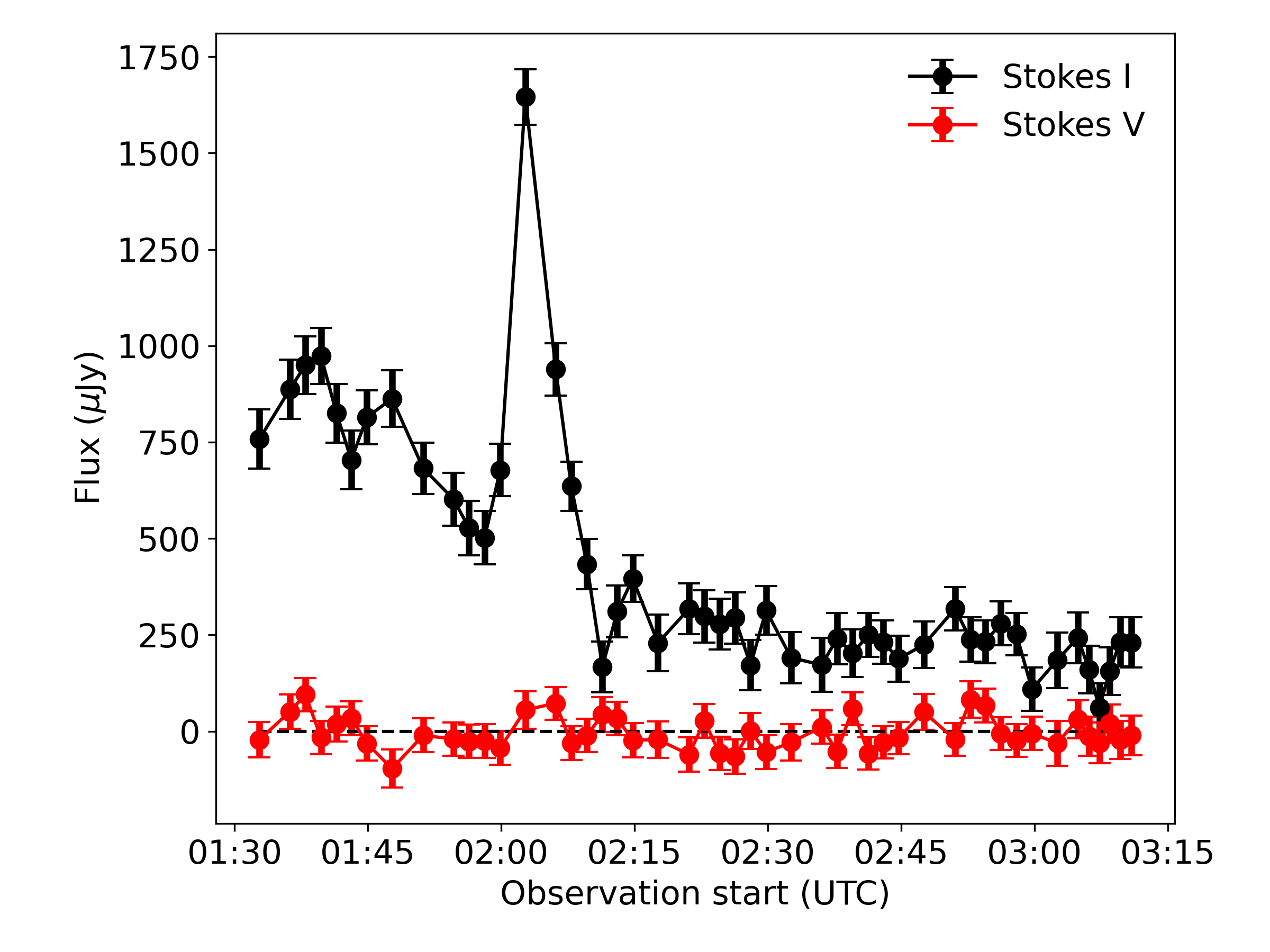}
        \caption{The 2--4 GHz light curve of V2400 Oph on August 9.}
        \label{fig:2a}
    \end{subfigure}
    \hspace{0.2em}
    \begin{subfigure}[b]{0.32\textwidth}
        \centering
        \includegraphics[width=\textwidth]{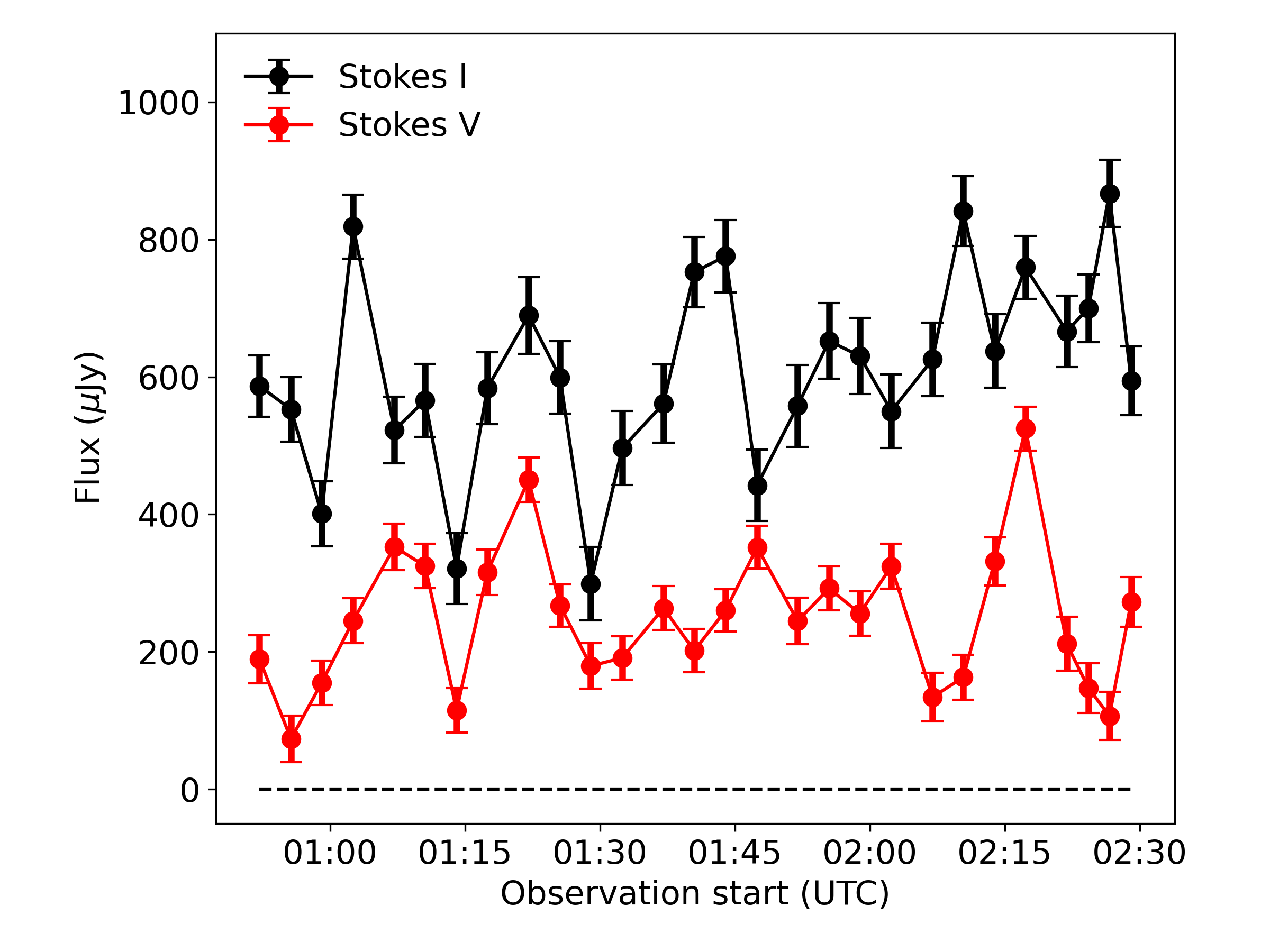}
        \caption{The 2--4 GHz light curve of V2400 Oph on August 15.}
        \label{fig:2b}
    \end{subfigure}

    \vspace{0.5em}

    \begin{subfigure}[b]{0.32\textwidth}
        \centering
        \includegraphics[width=\textwidth]{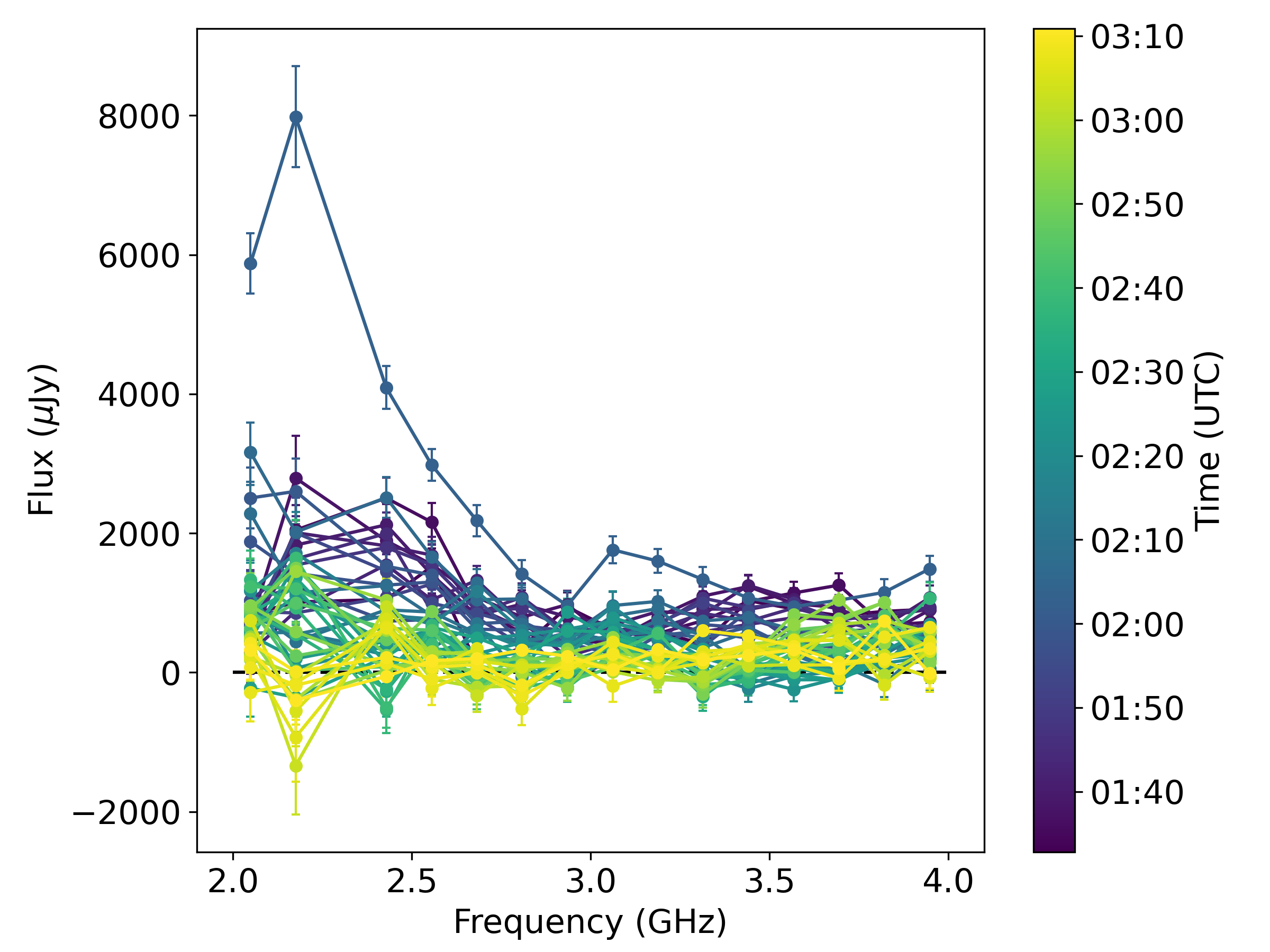}
        \caption{The 2--4 GHz dynamic SED of V2400 Oph in Stokes I on August 9.}
        \label{fig:3a}
    \end{subfigure}
    \hspace{0.2em}
    \begin{subfigure}[b]{0.32\textwidth}
        \centering
        \includegraphics[width=\textwidth]{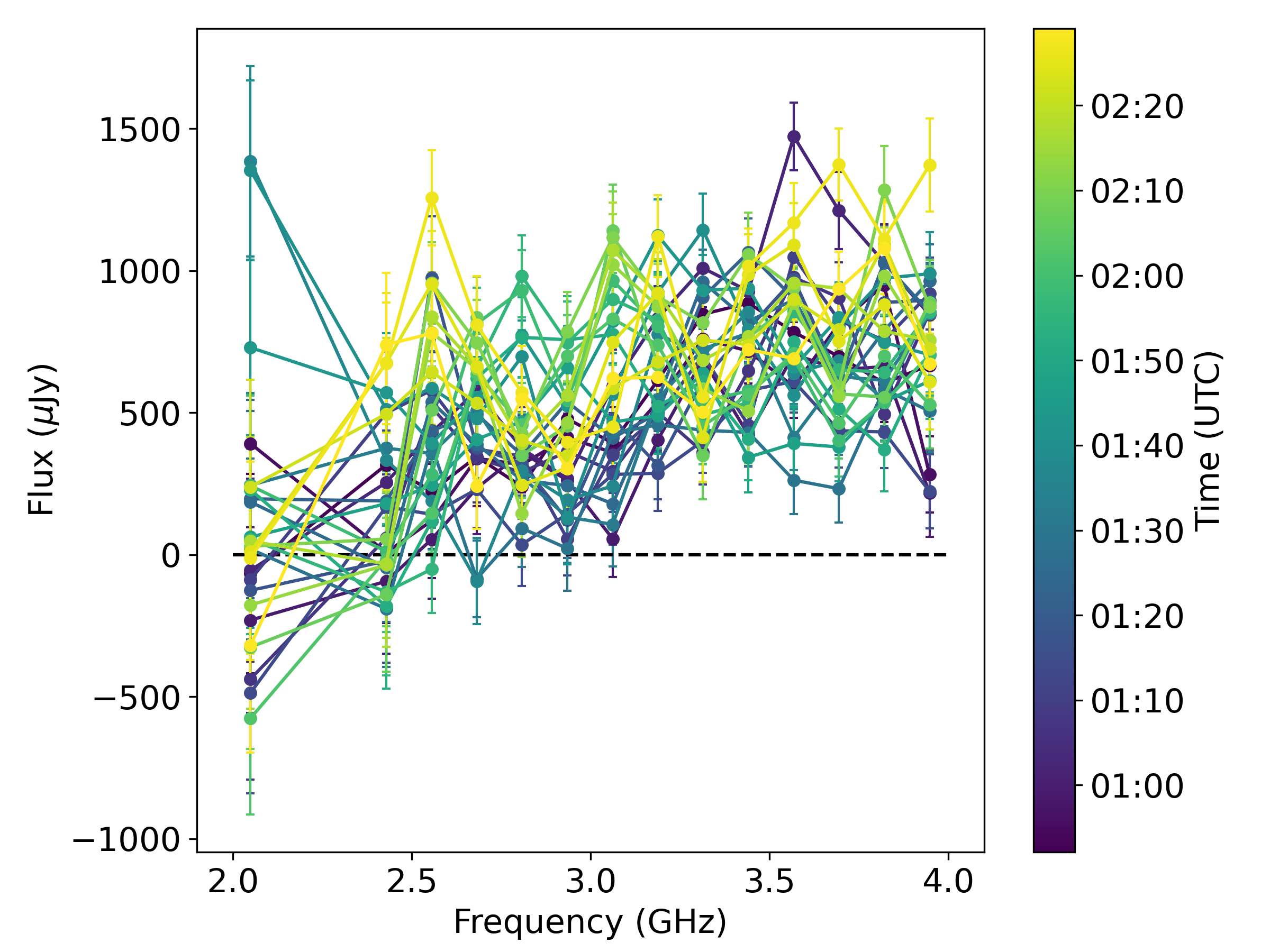}
        \caption{The 2--4 GHz dynamic SED of V2400 Oph in Stokes I on August 15.}
        \label{fig:3b}
    \end{subfigure}

    \caption{The 2--4 GHz data for V2400 Oph on 2025 August 9 (\textit{left column}) and 2025 August 15 (\textit{right column}). The top panels show the full-band SEDs, the middle panels show the full-band light curves, and the bottom panels show the dynamic SEDs in Stokes I. Each observation shows distinct behavior in V2400 Oph: at times emission can be unpolarized, but at others it can be moderately to highly circularly polarized.}
    \label{fig: v2400 25a}
\end{figure}

V2400 Oph was detected in Stokes I in both epochs of \textit{X}-band observations, but was undetected in Stokes V. Using a power law fit, assuming $S\propto\nu^\alpha$, we fit the SEDs to calculate the spectral index. Over the whole frequency band, the spectral index was $-2.5\pm0.3$ on 2019 April 25 and $-2.2\pm0.2$ on 2019 May 11. Despite some small variations (Fig.\ \ref{fig: v2400 19a}, it remained spectrally steep throughout the first observation. During the second observation, however, there is statistically significant variation in the spectral index (Fig.\ \ref{fig: v2400 oph spectral index 19}). At the point where V2400 Oph was brightest, the spectrum became almost flat and as it dimmed, the index dramatically steepened to $-4\pm1$. We searched for linear polarization, but found no source at the Stokes I position in Stokes Q or U.

On 2025 August 9, V2400 Oph was detected in Stokes I, but undetected in Stokes V, at 2-4 GHz. On 2025 August 15, it was detected in both Stokes I and Stokes V at 2-4 GHz(Fig.\ \ref{fig: v2400 25a}). The light curves appear to show a slow variation similar to the 8--12 GHz data, but also show bright, rapid flaring that is sometimes polarized.

\subsection{V603 Aql}


\begin{figure}[ht!]
    \centering

    \begin{subfigure}[b]{0.32\textwidth}
        \centering
        \includegraphics[width=\textwidth]{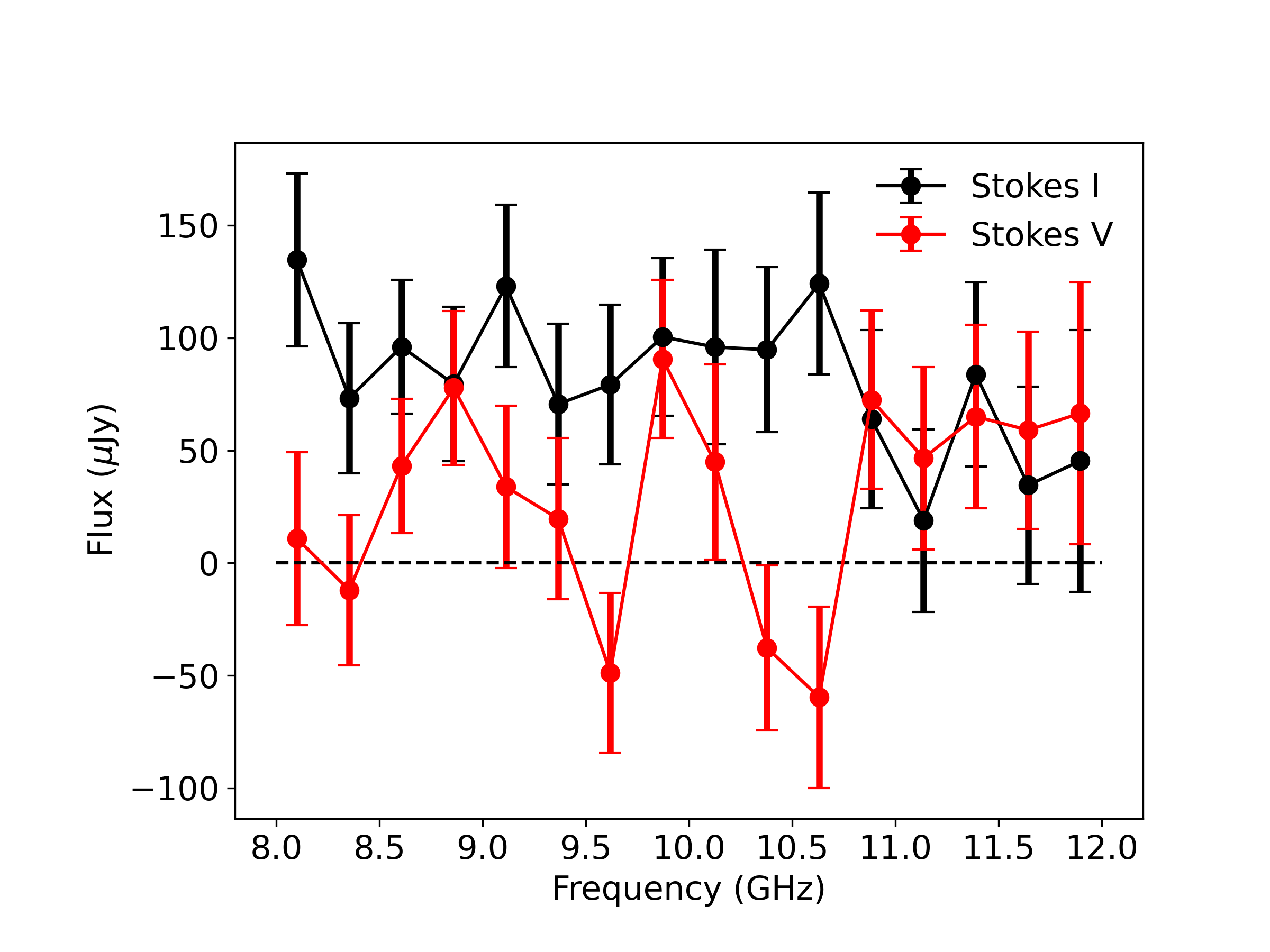}
        \caption{The 8--12 GHz SED of V603 Aql on December 18.}
        \label{fig:1a}
    \end{subfigure}
    \hspace{0.2em}
    \begin{subfigure}[b]{0.32\textwidth}
        \centering
        \includegraphics[width=\textwidth]{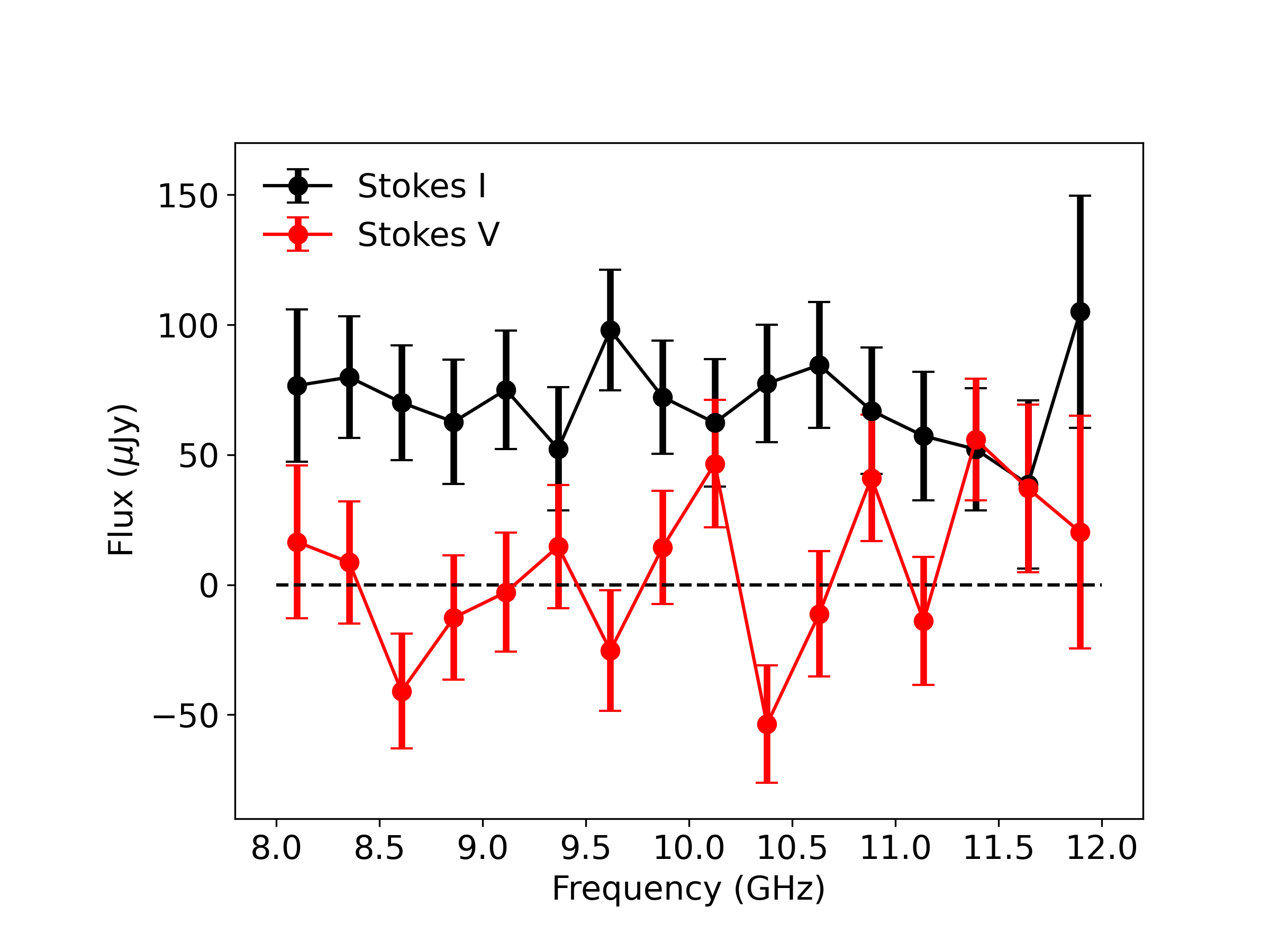}
        \caption{The 8--12 GHz SED of V603 Aql on January 25.}
        \label{fig:1b}
    \end{subfigure}

    \begin{subfigure}[b]{0.32\textwidth}
        \centering
        \includegraphics[width=\textwidth]{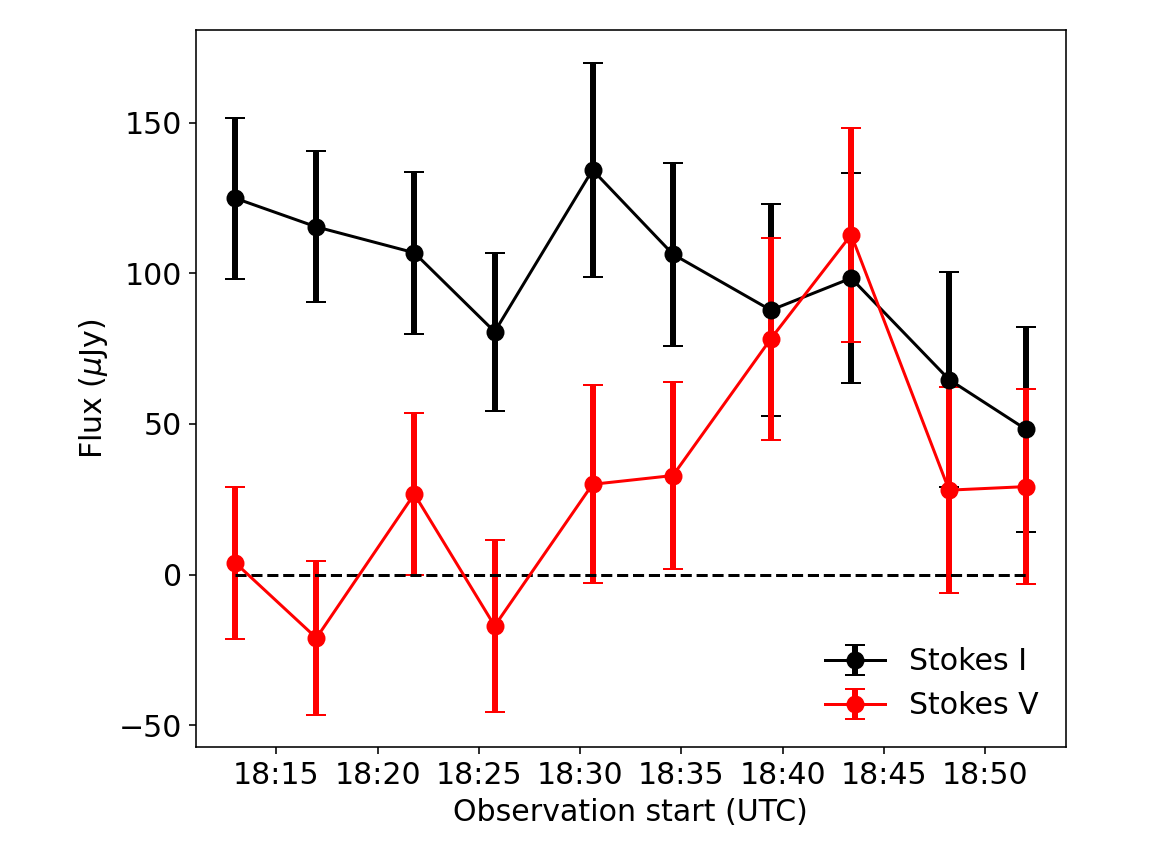}
        \caption{The 8--12 GHz light curve of V603 Aql on December 18.}
        \label{fig:2a}
    \end{subfigure}
    \hspace{0.2em}
    \begin{subfigure}[b]{0.32\textwidth}
        \centering
        \includegraphics[width=\textwidth]{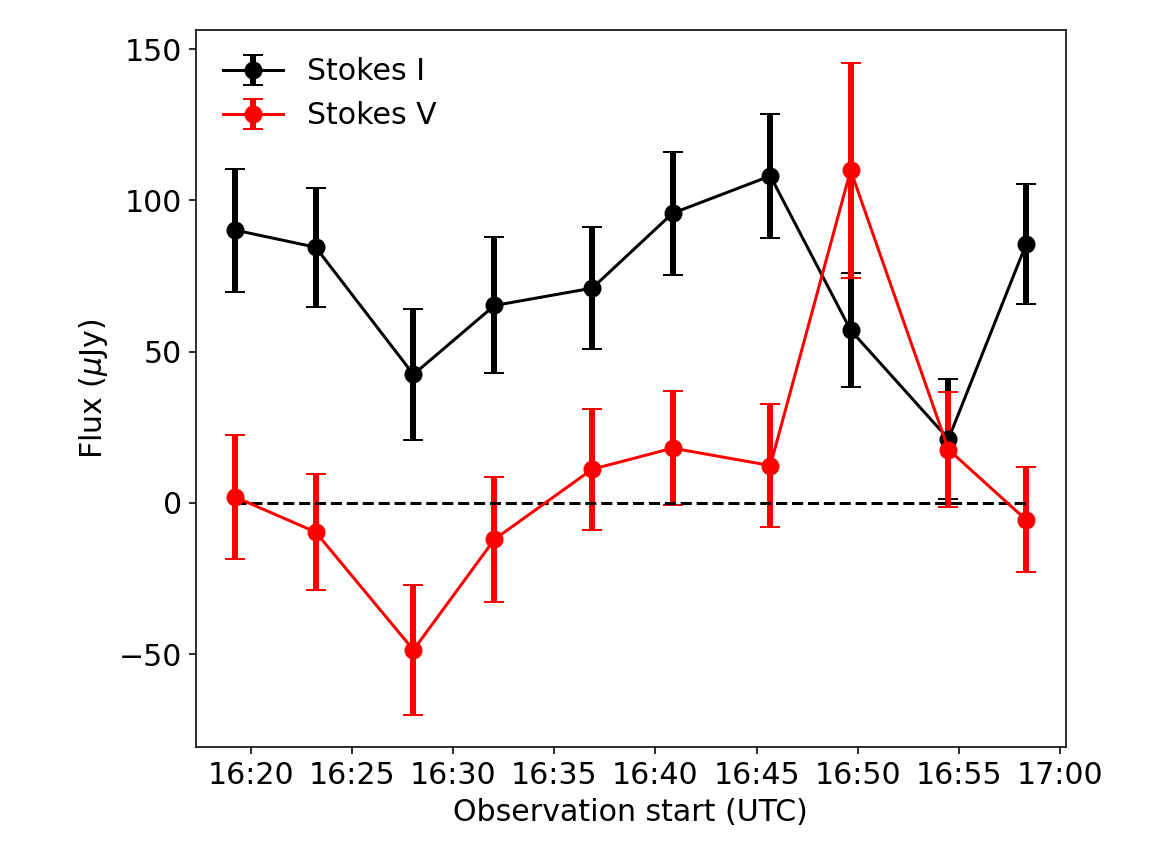}
        \caption{The 8--12 GHz light curve of V603 Aql on January 25.}
        \label{fig:2b}
    \end{subfigure}

    \caption{The 8--12 GHz data for V603 Aql on 2017 December 18 (\textit{left column}) and 2018 January 25 (\textit{right column}). The top panels show the full-band SEDs, while the bottom panels show the full-band light curves. There are no clear flares, but the light curves show slight variation.}
    \label{fig: v603 17b}
\end{figure}

Stokes V flux was only detected in the first epoch of X-band observations for V603 Aql. While there is some variation in the SED (Fig.\ \ref{fig: v603 17b}), there is a slight tendency for it to be right circularly polarized. However, there are no clear emission features that could help identify the emission mechanism in either observation. The light curves hint at a decrease in overall flux, but an increase in fractional circular polarization near the end of each observation. The highest Stokes V flux points are higher than the simultaneous Stokes I fluxes, which is impossible. However, because the simultaneous Stokes I and V fluxes are consistent with each other within their uncertainties, we can conclude that they are consistent with 100\% RCP at this time in the observation, but also consistent with 0\% at $2\sigma$ (the errors are 50\% and 90\% for the first and second observation, respectively).

\section{Discussion}

\begin{table*}
    \setlength{\tabcolsep}{1.5pt}
    \renewcommand{\arraystretch}{1.1}
    \centering
    \caption{The non-flaring flux ranges for all targets except for V2400 Oph. For observations of EF Eri and MR Ser that captured a flare, the 8--12 GHz fluxes surrounding the flare were used to calculate the mean flux. We present the fluxes alongside some relevant binary system parameters that may affect the radio emission (e.g. accretion rate). Some accretion rates were excluded if the literature did not have an estimate that corresponded with the accretion state of each target when the radio observations were taken. To the knowledge of the authors, there is currently no spectral type model for the donor V603 Aql. References for the final column are as follows: [1] \citet{beuermann2000}, [2] \citet{beuermann2007}, [3] \citet{ferrario1993}, [4] \citet{khangale2020}, [5] \citet{rousseau1996}, [6] \citet{ritter_kolb}, [7] \citet{sanad2015}, [8] \citet{schwope1993}, [9] \citet{sion2015}, [10] \citet{suleymanov1992}.}
    \label{tab: non-flare flux}    
    \begin{tabular}{c c c c c c c c}
    \hline \hline
         \rule{0pt}{2.5ex} 
         Name & $\text{P}_\text{orb}$ & Flux & Luminosity & WD $|\vec{B}|$ & $\dot{M}$ & Donor & Refs. \\
          & & & $\times10^{30}$ & & & & \\
          & (hr) & ($\mu$Jy) & (erg/s/cm$^2$) & (MG) & ($\text{M}_\odot / \text{yr}$) & &  \\
        \hline 
         EF Eri & 1.35 & 29--48 & 8.9--15 & 44 & 3$\times10^{-11}$ & L4--5 & [6], [1], [2] \\ 
         UZ For & 2.11 & 30--35 & 20--24 & 57 & $-$ & M4.5/5 & [5], [6], [4] \\ 
         ST LMi & 1.90 & 40--80 & 6.2--12 & 12 & (6--200)$\times10^{-14}$ & M6/5 & [6], [7], [3] \\ 
         MR Ser & 1.89 & 42--50 & 6.4--7.6 & 27 & $-$ & M8/5 & [6], [8] \\ 
         V603 Aql & 3.32 & 71--90 & 84--110 & $-$ & $\sim2\times10^{-9}$ & $-$ & [6], [9], [10] \\ 
         \hline
    \end{tabular}
\end{table*}

There is a wide variation in the radio emission from this sample of CVs. Notably, we saw three bright, highly circularly polarized flares from EF Eri and MR Ser at 8--12 GHz. At 2--4 GHz, we observed one circularly polarized flare from MR Ser, 2 polarized flares from ST LMi, as well as 2 unpolarized flares and rapidly varying polarized emission from V2400 Oph. We have also seen faint, highly circularly polarized candidate flares from UZ For and ST LMi at 8--12 GHz. 

The polarization fractions at the peaks of the \textit{X}-band flares from EF Eri and MR Ser ranged from 79--96\%, which suggests that a coherent emission process is involved, similar to that observed in \textit{S}-band observations of QS Vir \citet{ridder2025}. The peak emission frequency for MR Ser ($\sim$10 GHz) corresponds to a magnetic field strength of a few kG, appropriate for an M dwarf star similar to the donor star. ECME is typically restricted in frequency in the non-relativistic limit, so the broadband nature of the emission (see Figs. \ref{fig: eferi 17b} and \ref{fig: mrser 17b}) seems to argue against ECME as the origin of the radio emission. However, if there were multiple emission regions, each with a different magnetic field strength (e.g. along a flux tube), covering a range of several kG, ECME could reproduce the spectral behavior seen in the SEDs shown above. Unlike in the Jupiter-Io system, the motion powering the flux tube in a CV with a tidally locked donor would need to be the spin of the WD. In polar systems, the WD's spin is synchronized to the orbital period, making this scenario unlikely for most of our targets. Because the frequency of emission and the bandwidth are similar, we should consider the possibility that plasma radiation is responsible. This too should generate high fractions of circular polarization, but unlike weakly relativistic ECME, can be broadband, and the duration should be less than an hour \citet{callingham2024}.

Assuming all the above flares are coherent (as indicated by the circular polarization fraction), here we estimate an upper limit on the size of the emission region in the case of incoherent emission.
Using the form presented in \citet{vedantham2020}, the brightness temperature is,

\begin{equation}
    T_b = \frac{F c^2 d^2}{2 k_B \nu^2 \pi R^2},
\end{equation}

\noindent where $F$ is the flux density, $d$ is the distance to the object, $k_B$ is Boltzmann's constant, and $R$ is the size of the emitting region. We will assume that the brightness temperature for each flare is $\gtrsim10^{12}$ K since this is an approximate transition from incoherent to coherent emission. For EF Eri on 2017 September 28, $R\leq5\times10^9$ cm, and for 2017 October 2, $R\leq2\times10^9$ cm. For MR Ser on 2017 November 5, $R\leq4\times10^9$ cm. For ST LMi on 2025 March 23, the first flare corresponds to $R\leq7\times10^9$ cm and the second flare corresponds to $R\leq8\times10^9$ cm. Assuming that the size of a typical M dwarf is $\sim10^{10}$ cm, all of the flares we detected must be produced in regions that are less than the size of the donor star, ranging from about 1/5 to 4/5. Note that this does not rule out a region of similar size in between the WD and the donor star or near the WD itself.

The behavior of V2400 Oph is perhaps more difficult to explain. The steep spectral shape and low circular polarization in \textit{X}-band (Fig. \ref{fig: v2400 19a}) and \textit{S}-band on 2025 August 9 (Fig. \ref{fig: v2400 25a}) would suggest synchrotron emission, which could be produced by a jet launched by the WD in a similar way to SS Cyg \citet{Coppejans20}. Although we did not detect any linear polarization in any frequency band, this is not strong evidence against synchrotron emission. The steep spectral indices, however, indicate an optically thin synchrotron emission region, arguing against a compact jet. Another argument against a compact jet is the accretion geometry in V2400 Oph, which is hypothesized to undergo diamagnetic blob accretion, in which small blobs of material are accreted in short bursts \citet{langford_2022_blob}. We hypothesize that the electron population is accelerated at the interaction site with the WD's  rotating magnetic field, leading to the production of synchrotron radiation. Optical observations conducted by the Kepler K2 mission \citep{langford_2022_blob} show stochastic variation on the scale of $<$1 ks that has been attributed to interaction between blobs and the WD's magnetic field. We also see variations in the V2400 Oph VLA observations on timescales of $\sim$300 seconds (Figs.~\ref{fig: v2400 19a}). The evolution of the spectral index on 2019 May 11 may be explained by an initially energetic electron population that emits at higher frequencies, resulting in a flatter spectrum. At later times, the spectrum would steepen as the higher-energy electrons lose energy. This argument should not be taken as definitive, since the lowest spectral index of $-4\pm1$ corresponds to a physically implausible electron population spectral index of $-10$.

In diamagnetic blob accretion, blobs that gain enough energy from the WD's magnetic field may be flung out of the system, or can be flung back into the Roche lobe of the donor star \citet{langford_2022_blob}. In this way, V2400 Oph behaves in a somewhat similar way to the propeller system, AE Aqr, which is thought to expel material more energetically \citet{Wynn97}. Other IPs with an intact accretion stream (e.g. GK Per or DO Dra) that were previously observed by \citet{Barrett2017} remain undetected in the radio, despite being at a similar distance as V2400 Oph (700 pc). The way in which the WDs in V2400 Oph and AE Aqr interact with material from their donor stars may therefore drive the bright radio emission in these systems, which largely lacks the characteristic circular polarization seen from several other radio-bright CV systems. 

\citet{meintjes2003} assume a similar interaction with blobs for AE Aqr. Using their values for electron $\gamma$ (20) and magnetic field inside a blob (3 kG)\footnote{This estimate is the approximate surface magnetic field of the donor star. The magnetic field in a blob may be different.}, we obtain a synchrotron cooling time of

\begin{equation}
    t = \frac{5.16\times10^8}{B^2\gamma} = 3\text{ s}.
\end{equation}

\noindent Lowering $\gamma$ to $\sim$1 yields a cooling time of about a minute, which is similar to the minute-scale variation we see in Fig. \ref{fig: v2400 19a}. Since the spin period of the WD likely plays a role in the acceleration and V2400 Oph \citep[927 s; ][]{langford_2022_blob} has a much slower spin period than AE Aqr \citep[33 s; ][]{Wynn97}, an electron population with a lower average energy would make sense. If we compare the light curve of V2400 Oph at 8--12 GHz with that at 2--4 GHz (Fig. \ref{fig: v2400 25a}), the shorter timescale in the former vs. the latter seems to correspond to predictions of the van der Laan model \citep{vanderlaan1966}, in which initially optically thick blobs expand. In this model, the expansion causes the frequency at which the material is optically thick to decrease, allowing for faster variation in the light curve at the higher frequencies that are optically thin.

Synchrotron emission is proposed to be responsible for the radio emission from AE Aqr \citep{Bastian88, barrett2022}, but this is still debated \citep{jiang2024}. Indeed, the sharp rise in the C-band SED, of spectral index $\sim$0.7, presented in \citet{barrett2022} suggests the low frequency cutoff of a self-absorbed synchrotron spectrum \citep[e.g. ][]{Rybicki79}. The \textit{X}-band portion of the SED appears to decrease with an average spectral index of approximately $-1$. We see an even steeper (spectral index $-2.2$ to $-2.5$) spectral shape in V2400 Oph in \textit{X}-band. We do not, however, observe the discontinuity shown in AE Aqr's \textit{X}-band SED by \citet{barrett2022}, where the lower subband is spectrally flat at $\sim$5.75 $\mu$Jy and the upper subband is also spectrally flat but $\sim$5 $\mu$Jy. In K-band (though we note this was not taken simultaneously, \citet{barrett2022}), the flux continues to rise ($\alpha \sim 2$), which matches the simulated spectrum for AE Aqr produced by blob shocking by \citet{meintjes2003}. This spectrum was produced by simulating 10 blobs interacting with AE Aqr's magnetosphere. The differences between the SEDs of V2400 Oph and AE Aqr may be explained by differences in the properties of each system, including the WD spin period and magnetic field strength.

While this discussion has mainly been focused on the long-term variation in V2400 Oph, it is not clear whether we can apply the same explanation to the short-term variation. On the one hand, the bright flare starting at 2:00 UTC on 2025 August 9 also has a steep spectral shape apparent in the dynamic SED (Fig. \ref{fig: v2400 25a}) that is comparable to the shape corresponding to the slow-varying component (see the peak centered on 2.5 GHz). We may, in this case, be able to explain it with synchrotron emission. However, variation of a similar timescale in \textit{S}-band and \textit{X}-band does not agree with our application of the van der Laan model above as we would expect the timescale to be longer in \textit{S}-band. More confusing is the rapid variation in the 2025 August 15 data that at times is significantly circular polarized, and at other times is not. This behavior suggests that there are multiple emission mechanisms present in V2400 Oph, but it would be difficult to disentangle them. There may be one mechanism that produces circular polarization and one that does not or perhaps two circularly polarized sources of opposite handedness. Or perhaps a single mechanism which can either show circular polarization, or not. 

Of the other sources that showed mildly variable or non-variable, broadband, radio emission without any flaring behavior or pronounced circular polarization, gyrosynchrotron emission by moderately relativistic electrons seems appropriate. Here we expect a flat spectrum, like what is shown in many of the SEDs in Section \ref{res p3}, and moderate levels of circular polarization, depending on the viewing angle \citet{dulk1985}. For most observations, we do not have sufficient S/N ratio to identify measure the degree of circular polarization. However, for ST LMi on 2017 November 1 and V603 Aql on 2018 January 25, we can constrain the fractional circular polarization to $\leq30$\%, which is consitent with gyrosynchrotron emission \citet{callingham2024}.

We compare the non-flaring luminosities of all targets except V2400 Oph in Table \ref{tab: non-flare flux}. These are presented alongside the spectral type of the donor, the mass accretion rate, and the magnetic field of the WD, allowing for identification of potential correlations. The accretion rate was chosen by first estimating whether it was in a high or low state using optical observations by the American Association of Variable Star Observers (AAVSO) and then searching for an estimate of the accretion rate in the literature that corresponded with that state. Among the polars, there is no clear correlation between the intrinsic luminosity and the other values presented. However, V603 Aql has a radio luminosity $\sim10\times$ brighter and a mass accretion rate $\sim100\times$ larger than the rest of the sample. We suggest that radio emission in V603 Aql may be related to its accretion and hence it would be a good candidate for testing the jet hypothesis for CVs.

\section{Conclusions}

We infer a general paradigm for the radio emission from most of the CVs we have studied here, with the notable exception of V2400 Oph. Outside of radio flares, these CVs are clearly detected in Stokes I with fluxes of 40--80 $\mu$Jy, with circular polarization $\lesssim$30--70\%. We tentatively attribute this persistent radio emission to gyrosynchrotron emission, due to their flat spectra and low levels of circular polarization. The flaring emission seen from EF Eri, ST LMi, and MR Ser (and possibly, at lower levels, from UZ For and ST LMi at 8--12 GHz), is much better described by a coherent plasma process such as plasma radiation or ECME. Due to the broadband features in their SEDs and fractional circular polarization of 79--96\%, we can explain this emission either via plasma radiation, or from multiple regions (e.g. along a flux tube between the WD and donor star) that each produce ECME at their local electron gyrofrequencies. The steep radio spectrum and low circular polarization of V2400 Oph seem to indicate that synchrotron emission is responsible for its radio emission, but we note that the spectral index is exceptionally steep, meaning synchrotron may not be an appropriate explanation. The radio emission from V2400 Oph may be due to an interaction between blobs of material orbiting the WD and the WD's strong magnetic field. The time scale on which we see variation in its radio light curve and spectral index is plausibly consistent with a simple expanding-blob model. The similarities between V2400 Oph and AE Aqr suggest that the emission source is common to both systems.

\begin{acknowledgements}
The authors would like to thank Gregg Hallinan for providing detailed discussions and feedback on the results of this work. This project was supported in part by an appointment to the NRC Research Associateship Program at the U.S. Naval Research Laboratory (NRL), administered by the Fellowships Office of the National Academies of Sciences, Engineering, and Medicine. Basic research at the NRL is supported by 6.1 Base funding. C. O. Heinke is supported by NSERC Discovery Grant RGPIN-2023-04264.
\end{acknowledgements}

    

\facilities{VLA, AAVSO}

\software{CASA \citep{casa},  
          WSClean \citep{wsclean},
          breizorro \citep{breizorro}
          }

\bibliography{refs}
\bibliographystyle{aasjournalv7}

\end{document}